\documentclass[aps, prd,10pt,notitlepage,nofootinbib,superscriptaddress,showkeys]{revtex4-1}
\usepackage{amstext,amsmath,amssymb,amsfonts,amsthm,bbm}
\usepackage[latin1]{inputenc}
\usepackage{graphicx}
\usepackage{hyperref}
\usepackage{pstricks}

\usepackage{color}

\def\beq{\begin{equation}}
\def\be{\begin{equation}}
\def\ee{\end{equation}}
\def\bes{\begin{eqnarray}}
\def\ees{\end{eqnarray}}

\DeclareMathOperator{\Inv}{Inv}
\DeclareMathOperator{\tr}{tr}

\DeclareMathOperator{\Ad}{Ad}

\DeclareMathOperator{\id}{id}
\DeclareMathOperator{\SU}{SU}
\DeclareMathOperator{\SL}{SL}
\DeclareMathOperator{\SO}{SO}

\DeclareMathOperator{\U}{U}

\newcommand{\su}{\mathfrak{su}}

\renewcommand{\u}{\mathfrak{u}}

\newcommand{\unit}{\mathbbm{1}}
\newcommand{\hE}{\widehat{E}}
\newcommand{\hF}{\widehat{F}}

\def\f{\frac}

\def\mone{^{-1}}
\def\tl{\widetilde}

\def\eps{\epsilon}

\def\C{{\mathbbm C}}
\def\N{{\mathbbm N}}
\def\Z{{\mathbbm Z}}

\def\calH{{\mathcal H}}

\theoremstyle{definition}
\theoremstyle{definition}\newtheorem{theorem}{Theorem}
\theoremstyle{definition}\newtheorem{lemma}{Lemma}
\theoremstyle{definition}
\theoremstyle{definition}
\theoremstyle{definition}
\newtheorem{proposition}{Proposition}

\begin{document}

\title{\large \bf A new Hamiltonian for the Topological BF phase with spinor networks}

\author{{\bf Valentin Bonzom}}\email{vbonzom@perimeterinstitute.ca}
\affiliation{Perimeter Institute for Theoretical Physics, 31 Caroline St. N, ON N2L 2Y5, Waterloo, Canada}
\author{{\bf Etera R. Livine}}\email{etera.livine@ens-lyon.fr}
\affiliation{Laboratoire de Physique, ENS Lyon, CNRS-UMR 5672, 46 All\'ee d'Italie, 69007 Lyon, France EU}
\affiliation{Perimeter Institute for Theoretical Physics, 31 Caroline St. N, ON N2L 2Y5, Waterloo, Canada}

\begin{abstract}\noindent
We describe fundamental equations which define the topological ground states in the lattice realization of the $\SU(2)$ BF phase. We introduce a new scalar Hamiltonian, based on recent works in quantum gravity and topological models, which is different from the plaquette operator. Its gauge-theoretical content at the classical level is formulated in terms of spinors. The quantization is performed with Schwinger's bosonic operators on the links of the lattice. In the spin network basis, the quantum Hamiltonian yields a difference equation based on the spin 1/2. In the simplest case, it is identified as a recursion on Wigner 6j-symbols. We also study it in different coherent states representations, and compare with other equations which capture some aspects of this topological phase.





\end{abstract}

\maketitle

\section*{Introduction}

\subsection*{A practical guide through the equations of the topological BF phase}

Topological order \cite{wen-top-orders} is a recent key advance in our understanding of the phases of matter. It describes new phases in which ground states are characterized by the space(time) topology, irrespective to its metric properties, so that the physical degrees of freedom are only global. In particular, there is no local order parameter, but new, global characteristics, like the fact that the ground state degeneracy is determined by the topology. Topological phases have appeared in several condensed matter systems, like those exhibiting fractional quantum Hall effect \cite{wen-niu-fqhe, frohlich-kerler}, quantum spin liquids \cite{wen-spin-liquid91, kalmeyer-fqhe, wen-wilczek-zee, wen-spin-liquid02}, and also topological insulators \cite{hasan-colloquium, cho-bf-insulators} and topological superconductors \cite{hansson-superconductors, top-superconductors}. Among the new physical features, topological order is intimately related to anyons and exotic statistics.

Topological order can be generically framed into the language of topological gauge field theories describing the effective behavior, i.e. the low energy regime. Topological field theories are defined without using a metric and do not have propagating degrees of freedom. The most well-known is Chern-Simons theory, which is defined in $2+1$ dimensions. There is another similar theory, known as BF theory, which is defined in arbitrary dimensions\footnote{In 2+1 dimensions, it is a Chern-Simons theory. In particular, when the local symmetry group is $\SO(2,1)$, it is a Chern-Simons theory based on the Poincar\'e group.}, and also relies on special properties of the set of flat connections. It is actually quite well understood from the initial papers, by Horowitz \cite{horowitz-bf} and by Blau and Thompson \cite{blau-thompson-bf}. Further works interested in the symmetries and proving the renormalizability can be found in \cite{cattaneo-3d-4dBF, maggiore-sorella-perturbative-4dbf,symmetries-bf-bv, renormalisability-bf}.

After gauge-fixing, the energy-momentum tensor is in fact BRST-exact. Moreover, the space of non-gauge-equivalent solutions is the moduli space of flat connections, whose dimension is a topological invariant. Hence, there does exist a non-local order parameter determined by Wilson loops which are trivial except around non-contractible cycles in the BF phase.


The BF model first appeared in two dimensions, as the zero coupling limit of Yang-Mills theory \cite{witten-2dym}. Actually that relationship with Yang-Mills theory holds in any dimensions, and it was studied in four dimensions, in the continuum, in \cite{cattaneo-4dYM} (a long time after that phase was observed in lattice gauge theory). The BF TFT appears in relations to other interesting field theories. In three dimensions with the appropriate gauge group, it describes three-dimensional pure gravity, with or without cosmological constant, making it clear it is exactly soluble \cite{witten-3d}. Even better, the BF Plebanski actions \cite{freidel-depietri} offer a way to build four-dimensional general relativity as a BF theory supplemented with constraints. The latter are known as simplicity constraints and impose the non-Abelian electric field to come from a tetrad. Four-dimensional general relativity again can be formulated \`a la MacDowell-Mansouri, which is a BF theory for a larger gauge group together with a potential which explicitly breaks down the (topological) gauge symmetries to those of gravity \cite{bf-mdm}.

In addition to the effective description of condensed matter systems, and to its relations to the above field theories, topological orders provide us with a new way to think of fault-tolerant quantum computations \cite{dennis-top-memory}. The most celebrated example is the Kitaev code \cite{kitaev-code} which is based on a lattice version of the BF model with group $\Z_2$. It is also a model of importance to test background independent methods in arbitrary dimensions (aiming at a consistent quantization of gravity), and in particular the spin network quantization \cite{baez-spinnets, baez-intro-bf, perez-intro-lqg}.

Spin networks are excitations supported on graphs on the spatial manifold, whose links are colored with spins (or representations of a group) and nodes with intertwiners. In the context of Loop quantum gravity, they enable to make sense of the quantum geometry of space \cite{ashtekar-status-report}. They are the same as the string networks of \cite{levin-wen-condensation} which describe the condensation process leading to topological orders.

It is known (see for instance \cite{spinnets-marzuoli}), and we will further argue, that spin networks are key objects to get a global view on topological aspects which are observed in condensed matter, quantum information and background independent methods. For all of them, it is important to have a lattice description, and it is well-known that spin networks are well adapted to the BF model since they support an exact lattice version, at least in 2+1 dimensions \cite{freidel-louapre-PR1, barrett-naish, 3d-wdw}. Furthermore, it is a key result in loop quantum gravity that spin network states also span the kinematical Hilbert space of the continuum theory when one varies their graphs. Hence, spin networks will be central to our analysis.


The BF model is often studied with a finite and/or Abelian group (see \cite{baby-sf} for a presentation of the background independent point of view in the context of finite groups), or a quantum group. Here, we will focus on the (more) difficult case of $\SU(2)$, which is non-Abelian and in which the Fourier modes are unbounded. In spite of the difficulty, we will find that the topological order is encoded and identified by some well-known equations of representation theory, which can be handled explicitly by physicists, and which are free from open mathematical issues.

As the BF theory is a theory of flat connections, the objects which arise naturally in our quantization are re-coupling coefficients of the representation of theory of $\SU(2)$. These objects, together with spin networks themselves, have enjoyed a renewed interest in the last years, in loop quantum gravity, but also in mathematics \cite{costantino-generating, garoufalidis}, in quantum information \cite{spinnets-marzuoli}, and in semi-classical physics (see references in \cite{qm6j}). In particular, one challenge is the understanding of their large spin asymptotics. In addition to specific coefficients \cite{roberts, barrett-asym-summary, 6jnlo, pushing6j}, recent works have unraveled generic methods and asymptotic behaviors \cite{3nj-marzuoli, 3njsmall, Yu, dowdall-handlebodies} which have nice interpretations in terms of quantized flat simplices.

In addition to asymptotic approximations, exact results are interesting. A way to capture them is by using recursion relations on those re-coupling coefficients (which also provide a way to get the asymptotics \cite{SG1, 6jmaite}). As shown in \cite{recurrence-paper, recursion-semiclass, yetanother}, they give a way to probe geometric properties of quantized simplices. In this paper, by extending the seminal result of \cite{3d-wdw}, we show how to get all (well-known, \cite{varshalovich-book}) recursions on the Wigner 6j-symbol from a Hamiltonian for the topological BF phase, and present alternative forms in different bases. At the end of the day, we obtain a practical guide through the equations which describe and characterize that topological behavior.

\subsection*{A new Hamiltonian for the magnetic part}

Our analysis is based on a new lattice Hamiltonian. The usual Hamiltonian \cite{kitaev-code, levin-wen-condensation, noui-perez-ps3d} consists in two parts, a vertex operator and a plaquette operator. The equation for the ground state leads to two conditions: the vertex operator imposes the standard gauge invariance, and violations lead to electric charges, the plaquette operator imposes the curvature to vanish, and violations correspond to magnetic fluxes. We work with gauge invariant spin networks and hence only focus on the second constraint. The new Hamiltonian we introduce imposes flatness of the gauge field (hence the topological order) and has the following new features.
\begin{itemize}
 \item It is labeled not by a plaquette only but by a pair of a plaquette and a vertex on its boundary. We thus have several operators and hence constraints per plaquette. Remarkably, this echoes the proposal of \cite{smolin-ultralocality} where this feature was suggested to avoid the ultra-locality of Thiemann's quantization of the Hamiltonian of general relativity in loop quantum gravity.
 \item It is built out of bosonic operators acting on nodes, which create, destroy and exchange spins 1/2 between two half-lines meeting on a node.
 \item In the spin network representation, the quantum Hamiltonian leads to difference equations on the ground states. Here we study them in the case of triangular plaquettes, and find that they are recursion relations known from group representation theory.
 \item As an operator it can be written in any basis. This gives a variety of equations which all enable to identify the topological order in arbitrary dimensions.
\end{itemize}

Our proposal is based on recent progress in the context of background independent quantization using spin network states. In particular, the spin network formalism has been reformulated using spinor variables \cite{fine-structure, U(N)coherent, return-spinor, freidel-speziale-spinors, johannes-spinor1, johannes-spinor2, johannes-spinor3}, which have unraveled a local $\U(N)$ structure on each $N$-valent node. At the quantum level, it is possible to eliminate spins and to use spinors instead. We will call the resulting states {\bf spinor networks}.

Another progress \cite{3d-wdw} concerns the dynamics of the BF model, especially in $(2+1)$ dimensions where it also describes gravity. The usual way to solve 2+1 gravity emphasizes its topological behavior and makes use of the plaquette operator \cite{ooguri-3d, freidel-louapre-PR1, noui-perez-ps3d}. However, as a gravitational theory, it can be cast in the form of geometrodynamics, with a scalar Hamiltonian which generates time reparametrizations. In that case, it becomes unclear how to solve it. We think the plaquette operator is too simple, and solving the model as a gravitational theory is the best way to get insights for more complicated models, in particular 3+1 general relativity.

In \cite{3d-wdw} a new Hamiltonian was proposed which mimics as far as possible the scalar Hamiltonian of 2+1 Riemannian general relativity. It is less trivial than the usual projector and plaquette operators since it includes derivative operators corresponding to quantum triads. The quantum Hamiltonian equation, i.e. the Wheeler-DeWitt equation can be solved in some cases and shown to reproduce the results expected from the topological formulation. The final result is a {\bf quantization of flat Euclidean geometry}.


The present paper lifts the operator of \cite{3d-wdw} with the spinor variables to get a new, more fundamental Hamiltonian, based on the fundamental representation of $\SU(2)$.

The organization of the paper is as follows. After a short introduction to the BF model in the section \ref{sec:bfintro}, we present in the section \ref{sec:kin} the kinematical setting corresponding to $\SU(2)$ lattice gauge theory, or loop quantum gravity on a graph, in terms of spinors. The dynamics is described in the section \ref{sec:dyn}: the usual plaquette operator, the new Hamiltonian together with its Wheeler-DeWitt equation in the spin network basis, its geometric interpretation and the way to recover the usual constraint. In the section \ref{sec:new-bases}, we present the action of the new Hamiltonian on different bases of coherent states. We finally compare our key equation with other equations which also encode information on the topological phase in the section \ref{sec:other-eqs}.

\section{Introduction to the BF phase} \label{sec:bfintro}

Let us recall the basics of the Kitaev model \cite{kitaev-code} for topological order. Every link $e$ of a lattice is attached a spin which can be up or down. The Hamiltonian has two parts: star operators $H_s$ (or electric part) acting on nodes $s$, and plaquette operators (magnetic part) $H_p$ acting on plaquettes $p$,
\be
H_s = \prod_{e \supset s} \sigma^x_e,\qquad H_p = \prod_{e\subset p}\sigma^z_e\;.
\ee
$H_s$ flips the spins adjacent to a node, while $H_p$ measures the product of spins around a plaquette. One can check that $[H_s, H_p]=0$. The ground states are those states which are preserved by the all $H_s$ and $H_p$. On a Riemann surface of genus $h$, the ground state degeneracy is $2^{(2h)}$ and protected by a gap.

It is useful to understand the model as a lattice gauge theory for the group $\Z_2$. There is an element of $\Z_2$ on each link. Invariance under $H_s$ means local $\Z_2$ invariance, which the $\Z_2$ Gau\ss{} law. Invariance under $H_p$ means that the $\Z_2$ magnetic flux is trivial through each plaquette. This way of thinking generalizes to compact Lie group, such as $\SU(2)$. Let us put a $\SU(2)$ element $g_e$ on each link, called its {\bf holonomy}. Gauge invariance requires invariance under $\SU(2)$ translations on each node. Triviality of the magnetic fluxes means that the product of holonomies around each plaquette is the unit of the group,
\be \label{hol=1}
g_p\equiv\prod_{e\subset p} g_e = \unit\;.
\ee
The projector onto the solutions writes
\be
\prod_{\rm plaquettes} \delta\bigl( g_p\bigr)\;,
\ee
where $\delta$ is the Dirac function over the group (with respect to the Haar measure). It is well-defined for instance on Riemann surfaces of genus higher than two and is then independent of the lattice used to define it \cite{witten-2dym}: this is the meaning of saying it is topological. Conditions for convergence, divergence and invariance are given in \cite{twisted}.

This lattice model is actually an exact discrete realization of a field theory, called BF topological field theory, and known as the zero-coupling limit of Yang-Mills theory \cite{witten-2dym, cattaneo-4dYM}. Though it is a field theory, it only has a finite number of degrees of freedom at the classical level which are determined by spacetime topology. At the quantum level, it is mostly known from the fact that its partition function is the integral of the analytic torsion over the moduli space of flat connections, thus generalizing the two-dimensional case \cite{witten-2dym}. The continuum action is in dimension $d$,
\beq
S_{\rm BF} = \int \tr\bigl( B\wedge F(A)\bigr),
\ee
where $B$ is a $\su(2)$-valued $(d-2)$-form, and $A$ a $\su(2)$-valued connection 1-form, whose curvature tensor is $F(A)$. Quite clearly, $B$ is a Lagrange multiplier imposing the vanishing of $F(A)$. In addition to the standard $\SU(2)$ gauge invariance, there is another gauge invariance (which actually contains diffeomorphism invariance), in which $B$ transforms like $B\mapsto B+d_A\eta$. This symmetry is responsible for the disappearance of all local degrees of freedom. The path integral treatment was done in \cite{blau-thompson-bf}, and most of the Hamiltonian analysis and canonical quantization in \cite{horowitz-bf}. We refer to \cite{hasan-colloquium, cho-bf-insulators, hansson-superconductors, top-superconductors} for the relevance of this theory in the effective description of topological insulators and topological superconductors.

Thanks to its simplicity, it provides spin network states with a dynamics we are able to control \cite{noui-perez-ps3d, freidel-louapre-PR1} in the case $d=3$ \footnote{In three dimensions, the situation is well controlled for open manifolds, or topologies of the type a Riemann surface cross a real interval. But it is still unclear for generic closed manifolds because the gauge-fixing of \cite{freidel-louapre-PR1} then leaves a residual, non-compact gauge symmetry.}. The situation is more subtle in higher dimensions and under current investigations. And in spite of its apparent simplicity, it produces an interesting connection with a non-trivial equation from Lie group representation, the Biedenharn-Elliott identity.

The relation to the $\SU(2)$ lattice model is the following. The holonomies $g_e$ along the links are the Wilson lines of the gauge field $A$. The $\SU(2)$ action at each node of the lattice comes from the effect of gauge transformations on those Wilson lines: they transform it only on its end points. Hence, the Gau\ss{} law indeed asks for translation invariance at each node. The field strength $F(A)$ is regularized as usual in lattice gauge theory, by Wilson loops. In particular, the flatness equation $F(A)=0$ becomes the statement that Wilson loops around the plaquettes are trivial.

\section{Kinematics of lattice SU(2) gauge theory} \label{sec:kin}


\subsection{The lattice spinor variables}

The classical, kinematical setting we use is just the phase space of lattice gauge theory. It actually coincides with that inherited from loop quantum gravity restricted to a single graph. The formulation we present was proposed in \cite{freidel-speziale-spinors}. Further classical and quantum kinematical aspects can be found in \cite{fine-structure, U(N)coherent, return-spinor, johannes-spinor1, johannes-spinor2, johannes-spinor3}.


Let $\Gamma$ be a closed graph with $L$ oriented links and $V$ nodes. We introduce a classical phase space formed by attaching a spinor $z_e\in\C^2$ to the source vertex of each link, and another spinor $\tl{z}_e\in\C^2$ to the target vertex of each link. The Poisson brackets read:
\beq \label{bracket spinor}
\{ z_e^A, z_e^{*B} \} \,=\, i\,\delta^{AB},\quad \text{and}\qquad \{ \tl{z}_e^A, \tl{z}_e^{*B} \} \,=\, -i\,\delta^{AB}.
\ee
while all other brackets (between variables on different legs) vanish. The indices $A,B = \pm \f12$ are spinor indices, often short-handed to $\pm$. To make sure we construct $\SU(2)$ invariant quantities, we define the following, standard contractions of spinors. First define some notation:
\beq
\vert z\rangle = \begin{pmatrix} z^- \\ z^+\end{pmatrix},\qquad \langle z\vert = \begin{pmatrix} (z^{-})^* &(z^{+})^*\end{pmatrix},
\ee
together with the duality map $\varsigma$,
\beq \label{map sigma}
\vert z] \equiv \vert \varsigma z\rangle = \begin{pmatrix} -(z^+)^* \\ (z^-)^*\end{pmatrix},\quad \text{i.e.} \qquad (\varsigma z)^A = (-1)^{\f12-A}\,(z^{-A})^*.
\ee

The natural invariant contraction is: $\langle w \vert z\rangle$, and using the duality map we get a second invariant contraction:
\beq
\langle w \vert z\rangle = \sum_{A} (w^A)^*\,z^A,\quad \text{and}\qquad [w\vert z\rangle = \sum_A (-1)^{\f12-A}\,w^{-A}\,z^A.
\ee
One checks easily that: $\langle gw \vert gz\rangle = \langle w \vert z\rangle$, for any $g\in\SU(2)$. The invariance of $[w \vert z\rangle$ is ensured by the fact that: $(\varsigma gz)^A = \sum_B g_{AB} (\varsigma z)^B$, which means that $\varsigma z$ transforms like $z$ under $\SU(2)$ \footnote{The reader may have noticed that this second contraction uses the antisymmetric $\epsilon$-tensor in the fundamental representation, $\epsilon = \left(\begin{smallmatrix} 0&1\\ -1 &0\end{smallmatrix}\right)$. Its matrix elements in the irreducible representation of spin $j$ are: $\langle j,n\vert \epsilon\vert j,m\rangle = (-1)^{j-n}\delta_{-n,m}$, and they will be used in a systematic way in the remaining of the paper to contract Wigner 3jm-symbols.}.

Forming scalars with these inner products enables to define invariant observables, but we are also interested in covariant objects. Let us consider vectors. We need obviously to tensor two spinors, and then use an intertwiner $\calH_{\f12}\otimes \calH_{\f12}\rightarrow\calH_1$ to map covariantly the tensor product to the representation of spin 1. Such an intertwiner is well-known and its components are basically the matrix elements of the Pauli matrices. More precisely, take the matrix-valued vector $\vec{\sigma}$ which reads in the fundamental representation: $\sigma_x = \left(\begin{smallmatrix} 0&1\\1&0\end{smallmatrix}\right), \sigma_y = \left(\begin{smallmatrix} 0&-i\\i&0\end{smallmatrix}\right), \sigma_z = \left(\begin{smallmatrix} 1&0\\0&-1\end{smallmatrix}\right)$. The commutation relations are: $[\sigma^i, \sigma^j]=2i\epsilon^{ij}_{\phantom{ij}k} \sigma^k$.
Then, vectors, known as the {\bf flux} variables, are given by:
\beq
X^i = -\f12\ \langle z\vert\, \sigma^i\,\vert z\rangle = -\f12 \ \sum_{A,B} z^{A*}\,\sigma^i_{AB}\,z^B,
\ee
and they satisfy the $\su(2)$ algebra.

More than vectors, we can actually reconstruct the classical phase space $T^*\SU(2)^L$ from the spinors (and this is the lattice gauge theory phase space). The idea is to introduce a group element $g_e$ to each link which sends the spinor on the source to that on the target,
\beq \label{spinor transport}
g_e \frac{\vert z_e\rangle}{\sqrt{\langle z_e\vert z_e\rangle}} = \frac{\vert \tl{z}_e\rangle}{\sqrt{\langle \tl{z}_e\vert \tl{z}_e\rangle}},\quad \text{and} \qquad g_e \frac{\vert z_e]}{\sqrt{\langle z_e\vert z_e\rangle}} = \frac{\vert \tl{z}_e]}{\sqrt{\langle \tl{z}_e\vert \tl{z}_e\rangle}}.
\ee
The spinors $z_e, \tl{z}_e$ actually completely determine $g_e$ \cite{freidel-speziale-spinors},
\beq \label{holonomy spinor}
g_e = \frac{\vert \tl{z}_e \rangle\, \langle z_e\vert\,+\,\vert \tl{z}_e]\, [z_e\vert}{\sqrt{\langle z_e\vert z_e\rangle\,\langle \tl{z}_e\vert \tl{z}_e\rangle}}.
\ee
Those objects are the holonomies (they can be seen as parallel transport operators, from the source to the target, arising as Wilson lines of a gauge field). The full underlying symplectic geometry has been described in \cite{freidel-speziale-spinors} (though with different conventions). The Poisson brackets are:
\beq
\bigl\{X_e^i,X_e^j\bigr\} = \eps^{ij}_{\phantom{ij}k}\,X_e^k,\qquad \bigl\{X_e^i, g_e\bigr\} = g_e\,\tau^i,
\ee
where the matrices $(\tau_i)_{i=1,2,3}$ are anti-Hermitian generators \footnote{They read: $\tau^i = -\f{i}{2}\sigma^i$, in terms of the Pauli matrices $(\sigma^i)$.} satisfying $[\tau^i,\tau^j] = \eps^{ij}_{\phantom{ij}k}\tau^k$. All other brackets vanish. Hence, fluxes $X_e^i$ act as left invariant derivatives, and should be thought as attached to the half-link given by $e$ and its source vertex.

Right invariant derivatives are obtained by transporting $X_e$ to the target vertex of $e$ via the adjoint action:
\beq \label{Xtilde}
\widetilde{X}_e = \Ad(g_e)\, X_e = g_e\,X_e\,g_e\mone,
\ee
together with:
\beq
\bigl\{\widetilde{X}_e^i,\widetilde{X}_e^j\bigr\} = -\eps^{ij}_{\phantom{ij}k}\,\widetilde{X}_e^k,\qquad \bigl\{\widetilde{X}_e^i, g_e\bigr\} = \tau^i\,g_e.
\ee
Notice that the adjoint representation on the algebra is exactly the representation of spin 1, acting on 3-vectors.

That holonomy-flux algebra is the starting point of lattice gauge theory. In this context, one introduces a $\SU(2)^V$-action, by group translation on each node $v$ of $\Gamma$. The Gau\ss{} law generates those transformations through the Poisson brackets, and local $\SU(2)$ invariance is imposed by the constraint
\beq \label{closure}
\sum_{{\rm outgoing}\ e \supset v}  X_e - \sum_{{\rm ingoing}\ e \supset v}  \tl{X}_e= 0.
\ee



Straightforward quantization on $T^*\SU(2)^L$ leads to the Hilbert space $L^2(\SU(2)^L)$, equipped with the Haar measure (or $L^2(\SU(2)^L/\SU(2)^V)$ when gauge invariance is required). This is the kinematical Hilbert space of loop quantum gravity restricted to the graph $\Gamma$ in 3+1 dimensions (see \cite{baez-spinnets, perez-intro-lqg, ashtekar-status-report} for reviews, or more specifically \cite{freidel-speziale-twisted-geom}), but also in 2+1 dimensions \cite{noui-perez-ps3d,freidel-louapre-PR1}.



The phase space $T^*\SU(2)^L$ assigns six real variables to each link, while the spinor setting gives eight variables. The relation between both frameworks is given in \cite{freidel-speziale-spinors}. The idea is to perform symplectic reduction with respect to a $\U(1)^L$-group action. As spinors and their quantization are well known, it is very appealing to reformulate (at least the kinematics of) loop quantum gravity this way, as done in \cite{return-spinor}, and hopefully get new insights on the dynamics.

A key difference between spinors and holonomies/fluxes is that the latter encode the information on the links of the graph, while spinors enable to factorize it on the nodes. Therefore, we now focus on a single $N$-valent node $v$ and consider without loss of generality that all the links which meet there are outgoing. Following \cite{return-spinor}, we introduce elementary $\SU(2)$-invariant local observables (i.e. acting on the node). We use here latin indices $a,b,c,d$ to denote the legs of the node.
\begin{alignat}{2} \label{E}
E_{ab} &\equiv  \langle z_a\vert z_b\rangle, &\qquad \qquad& E_{ba} = E_{ab}^*,\\
F_{ab} &\equiv \langle z_a\vert z_b], && F_{ba}=-F_{ab} \label{F}
\end{alignat}
That set of observables forms a closed algebra. Interestingly, the observables $E_{ab}$ satisfy a $\u(N)$ sub-algebra, and this is the reason why this formalism has been coined the $\U(N)$ formalism for loop quantum gravity \cite{U(N)coherent, fine-structure, return-spinor},
\begin{alignat}{2}
\nonumber &\{ E_{ab}, E_{cd}\} &\ =& -i\bigl(\delta_{bc}\,E_{ad} - \delta_{ad}\,E_{cb}\bigr),\\
\nonumber &\{ E_{ab}, F_{cd}\} &=& -i\bigl(\delta_{bc}\,F_{ad} - \delta_{bd}\,F_{ac}\bigr),\\
\nonumber &\{ F_{ab}, F_{cd}\} &=&\ 0,\\
&\{ F_{ab}^*, F_{cd}\} &=& -i\bigl(\delta_{ac}\,E_{db} + \delta_{bd}\,E_{ca} - \delta_{bc}\,E_{da} - \delta_{ad}\,E_{cb}\bigr).
\end{alignat}

\subsection{Quantization: Intertwiners and Schwinger's boson operators}

Let $\calH_j$ denote the carrier space of the irreducible representation of $\SU(2)$ with spin $j\in\N/2$, of dimension $d_j\equiv 2j+1$. From the Peter-Weil theorem we known that $L^2(\SU(2)) = \oplus_{j\in\frac{\N}{2}} \calH_j\otimes \calH_j^*$.

The Hilbert space on $\Gamma$ is $L^2(\SU(2)^L)$, and thus it is spanned by the products of the matrix elements of the holonomies in any set of $L$ irreducible representations $(j_e)$ attached to the links of $\Gamma$. Therefore, a matrix element of the holonomy $g_e$ along $e$ in a representation $k_e$ acts as an operator on the Hilbert space by a simple multiplication,
\begin{multline}
\widehat{\langle k_e, p\vert g_e\vert k_e, q\rangle} \langle j_e, n\vert g_e \vert j_e, m\rangle \equiv \langle k_e, p\vert g_e\vert k_e, q\rangle\ \langle j_e, n\vert g_e \vert j_e, m\rangle \\
= \sum_{l_e=\vert j_e-k_e\vert}^{j_e+k_e} \sum_{M,N=-l_e}^{l_e} d_{l_e}
(-1)^{l_e-N}\begin{pmatrix} k_e &j_e & l_e\\ p &n &-N\end{pmatrix}\, (-1)^{l_e-M}\begin{pmatrix} k_e &j_e & l_e\\ q &m &-M\end{pmatrix} \langle l_e, N\vert g_e\vert l_e, M\rangle\;.
\end{multline}
The second line just comes from Clebsch-Gordan re-coupling, to map the reducible tensor product $\calH_{k_e}\otimes \calH_{j_e}$ to the sum of irreducible representations $\oplus_{l_e=\vert j_e-k_e\vert}^{j_e+k_e} \calH_{l_e}$. The quantities into brackets are Wigner 3jm-symbols (equivalent to Clebsch-Gordan coefficients but more convenient for our purposes) \cite{varshalovich-book}.

The flux variables become insertions of anti-hermitian generators (left or right derivatives),
\beq
\widehat{X}_e^i\ \langle j_e,n\vert g_e\vert j_e,m\rangle = \langle j_e,n\vert g_e\,\tau^i\,\vert j_e,m\rangle,\qquad
\widehat{\tl{X}_e^i}\ \langle j_e,n\vert g_e\vert j_e,m\rangle = \langle j_e,n\vert\,\tau^i\, g_e \vert j_e,m\rangle,
\ee
with $\tau^i = -\f{i}{2}\sigma^i$.

A basis of the gauge invariant Hilbert space on $\Gamma$, $L^2(\SU(2)^L/\SU(2)^V)$ is given by spin network functions. The latter are formed by contracting the magnetic indices of the Wigner matrices along each link with a specific tensor at each node. This tensor has to ensure the invariance of the function under (left or right) translation of the holonomies and it must then be an intertwiner between all representations meeting on the node. Consider typically a 3-valent node, where the links are outgoing and carry the spins $j_1,j_2,j_3$. There is a single intertwiner $\iota_{j_1 j_2 j_3}$, i.e. a single invariant vector in $\calH_{j_1}\otimes \calH_{j_2}\otimes \calH_{j_3}$, which is up to normalization the Wigner 3jm-symbol \footnote{Being an invariant vector in $\calH_{j_1}\otimes \calH_{j_2}\otimes \calH_{j_3}$ means
\be
\sum_{m_1,m_2,m_3} \langle j_1,n_1\vert g\vert j_1,m_1\rangle \langle j_2,n_2\vert g\vert j_2,m_2\rangle \langle j_3,n_3\vert g\vert j_3,m_3\rangle \begin{pmatrix} j_1 &j_2 &j_3\\m_1 &m_2 &m_3\end{pmatrix} = \begin{pmatrix} j_1 &j_2 &j_3\\n_1 &n_2 &n_3\end{pmatrix}\;,
\ee
for any $g\in\SU(2)$.}. It is convenient to see it as a map and write its tensor elements like:
\beq
\langle j_1,m_1 ; j_2, m_2; j_3, m_3\vert \iota_{j_1 j_2 j_3}\vert 0\rangle = \begin{pmatrix} j_1 &j_2 &j_3\\m_1 &m_2 &m_3\end{pmatrix}\;,
\ee
where $\vert 0\rangle$ is the normalized vector in the trivial representation.

\begin{figure}\begin{center}
\includegraphics[scale=0.55]{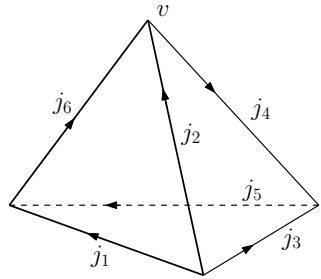}
\caption{ \label{fig:tet} The tetrahedral graph we consider throughout the paper. The orientations are the same as those of the graphical representation of the 6j-symbol \cite{varshalovich-book}. The three fat lines form the cycle (126) we will consider in order to explain the action of the new Hamiltonian.
}
\end{center}
\end{figure}

If some links are ingoing, as it happens on the node where $(e_2,e_6,e_4)$ meet in the figure \ref{fig:tet}, one dualizes the representations, here $j_2, j_4$, and writes\footnote{Note that with these conventions the standard group averaging formula for the tensor product of three representations is:
\beq
\int dg\ \otimes_{e=1}^3 D^{(j_e)}(g) = \iota_{j_1 j_2 j_3}\vert 0\rangle \langle 0\vert \iota_{j_1* j_2* j_3*}\;,
\ee
where $D^{(j)}(g)$ is the Wigner matrix.}
\beq
\langle j_4,m_4\vert \iota_{j_2^* j_6^* j_4}\vert j_2,m_2 ; j_6,m_6\rangle = (-1)^{j_2-m_2}(-1)^{j_6-m_6}\begin{pmatrix} j_2& j_6 &j_4\\-m_2 &-m_6 &m_4\end{pmatrix}\;.
\ee

Gauge invariant observables are obtained from traces of holonomies, and more generally from full, rotation invariant contractions of flux indices and magnetic indices of Wigner matrices (see \cite{3d-wdw} for a recent use of such operators).

Since the flux $X_e^i$ carries a vector index, any gauge invariant observables containing it will add a link to $\Gamma$ in the representation of spin 1. This operation is known as a {\bf grasping} and enables to extract quantum geometric information from spin networks \cite{3d-wdw}. Taking tensor product, one easily creates graspings with arbitrary integral spins. However, we would like to perform graspings with half-integers also. This is what lifting the phase space to spinors allows. Let us proceed to their quantization.

The spinor we have introduced on each half-leg of $\Gamma$ is obviously the classical version of the Schwinger representation of the angular momentum using two harmonic oscillators. Upon quantization, the latter become two annihilation operators $(a^A)_{A=-1/2,+1/2}$, such that:
\beq
\left[a^A, a^{B\dagger} \right] = \delta^{AB}\,\id,\qquad \left[a^A,a^B\right]=0.
\ee
From them, the $\su(2)$ generators can be constructed as quadratic operators:
\beq \label{su2 gen}
\tau^i = -\f{i}{2}\,\langle a\vert\,\sigma^i\,\vert a\rangle.
\ee
These objects are anti-hermitian: $(\tau^i)^\dagger = -\tau^i$, they indeed generate the expected algebra $[\tau_i,\tau_j] = \epsilon_{ij}^{\phantom{ij}k}\ \tau_k$, for $i,j,k = x,y,z$. $(a^A)$ can be thought as an operator-valued spinor which transforms under the fundamental representation of $\SU(2)$.

A natural basis $(\vert n^A\rangle)$ is that obtained by diagonalizing the occupation number operators, $(a^{-})^\dagger a^{-}$ and $(a^+)^\dagger a^{+}$. This basis is exactly the standard basis of spins and magnetic numbers $(\vert j,m\rangle)$ on the sum over all irreducible representations $\oplus_{j\in\f{\N}{2}} \calH_j$. The relation between both sets of eigenvalues is:
\beq
j = \f12(n^{+} + n^{-}),\qquad m = \f12(n^{+} - n^{-}).
\ee
Hence, we can write the action of $(a^A, a^{B\dagger})$ on the standard basis. Compared to the generators $(\tau_i)$ of the algebra, the new feature is that they enable to go up and down between different irreducible representations. Indeed,
\beq \label{spin1/2op}
a^A\,\vert j,m\rangle = \sqrt{j + (-1)^{\f12 -A}m}\,\vert j-\f12, m-A\rangle,\qquad a^{A\dagger}\,\vert j,m\rangle = \sqrt{j + (-1)^{\f12 -A}m +1}\,\vert j+\f12, m+A\rangle.
\ee

It is well-known that the matrix elements of the generators $(\tau_i)$ can be re-expressed as Wigner 3jm-symbols with a spin 1 (that is the spin carried by the generators). This is the way one could construct a Hamiltonian operator generating a recursion relation on the 6j-symbol, well-known in re-coupling theory, in \cite{3d-wdw}.
Actually, a similar result holds in the present case:
\begin{alignat}{3}
\nonumber \langle k,n \vert\  &a^A\ \vert j,m\rangle &\ =\ & \sqrt{d_j\,d_k}\ (-1)^{k-n+\f12-A+1}&\ \delta_{k,j-\f12}\ &\begin{pmatrix} \f12 &j &k\\ -A &m &-n\end{pmatrix}\;, \label{a^A 3jm} \\
\langle k,n \vert\ &a^{A\dagger} \,\vert j,m\rangle &\,=\,& \sqrt{d_j\,d_k} \ (-1)^{j-m+\f12-A+1}&\ \delta_{k,j+\f12}\ &\begin{pmatrix} \f12 &k &j\\ -A &n &-m\end{pmatrix}\;,
\end{alignat}
with the notation $d_j=2j+1$. It also suggests that their action can be translated into re-coupling equations.

Now we equip the graph $\Gamma$ with these operators. At the quantum level, to take into account $\SU(2)$ invariance (the Gau\ss{} law), the degrees of freedom at each $N$-valent node are intertwiners. However, as the Schwinger's bosons change the spin, we cannot consider the spin on each leg to be fixed. Instead we take the invariant tensors living on $N$ copies of $\oplus_{j\in\f\N2}\calH_j$,
\beq
\calH_N = \bigoplus_{\{j_e\}} \Inv\left[\otimes_{e=1}^N \calH_{j_e}\right]\;,
\ee
where one sums over all sets of irreducible representations attached to legs. We have assumed that all legs are oriented outwards\footnote{Changing the orientation of a link amounts to changing the corresponding spin $j$ with its dual representation, denoted $j^*$.}. Schwinger boson operators enable to define operators on that space which were out of reach in the usual loop quantum gravity framework. The operators we are interested in are quadratic and create, annihilate or exchange quanta between two given legs of a node:
\begin{alignat}{2}
\nonumber \hE_{ee'} &\equiv \langle a_e\,\vert\, a_{e'} \rangle = \sum_{A=\pm\f12} a_e^{A\dagger}\ \otimes\ a_{e'}^A, &\qquad \qquad& \hE_{e'e} = \hE_{ee'}^\dagger\;,\\
\hF_{ee'} &\equiv [ a_e \,\vert\,a_{e'}\rangle = \sum_{A=\pm\f12} (-1)^{\f12-A}\, a^{-A}_e\ \otimes\ a^A_{e'}, && \hF_{e'e}=-\hF_{ee'}\;.
\end{alignat}
These operators are clearly the quantization of the observables $E_{ee'}$ and $\bar{F}_{ee'}$, \eqref{E}, \eqref{F}. Due to the special contractions of the spinor index, they are invariant under the global $\SU(2)$ transformations acting on the node, i.e. they commute with the generator $\vec{J} = \sum_{e=1}^N \vec{\tau}_e$. The set of operators $\hE, \hF, \hF^\dagger$ form a closed algebra and like in the classical case, the operators $\hE_{ee'}$ satisfy a $\u(N)$ algebra.

Putting the Schwinger operators on spin networks will just mean for us that we have that structure on each node, acting on $\calH_N$. One link $e$ has two sets of operators, $(a_e^A)$ on the source vertex, and $(\tl{a}^A_e)$ on the target vertex of $e$.

As we see in the definition \eqref{bracket spinor}, the symplectic structure on the incoming legs has the opposite sign to that of outcoming legs. This is reflected in the quantization by the fact that $\tl{a}^A$ and $\tl{a}^{A\dagger}$ act on the right, i.e. on bras. This is due to the dualization of the representation. Thus, our conventions imply that those operators satisfy the usual algebra:
\beq
\left[\tl{a}^A, \tl{a}^{B\dagger}\right]=\delta^{AB}\id\;.
\ee
Hence, we can forget the tilde when considering their matrix elements. But the interpretation is reversed: since they act on the right, the creation operator, i.e.  that shifting the spin on the leg by $+\f12$, is $\tl{a}^A$, while $\tl{a}^{A\dagger}$ is the annihilator.


\subsection{Graspings with spin 1/2 and holonomy operator} \label{sec:grasping}

We show here that the operators $\hE, \hF, \hF^\dagger$ generalize the usual operators used in the context of loop quantum gravity, by producing graspings with the spin $1/2$, and we collect the results of their action.

One way to evaluate these operators is by using the standard action of the creators and annihilators \eqref{spin1/2op}. A brief examination shows that one has to use some recursion relations on the 3jm-symbols to carry out the calculation. Another way through, which instead highlights the re-coupling nature of those operators and their interpretation as graspings, can be used thanks to \eqref{a^A 3jm}. From the results below, closed forms are obtained by taking the values of the re-coupling coefficients in a textbook \cite{varshalovich-book}, which are reported in appendix for completeness.

\begin{figure} \begin{center}
\includegraphics[scale=0.65]{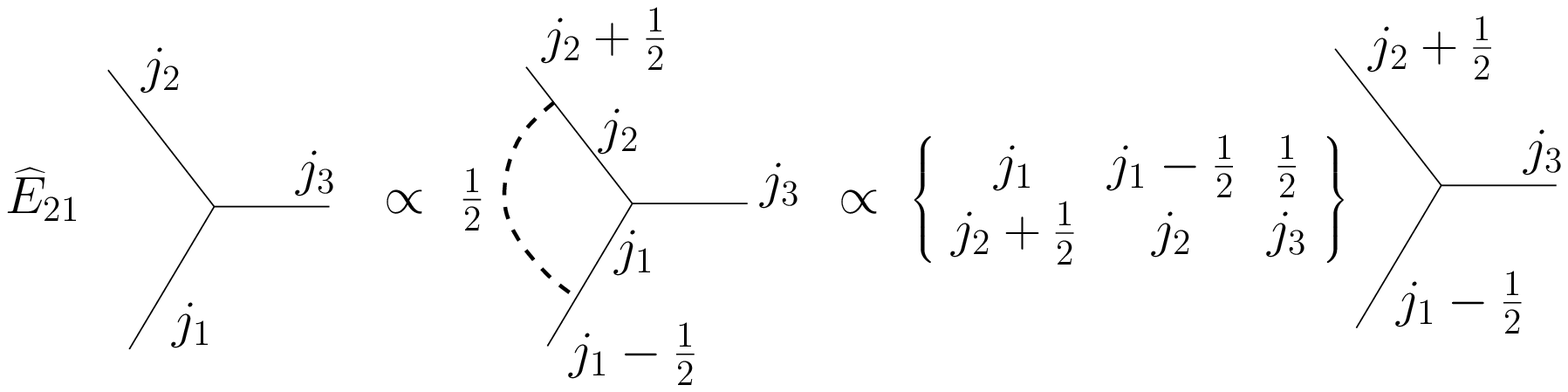}
\caption{ \label{fig:1halfgrasping} The 3-valent nodes represent 3jm-symbols, where legs carry the spins. A magnetic index is summed when there is a link joining two nodes. The action of $\widehat{E}_{21}$ is a grasping between $e_1$ and $e_2$, denoted by a dashed line which carries the spin 1/2. The final result is proportional to the 3jm-symbol with the spins $k_1=j_1-\f12, k_2=j_2+\f12$.}
\end{center}
\end{figure}

Let us write the action of, say, the operator $\hE_{21}$ on the node where $e_1, e_2, e_3$ meet in the figure \ref{fig:tet},
\begin{align}
\hE_{21}\ \iota_{j_1 j_2 j_3}\vert 0\rangle &= \sum_{A,m_1,m_2,m_3} a_1^A \vert j_1,m_1\rangle \otimes\, a_2^{A\dagger}\vert j_2, m_2\rangle\otimes\, \vert j_3, m_3\rangle \begin{pmatrix} j_1 &j_2 &j_3\\m_1 &m_2 &m_3\end{pmatrix}\;,\\
& = (-1)^{j_1+j_2+j_3+1}\sqrt{d_{j_1} d_{k_1}\,d_{j_2} d_{k_2}}\ \delta_{k_1,j_1-\f12} \delta_{k_2,j_2+\f12}\ \begin{Bmatrix} j_1 &k_1 &\f12 \\k_2 &j_2 &j_3\end{Bmatrix}\ \iota_{k_1 k_2 j_3}\vert 0\rangle\;.
\end{align}
The last line is obtained after some re-coupling\footnote{ First rewrite the action like
\be
\hE_{21}\ \iota_{j_1 j_2 j_3}\vert 0\rangle = \sqrt{d_{j_1} d_{k_1}\,d_{j_2} d_{k_2}}\ \delta_{k_1,j_1-\f12} \delta_{k_2,j_2+\f12} \sum_{n_1,n_2,m_3} \vert k_1,n_1\rangle\otimes\, \vert k_2,n_2\rangle\otimes\, \vert j_3, m_3\rangle \times \mathcal{I}^{k_1 k_2 j_3}_{n_1 n_2 m_3}\;,
\ee
where the object $\mathcal{I}^{k_1 k_2 j_3}$ is
\beq
\mathcal{I}^{k_1 k_2 j_3}_{n_1 n_2 m_3} = \sum_{A,m_1,m_2}(-1)^{k_1-n_1} \begin{pmatrix} \f12 &j_1 &k_1\\-A& m_1& -n_1\end{pmatrix} (-1)^{j_2-m_2}\begin{pmatrix} \f12 &k_2 &j_2\\-A &n_2 &-m_2\end{pmatrix} \begin{pmatrix} j_1 &j_2 &j_3\\m_1 &m_2 &m_3\end{pmatrix}\;.
\ee
We have used first the definition, and in the second equality the expression of the Schwinger's operators in terms of 3jm-symbols \eqref{a^A 3jm}. $\mathcal{I}^{k_1 k_2 j_3}_{n_1 n_2 m_3}$ is a tensor with three indices, $n_1, n_2, m_3$, respectively in the representations of spin $k_1, k_2, j_3$. It is easy to check that it is $\SU(2)$ invariant, and hence proportional to the 3jm-symbol, $\mathcal{I}^{k_1 k_2 j_3}_{n_1 n_2 m_3} \propto \left(\begin{smallmatrix} k_1 &k_2 &j_3\\ n_1 &n_2 &m_3\end{smallmatrix}\right)$. More precisely, the pre-factor comes from:
\beq
\mathcal{I}^{k_1 k_2 j_3}_{n_1 n_2 m_3} = \left[\sum_{p_1,p_2,p_3} \mathcal{I}^{k_1 k_2 j_3}_{p_1 p_2 p_3} \begin{pmatrix} k_1 &k_2 &j_3\\ p_1 &p_2 &p_3 \end{pmatrix} \right]\ \langle k_1, n_1; k_2, n_2; j_3, m_3\vert \iota_{k_1 k_2 j_3}\vert 0\rangle.
\ee
That pre-factor is built with four 3jm-symbols, which are contracted following the pattern of the tetrahedron, and thus produces a 6j-symbol. That 6j-symbol is special since it has a fixed spin being $1/2$. The exact evaluation yields the desired result.}.
The operator $\hE_{12}$ thus performs a grasping on the intertwiner $\iota_{j_1 j_2 j_3}$ with a spin $1/2$ inserted between the spins $j_1$ and $j_2$. The shift of $\pm \f12$ on $j_1, j_2$ comes from selecting the value $k_1$ in the tensor product:
\beq
\calH_{j_1}\otimes \calH_{\f12} = \calH_{j_1-\f12}\oplus \calH_{j_1+\f12},
\ee
and the value $k_2$ in $\calH_{j_2}\otimes \calH_{\f12}$. The values of $k_1, k_2$ are determined by the operators, here $a_1$ and $a_2^\dagger$. The contraction of the magnetic indices $A, m_1, m_2$ results in an invariant tensor in $\calH_{k_1}\otimes \calH_{k_2}\otimes \calH_{j_3}$, which is obviously proportional to the normalized intertwiner $\iota_{k_1 k_2 j_3}$. The pre-factor involves the quantity into curly brackets which appears in the final form and known as a Wigner 6j-symbol with a spin fixed to $1/2$. This is summarized in the figure \ref{fig:1halfgrasping}.

This calculation produces a spin $1/2$ grasping which generalizes the standard spin 1 grasping, well-known in loop quantum gravity. The spin 1 grasping simply comes from the action of the fluxes. Instead of considering $E_{21} = \langle z_2\vert z_1\rangle$, one can just consider the dot product of the corresponding fluxes, $\vec{X}_1\cdot\vec{X}_2$. The quantization of this observable is obtained by promoting each $X^i$ to an anti-hermitian generator $\tau^i$ acting on the node. The calculation has been reported precisely in \cite{3d-wdw}, and leads to
\beq \label{quantum dihedral}
\bigl(\vec{\tau}_1\cdot\vec{\tau}_2\bigr)\ \iota_{j_1 j_2 j_3}\vert 0\rangle = (-1)^{j_1+j_2+j_3+1}\sqrt{j_1(j_1+1)d_{j_1}\,j_2(j_2+1)d_{j_2}}\ \begin{Bmatrix} j_1&j_1 &1\\ j_2 &j_2 &j_3\end{Bmatrix}\ \iota_{j_1j_2 j_3}\vert 0\rangle.
\ee
The fact that no spin gets shifted in this case is clear from the fact that the operator $(\vec{\tau}_1\cdot\vec{\tau}_2)$ selects in the tensor product $\calH_{j_1}\otimes \calH_1$ the factor $\calH_{j_1}$. The operator $(\vec{\tau}_1\cdot\vec{\tau}_2)$ is quadratic in the generators and thus quartic in the operators $a^A$. Hence, it can also be obtained from some square of the spinor contraction $E_{12}$. A similar process was detailed in \cite{3d-wdw} to get graspings with a spin 2 from those with the spin 1. It obviously involves the square of the spin 1 operator together with some symmetrization. At the quantum level, it can be derived using the recursion formula on the 6j-symbol to relate the different sets of eigenvalues and pre-factors, i.e. here to express $\{\begin{smallmatrix} j_1 &j_1 &1\\ j_2 &j_2 &j_3\end{smallmatrix}\}$ in terms of $\{\begin{smallmatrix} j_1 &k_1 &\f12\\ k_2 &j_2 &j_3\end{smallmatrix}\}$.

Let us compute in addition the action of a $\hF$ operator on the same node, $\hF_{21}^\dagger = \langle a_1\vert a_2]$,
\begin{align}
\nonumber \hF^\dagger_{21}\ \iota_{j_1 j_2 j_3}\vert 0\rangle &= \sum_{A,m_1,m_2,m_3} a_1^{A\dagger} \vert j_1,m_1\rangle \otimes\, (-1)^{\f12-A} a_2^{-A\dagger}\vert j_2, m_2\rangle\otimes\, \vert j_3, m_3\rangle \begin{pmatrix} j_1 &j_2 &j_3\\m_1 &m_2 &m_3\end{pmatrix},\\
&= (-1)^{j_1+j_2+j_3+1}\sqrt{d_{j_1} d_{k_1}\,d_{j_2} d_{k_2}}\ \delta_{k_1,j_1+\f12} \delta_{k_2,j_2+\f12}\ \begin{Bmatrix} j_1 &k_1 &\f12 \\k_2 &j_2 &j_3\end{Bmatrix}\ \iota_{k_1 k_2 j_3}\vert 0\rangle.
\end{align}
The spins $j_1, j_2$ are both raised to $k_1=j_1+\f12, k_2 =j_2+\f12$ due to the creators.

Quite clearly, all operators $\hE, \hF, \hF^\dagger$ will generate a 6j-symbol with a spin $1/2$, of the form $\left\{\begin{smallmatrix} j_1 &k_1 &\f12\\ k_2 &j_2 &j_3\end{smallmatrix}\right\}$, with $k_e = j_e\pm\f12$. The main difficulty is however to establish properly the overall signs, which are so important to get the correct recursion formula in the end. Let us here report the results of a few actions, in particular on nodes with incoming and/or outcoming legs. The conventions and notations correspond to the situation of the figure \ref{fig:tet}.
\begin{gather}
\hF_{12}\,\iota_{j_1 j_2 j_3}\vert 0\rangle \equiv [a_1\vert a_2\rangle\,\iota_{j_1 j_2 j_3}\vert 0\rangle = (-1)^{j_1+j_2+j_3+1}\sqrt{d_{j_1}d_{k_1}\,d_{j_2}d_{k_2}}\ \delta_{k_1, j_1-\f12} \delta_{k_2, j_2-\f12}\ \begin{Bmatrix} j_1 &k_1 &\f12 \\k_2 &j_2 &j_3\end{Bmatrix}\ \iota_{k_1 k_2 j_3}\vert 0\rangle\;, \label{F12}\\
\hE_{12}\,\iota_{j_1 j_2 j_3}\vert 0\rangle \equiv \langle a_1\vert a_2\rangle\,\iota_{j_1 j_2 j_3}\vert 0\rangle = (-1)^{j_1+j_2+j_3+1}\sqrt{d_{j_1}d_{k_1}\,d_{j_2}d_{k_2}}\ \delta_{k_1, j_1+\f12} \delta_{k_2, j_2-\f12}\ \begin{Bmatrix} j_1 &k_1 &\f12 \\k_2 &j_2 &j_3\end{Bmatrix}\ \iota_{k_1 k_2 j_3}\vert 0\rangle\;, \label{E12}\\
\hF_{61}\,\iota_{j_1^* j_5^* j_6} \equiv [a_6\vert \tl a_1\rangle\,\iota_{j_1^* j_5^* j_6} = (-1)^{j_1+j_5+j_6+1}\sqrt{d_{j_1} d_{k_1}\,d_{j_6} d_{k_6}}\ \delta_{k_1,j_1+\f12} \delta_{k_6,j_6-\f12}\ \begin{Bmatrix} j_1 &k_1 &\f12 \\k_6 &j_6 &j_5\end{Bmatrix}\ \iota_{k_1^* j_5^* k_6}\;,\label{F61}\\
\hF_{16}^\dagger\,\iota_{j_1^* j_5^* j_6} \equiv \langle a_6\vert \tl{a}_1]\,\iota_{j_1^* j_5^* j_6} = (-1)^{j_1+j_5+j_6}\sqrt{d_{j_1} d_{k_1}\,d_{j_6} d_{k_6}}\ \delta_{k_1,j_1-\f12} \delta_{k_6,j_6+\f12}\ \begin{Bmatrix} j_1 &k_1 &\f12 \\k_6 &j_6 &j_5\end{Bmatrix}\ \iota_{k_1^* j_5^* k_6}\;, \label{F16}\\
\hE_{61}\,\iota_{j_1^* j_5^* j_6} \equiv \langle a_6\vert \tl{a}_1\rangle \,\iota_{j_1^* j_5^* j_6} = (-1)^{j_1+j_5+j_6+1}\sqrt{d_{j_1} d_{k_1}\,d_{j_6} d_{k_6}}\ \delta_{k_1,j_1+\f12} \delta_{k_6,j_6+\f12}\ \begin{Bmatrix} j_1 &k_1 &\f12 \\k_6 &j_6 &j_5\end{Bmatrix}\ \iota_{k_1^* j_5^* k_6}\;, \label{E61}\\
\hE_{16}\,\iota_{j_1^* j_5^* j_6} \equiv \langle \tl a_1\vert a_6\rangle \,\iota_{j_1^* j_5^* j_6} = (-1)^{j_1+j_5+j_6}\sqrt{d_{j_1} d_{k_1}\,d_{j_6} d_{k_6}}\ \delta_{k_1,j_1-\f12} \delta_{k_6,j_6-\f12}\ \begin{Bmatrix} j_1 &k_1 &\f12 \\k_6 &j_6 &j_5\end{Bmatrix}\ \iota_{k_1^* j_5^* k_6}\;, \label{E16}\\
\hF_{62}\,\iota_{j_2^* j_6^* j_4} \equiv [\tl{a}_6\vert \tl{a}_2\rangle\,\iota_{j_2^* j_6^* j_4} = (-1)^{j_2+j_6+j_4+1}\sqrt{d_{j_2} d_{k_2}\,d_{j_6} d_{k_6}}\ \delta_{k_2,j_2+\f12} \delta_{k_6,j_6+\f12}\ \begin{Bmatrix} j_2 &k_2 &\f12 \\k_6 &j_6 &j_4\end{Bmatrix}\ \iota_{k_2^* k_6^* j_4}\;, \label{F62}\\
\hF_{26}^\dagger\,\iota_{j_2^* j_6^* j_4} \equiv \langle \tl{a}_6\vert \tl{a}_2]\,\iota_{j_2^* j_6^* j_4} = (-1)^{j_2+j_6+j_4+1}\sqrt{d_{j_2} d_{l_2}\,d_{j_6} d_{l_6}}\ \delta_{l_2,j_2-\f12} \delta_{l_6,j_6-\f12}\ \begin{Bmatrix} j_2 &l_2 &\f12 \\l_6 &j_6 &j_4\end{Bmatrix}\ \iota_{l_2^* l_6^* j_4}\;. \label{F26dagger}
\end{gather}

It is essential that all usual loop quantum gravity operators on a fixed graph can be expressed using the operators $\hE, \hF, \hF^\dagger$. We have already discussed the case of flux operators acting on a single node, as quartic operators in terms of spinors. Since the loop quantum gravity phase space $\{X_e,g_e\}$ can be recast into a (constrained) phase space based on spinors, one should also be able to quantize the holonomies $(g_e)$ with the Schwinger's boson operators.

On the kinematical Hilbert space $L^2(\SU(2)^L)$ over $\Gamma$, a matrix element of any holonomy in a any irreducible representation acts by simple multiplication. We consider the elementary situation: the holonomy in the fundamental representation along a single link $e$ carrying the spin $j_e$. The multiplication is re-expanded onto the matrix elements of irreducible representations as follows,
\begin{multline}
\langle 1/2, A\vert g_e\vert 1/2, B\rangle \langle j_e, n\vert g_e \vert j_e, m\rangle\\
 = \sum_{\substack{J_e=j_e\pm\f12\\N,M}} d_{J_e}
(-1)^{J_e-N}\begin{pmatrix} \f12 &j_e & J_e\\ A &n &-N\end{pmatrix}\, (-1)^{J_e-M}\begin{pmatrix} \f12 &j_e & J_e\\ B &m &-M\end{pmatrix} \langle J_e, N\vert g_e\vert J_e, M\rangle\;.
\end{multline}
There are only two terms in the sum, $J_e=j_e\pm\f12$, and for each of them the 3jm-symbols can be understood as matrix elements of Schwinger boson operators thanks to \eqref{a^A 3jm},
\beq
\langle 1/2, A\vert g_e\vert 1/2, B\rangle \langle j_e, n\vert g_e \vert j_e, m\rangle = \f1{d_{j_e}} \left[ \langle j_e,n\vert\, a_e^A\, g_e\, a_e^{B\dagger}\, \vert j_e, m\rangle + \langle j_e, n\vert\,(-1)^{\f12-A} a_e^{-A^\dagger}\, g_e\, (-1)^{\f12-B}a_e^{-B}\, \vert j_e, m\rangle \right].
\ee
This may be written in a more compact, symbolic form,
\beq \label{quantum hol}
\widehat{g}_e\,D^{(j_e)}(g_e) =  \Bigl( \vert \tl{a}_e\rangle \langle a_e \vert + \vert \tl{a}_e] [a_e\vert\Bigr)\ \f1{d_{j_e}}\,D^{(j_e)}(g_e).
\ee
This is obviously the quantization of the classical expression of the holonomy in terms of spinors \eqref{holonomy spinor}. Notice that the factor $1/d_j$ which enters here is nothing but the quantum translation of the holonomy normalization $1/\sqrt{\langle \tl{z}\vert \tl{z}\rangle \langle z\vert z\rangle}$. However, one has ordering ambiguities when promoting the classical expression \eqref{holonomy spinor} to an operator. What we find here is that the result of the correct ordering is $1/d_j$.

The relation \eqref{quantum hol} enables to get the translation of the parallel transport of spinors \eqref{spinor transport} at the quantum level:
\begin{gather}
\nonumber \sum_B (g_e)_{AB}\ \langle j_e-1/2, n\vert g_e\ a^B \vert j_e,m\rangle = \langle j_e-1/2,n \vert a^A\ g_e\vert j_e,m\rangle,\\
\sum_B (g_e)_{AB}\ \langle j_e, n\vert g_e\ (-1)^{\f12-B}\,a^{-B\dagger} \vert j_e-1/2,m\rangle = \langle j_e,n \vert (-1)^{\f12-A}\,a^{-A\dagger}\ g_e\vert j_e-1/2,m\rangle\;. \label{quantum transport}
\end{gather}
The analogous formula for the spin 1 is equivalent to the well-known fact that the adjoint action on a Pauli matrix is a rotation on the matrix-valued vector $\vec{\tau}$. Indeed from: $g\mone\tau^i g= \sum_j R(g)^i_{\phantom{i}j} \tau^j$, we get: $\sum_j R(g)^i_{\phantom{i}j} \langle j,n\vert g\,\tau^j\vert j,m\rangle = \langle j,n\vert \tau^i\,g\vert j,m\rangle$.

\section{Dynamics of BF theory} \label{sec:dyn}

The previous section presents the kinematics of loop quantum gravity on a single graph, or equivalently the framework of lattice gauge theory, formulated with spinors. We now study the dynamics of the BF model. The Hamiltonian consists in two constraints, one imposing the Gau\ss{} law, which we have treated as part of the kinematics, and the other the vanishing of the curvature, i.e. of magnetic fluxes.

First, we present the direct equivalent to the plaquette operator of the Kitaev model, which is based on characters of the Wilson loops. Those calculations are not really new, but the geometric interpretation as tent moves is and was suggested to us by B. Dittrich. Then we summarize the ideas coming from the new Hamitlonian introduced in \cite{3d-wdw}, before presenting its extension to spinors.




\subsection{The plaquette operator: Tent move evolution} \label{sec:tent}

We consider a plaquette $p$ bounded with $n$ links. The constraint $g_p=\unit$, \eqref{hol=1}, has three real components, since $\dim \SU(2)=3$, on each plaquette. This can also be seen by gauge fixing all holonomies around $p$ to the unit but one, which is then set to $\unit$ by the constraint.

The set of constraints \eqref{hol=1} is sufficient and the projector \eqref{projector} is well-defined for example on Riemann surfaces of genus higher than two \cite{witten-2dym}. To implement it on the lattice, it is standard to expand it over the $\SU(2)$ modes of the delta function,
\be
\delta(g_p) = \sum_{j_p\in\f{\N}2} d_{j_p}\ \chi_{j_p}(g_p)\;,
\ee
where $\chi_j$ is the character (the trace) in the spin $j$, normalized to $\chi_j(\unit)=d_j$. All terms in the above sum are well-defined as $\SU(2)$ invariant operators on spin networks, and this is the way Noui and Perez solved 2+1 gravity in \cite{noui-perez-ps3d}.

However, the set of constraints on the Wilson loops usually has redundancies which make the projector ill-defined, in the absence of a suitable regularization, as it would involve divergent products of Dirac deltas. One can still look for physical states directly as solutions of a well-defined Wheeler-DeWitt equation. As suggested by the above decomposition, one can use the character of the holonomy $g_p$ around $p$, for any irreducible representation of spin $j$, to enforce the constraint
\beq
\widehat{\chi_j(g_p)}\ \vert \psi\rangle_{\rm phys} - d_j\,\vert \psi\rangle_{\rm phys} = 0\;, \qquad j\in\f{\N}{2}.
\ee


\begin{figure}
\begin{center}
\includegraphics[width=11cm]{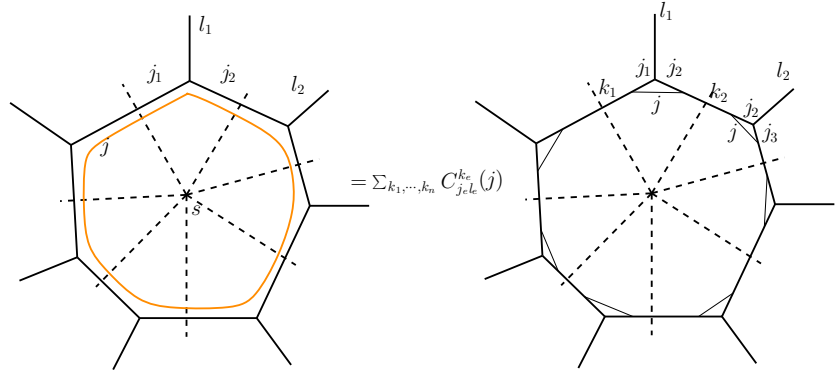}
\end{center}
\caption{ \label{fig:dualtent-nvalent} A pictorial representation of \eqref{recholgen}. The character $\chi_j$ along the closed loop acts on the left. On the right we have depicted the situation after re-coupling. A specific 6j-symbol is extracted on each node, and one has to sum over the colorings $k_1,\dotsc,k_n$. The dashed lines correspond to the dual 2d triangulation to the plaquette if we think of the latter as embedded in flat 3-space. The vertex $s$ of the 2d triangulation is then dual to the plaquette.}
\end{figure}

\begin{figure}
\begin{center}
\includegraphics[width=6cm]{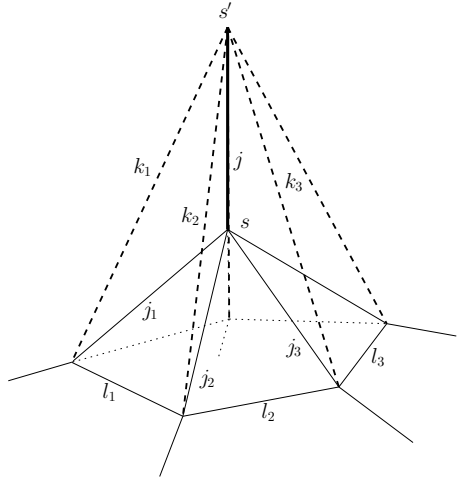}
\end{center}
\caption{ \label{fig:tentnvalent} Here we have displayed the geometric interpretation of the character operator on the plaquette as a tent move. The vertex $s$ is evolved to a new vertex $s'$, with an edge of length $j+\f12$, the tent pole. Between the initial and the final triangulations we have a piece of 3d triangulation. The character operator then generates the evaluation of the Ponzano-Regge amplitude on this triangulation.}
\end{figure}

When written in the spin network basis, this constraint leads to a difference equation \cite{recurrence-paper, mathese} which has some interesting features. The matrix elements of each holonomy in the character can be re-coupled with those of the same holonomy in the spin network function. This shifts the spins of the cycle $(j_1,\dotsc,j_n)$ to $(k_1,\dotsc,k_n)$ with sums over $k_i = \vert j_i-j\vert,\dotsc,j_i+j$. One is left with graspings in the spin $j$ at each node of the cycle, and they can be factorized from the function as 6j-symbols with a spin $j$. Precisely, this gives
\beq \label{recholgen}
\chi_j(g_1\dotsm g_n)\,s_{\{j_e\}}(g_e) = \sum_{k_1,\dotsc,k_n}  (-)^{j+\sum_{e=1}^n j_e+k_e+l_e}\prod_{e=1}^n d_{k_e} \begin{Bmatrix} k_2& j_2 & j \\ j_1& k_1 & l_1\end{Bmatrix}
\begin{Bmatrix} k_3& j_3 & j \\ j_2& k_2 & l_2\end{Bmatrix} \dotsm
\begin{Bmatrix}k_1& j_1 & j \\ j_n& k_n & l_n\end{Bmatrix}\,
s_{\{k_e\}}(g_e)\;.
\ee
This is represented in the figure \ref{fig:dualtent-nvalent}. Note that this enables to write the projector on the topological ground state restricted to a single plaquette in the spin network basis,
\beq \label{projector}
\delta(g_1\dotsm g_n)\,s_{\{j_e\}}(g_e) = \sum_{j,k_1,\dotsc,k_n}  (-)^{j+\sum_{e=1}^n j_e+k_e+l_e} d_j \prod_{e=1}^n d_{k_e}
\begin{Bmatrix} k_2& j_2 & j \\ j_1& k_1 & l_1\end{Bmatrix}
\begin{Bmatrix} k_3& j_3 & j \\ j_2& k_2 & l_2\end{Bmatrix} \dotsm
\begin{Bmatrix}k_1& j_1 & j \\ j_n& k_n & l_n\end{Bmatrix}\,
s_{\{k_e\}}(g_e)\;.
\ee


It has a very nice geometric interpretation in terms of a $(2+1)$-dimensional evolution. The reason for the $(2+1)$ dimensions is that it is natural and always possible to consider a plaquette to be dual to a piece of a 2-dimensional triangulation embedded in flat 3-space. The plaquette itself is dual to a vertex $s$, and the links on its boundary are dual to edges which connect $s$ to vertices $(s_i)$ dual to the surrounding plaquettes.

Then we use the Ponzano-Regge interpretation \cite{freidel-louapre-PR1} of the 6j-symbol: it is a weight associated a tetrahedron whose edge lengths are given by $(j_e+1/2)$. A Ponzano-Regge amplitude is then obtained by taking products of 6j-symbols corresponding to a gluing of tetrahedra, and summing over the spins of the internal dual edges, keeping those on the boundary fixed.

This interpretation shows that the equation \eqref{recholgen} is actually computing the Ponzano-Regge amplitude for a piece of a 3d triangulation built as follows. Draw from $s$ an edge outside of the 2d triangulation, with length $(j+1/2)$ going to a new vertex $s^*$, and connect $s^*$ to the vertices $(s_i)$. This process creates a piece of 3d triangulation with $n$ tetrahedra, all sharing the edge $(ss^*)$. Such an evolution process of the canonical surface is known as a {\bf tent move}, and it is depicted in the figure \ref{fig:tentnvalent}. It gives an evolution in discrete time steps, where the triangulations of the constant time slices are all the same. This evolution has been analyzed in background independent approaches to lattice gravity in \cite{bahr-broken-sym}. The tent pole is here $(ss^*)$ and has length $(j+1/2)$. By Fourier transforming \eqref{recholgen}, one sees that the evolved state $\psi(k_e)$ is obtained by summing over all admissible values of the spins on the initial surface,
\beq
\bigl(\operatorname{Tent}_j\,\psi\bigr)(k_e) =  \sum_{j_1,\dotsc,j_n}  (-)^{j+\sum_{e=1}^n j_e+k_e+l_e}\prod_{e=1}^n d_{j_e}
\times \begin{Bmatrix} k_2& j_2 & j \\ j_1& k_1 & l_1\end{Bmatrix}
\begin{Bmatrix} k_3& j_3 & j \\ j_2& k_2 & l_2\end{Bmatrix} \dotsm
\begin{Bmatrix}k_1& j_1 & j \\ j_n& k_n & l_n\end{Bmatrix}\ \psi(j_e)\;.
\ee
In particular, a physical solution to the flatness constraint must satisfy
\beq
\bigl(\operatorname{Tent}_j\,\psi\bigr)(k_e) = d_j\ \psi(k_e)\;,
\ee
that is it must be invariant under tent moves weighted by their Ponzano-Regge amplitudes. A way to enforce this requirement is to use the  projector on the topological sector \eqref{projector} which is geometrically interpreted as a summation over all possible tent pole lengths.

\subsection{The spin 1 Hamiltonian}

However, there are some drawbacks. In particular, the difference equation one gets shifts the spins on each link in the boundary of $p$. This means that one cannot solve the Wheeler-DeWitt equation on each single face independently of each other. And yet we know that in some situations, one should be able to get and solve an equation on each spin independently.

The simplest example is that of the canonical surface $S^2$ triangulated by the boundary of a tetrahedron. Let us write explicitly the spin network function, labelled by six spins $(j_1,\dotsc,j_6)$, following the figure \ref{fig:tet}:
\begin{multline} \label{defspinnet}
s^{\{j_e\}}_{\rm tet}(g_1,\dotsc,g_6) = \sum_{\substack{ m_1,\dotsc,m_6 \\n_1,\dotsc,n_6}} \begin{pmatrix} j_1 &j_2 &j_3\\ m_1 &m_2 &m_3\end{pmatrix} \begin{pmatrix} j_1 &j_5 &j_6\\ -n_1 &-n_5 &m_6\end{pmatrix} \begin{pmatrix} j_3 &j_4 &j_5\\-n_3 &-n_4 &m_5\end{pmatrix} \begin{pmatrix} j_2 &j_6 &j_4\\ -n_2 &-n_6 &m_4\end{pmatrix} \\
\times \left[\prod_{e=1}^6 (-1)^{j_e-n_e} \langle j_e, n_e \vert g_e \vert j_e,m_e\rangle\right] .
\end{multline}
The range of summation is $-j_e\leq m_e,n_e\leq j_e$ for each link. The natural inner product on $L^2(\SU(2)^L/\SU(2)^V)$ is evaluated with the Haar measure on $\SU(2)^L$, and given the chosen normalization for spin network functions we get here
\beq \label{ps}
\langle s^{\{j_e\}}_{\rm tet}\,\vert\, s^{\{k_e\}}_{\rm tet}\rangle = \prod_{e=1}^6 \frac{\delta_{j_e, k_e}}{d_{j_e}}\;.
\ee

Given the trivial topology on the 2-sphere and the chosen fixed graph, there is a single `physical', i.e. topological, state, which is peaked on flat connections with full measure:
\beq
\psi_{\rm phys}(g_1,\dotsc,g_6) = \delta(g_6 g_5 g_4)\,\delta(g_6 g_1 g_2\mone)\,\delta(g_3\mone g_4 g_2).
\ee
Here $\delta(g)$ is the Dirac delta on $\SU(2)$ with respect to the Haar measure. To get its expansion on the spin network basis, we take the inner product with a spin network function $s^{\{j_e\}}_{\rm tet}(g_1,\dotsc,g_6)$,
\begin{align}
\psi_{\rm phys}(j_1,\dotsc,j_6) &= \int \prod_{e=1}^6 dg_e\ s^{\{j_e\}}_{\rm tet}(g_1,\dotsc,g_6)\ \psi_{\rm phys}(g_1,\dotsc,g_6),\\
&= \begin{Bmatrix} j_1 &j_2 &j_3 \\j_4 &j_5 &j_6 \end{Bmatrix}.
\end{align}
This result simply comes from the fact that the physical state enforces the evaluation of the integral on the flat connections (up to gauge transformations), while there: $s_{\{j_e\}}(\mathbbm{1})=\{6j_e\}$.

Thus we could call Wheeler-DeWitt equation in that situation any equation which characterizes the 6j-symbol. As a re-coupling coefficient, the latter is known to satisfy a special identity, the Biedenharn-Elliott, or pentagon, identity \cite{varshalovich-book}. From it, one derives \cite{SG1} a recursion formula which enables the evaluation of the 6j-symbol,
\be \label{rec1 6j}
A_{+1}(j_1)\,\begin{Bmatrix} j_1+1 &j_2 &j_3 \\ j_4 &j_5 &j_6 \end{Bmatrix} +
A_{0}(j_1)\,\begin{Bmatrix} j_1 &j_2 &j_3 \\ j_4 &j_5 &j_6 \end{Bmatrix} +
A_{-1}(j_1)\,\begin{Bmatrix} j_1-1 &j_2 &j_3 \\ j_4 &j_5 &j_6 \end{Bmatrix} = 0.
\ee
The coefficients are given in \cite{3d-wdw} for example, and they are built on special 6j-symbols which have one spin equal to 1. It turns out that the coefficient $A_{+1}$ (respectively $A_{-1}$) vanishes when the spin $j_1$ reaches the highest admissible value (respectively the lowest admissible value). This means that the recursion is initially first order, i.e. it can be iterated from a single initial condition.

The relationship to the topological, flatness constraint \eqref{hol=1} was unraveled in \cite{3d-wdw}. We want an equation acting an a single face which selects the 6j-symbol in the case of a cycle with three links. It was shown in \cite{3d-wdw} that the above recursion can be derived from the quantization of a constraint on the phase space of loop quantum gravity. To build the constraint the idea is to extract the matrix elements of the holonomy around a plaquette $g_p$ by projecting onto a flux variable on the left and on the right. Consider the situation depicted in the figure \ref{fig:tet}, with $g_p = g_6 g_1 g_2\mone$ based on the vertex where $e_2$ and $e_6$ meet. We obtain a spin 1 Hamiltonian by projecting the rotation matrix $R(g_p)$ onto the fluxes $\tl{X}_2, \tl{X}_6$,
\beq \label{spin1H}
H^{(1)}_{e_6 e_1 e_2} = \sum_{i,j} \tl{X}_6^i\,\bigl( \delta_{ij} - R(g_6 \,g_1\,g_2\mone)_{ij}\bigr)\,\tl{X}_2^j = \tl{X}_{6}\cdot \tl{X}_{2} - \tl{X}_{6}\cdot \Ad(g_6 \,g_1\,g_2\mone)\,\tl{X}_{2}.
\ee
$\Ad$ is the adjoint representation on fluxes seen as Lie algebra variables, which is of course equivalent to the standard vector representation of $\SU(2)$. Thus, we look at the rotation $R(g_p)$ in the basis spanned by the flux variables. The constraint can be re-written in a more convenient way using the parallel transport relation between left and right fluxes \eqref{Xtilde}:
\beq
H^{(1)}_{e_6 e_1 e_2} = \tl{X}_{6}\cdot \tl{X}_{2} - X_6\,\cdot \Ad(g_1)\,X_2.
\ee
On a spin network function, the first term $\tl{X}_{6}\cdot \tl{X}_{2}$ is diagonal and given by \eqref{quantum dihedral}. The second term is a bit more involved but an essential point is that there is only one holonomy $g_1$, and hence only one spin, $j_1$ gets shifted by $-1,0,+1$. This leads to the difference equation \cite{3d-wdw}:
\be \label{spin 1 wdw}
A_{+1}(j_1)\,\psi(j_1+1) +
A_{0}(j_1)\,\psi(j_1) +
A_{-1}(j_1)\,\psi(j_1-1) = 0.
\ee
The coefficients are:
\begin{align} \label{A_0}
A_0(j_1) &= (-1)^{j_2+j_4+j_6}\begin{Bmatrix} j_2 & j_2 &1 \\ j_6 &j_6 &j_4\end{Bmatrix} + (-1)^{2j_1+j_2+ j_3+j_5+j_6}(2j_1+1)\ \begin{Bmatrix} j_1 & j_1 &1\\ j_2 &j_2 &j_3\end{Bmatrix}\,\begin{Bmatrix} j_1 & j_1 &1\\ j_6 &j_6 &j_5\end{Bmatrix},\\
A_{\pm 1}(j_1) &= (-1)^{2j_1+j_2+ j_3+j_5+j_6+1}\bigl(2(j_1\pm1)+1\bigr)\ \begin{Bmatrix} j_1\pm1 & j_1 &1\\ j_2 &j_2 &j_3\end{Bmatrix}\,\begin{Bmatrix} j_1\pm 1 & j_1 &1\\ j_6 &j_6 &j_5\end{Bmatrix}. \label{Apm1}
\end{align}
This difference equation is exactly the recursion relation satisfied by the 6j-symbol.

\begin{figure}
\begin{center}
\includegraphics[scale=0.65]{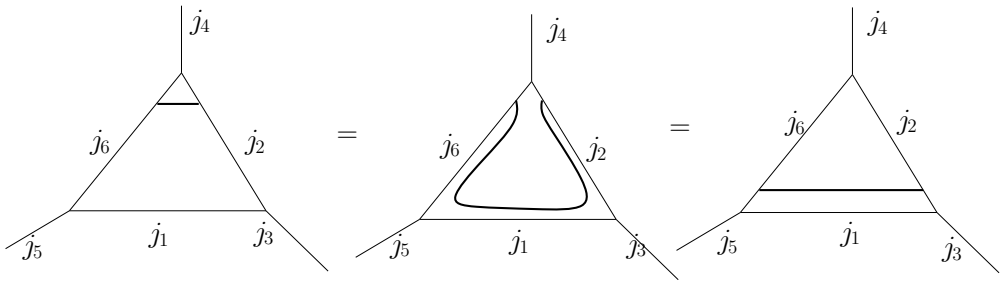}
\end{center}
\caption{ \label{fig:newH} A graphical representation of the action of the new Hamiltonian. The basic idea is that the holonomy around a closed loop in the topological sector only depends on its homotopy type, so that we can deform the grasping on the left to that on the right, picking up this way some holonomy which must be trivial.}
\end{figure}

A graphical representation of the action of $H^{(1)}$ is given in the figure \ref{fig:newH}. The idea is that on the topological sector, the holonomy along a closed loop only depends on its homotopy type. Hence, we can make a grasping, i.e. insert a line inside the plaquette whose ends are glued on the boundary of the plaquette, and being in the topological sector means that the curve can be deformed, since the plaquette is homotopically trivial. The Hamiltonian proceeds like this: it inserts a infinitesimal curve (i.e. with trivial holonomy) at node between two links ($e_2, e_6$ on the figure \ref{fig:newH}). The constraint is just that this curve can be deformed to run all along the plaquette, picking up the corresponding holonomies. Hence, it indeed produces a constraint on these holonomies, whose product has to be trivial.


In addition to this topological interpretation, it enjoys the following properties.
\begin{itemize}
 \item It is labeled by a cycle (which bounds $p$) and a vertex $v$ in the cycle. Since a cycle has at least three links, this gives a sufficient number of constraints to impose the topological condition $R(g_p)=\unit \in \operatorname{SO}(3)$ (i.e. $g_p=\pm\unit\in\SU(2)$).
 \item $H^{(1)}_{v,p}$ appears as a regularization of the scalar Hamiltonian in the 2+1 gravity,
 \be
 H_{\rm GR} = E^a_i E^b_j \epsilon^{ij}_{\phantom{ij}k}F_{ab}^k\;.
 \ee
 $F$ is the field strength, and $E$ the pull-back of the triad to the canonical surface, and $a,b$ are space indices. It comes from projecting the constraint $F=0$ onto the lapse and the shift. This splits the full 3d diffeomorphism algebra into a constraint which generates spatial diffeomorphisms, plus $H_{\rm GR}$ which generates time reparametrization. However, it is unclear how the same splitting can be done on the lattice.

 The answer provided by $H^{(1)}$ is to project the curvature around a node, say $s$ in the figure \ref{fig:dualtent-nvalent}, onto the normals to each of the triangles sharing $s$. Generically, if there are at least three triangles, all directions of 3-space will be spanned. This way we get an evolution of the vertex $s$ in at least three ``time'' directions.
 \item In 3d, the phase space of loop quantum gravity consists in Regge geometries, which means they have intrinsic geometries determined by the spins interpreted as lengths. The constraint $H^{(1)}_{v,p}=0$ imposes a relation between the intrinsic, spatial geometry and its extrinsic curvature characterizing the embedding into flat spacetime.
 \item In 4d BF theory, the geometric interpretation is similar in the Regge sector, but they are also non-Regge, discontinuous intrinsic geometries in the phase space \cite{dittrich-ryan1}, known as twisted geometries. $H^{(1)}_{v,p}$ provides a well-defined constraint and well-defined recursion relations for the twisted geometries \cite{recursion-semiclass}.
 \item The asymptotics of the equation can be studied, for example, using coherent intertwiners. The solutions to the lowest order in the WKB approximation in the Regge sector are linear combinations of $\exp(\pm iS_{\rm R})$, where $S_{\rm R}$ is the Regge action of the simplex. We thus get a criterion to tell whether the Regge action is part of the asymptotics of a lattice model \cite{recursion-semiclass}. The Hamiltonian $H^{(1)}_{v,p}$ is an example which generates exactly solvable recursions beyond this asymptotic behavior.
\end{itemize}
Those are features we would like to reproduce using spinors, for the group $\SU(2)$ instead of $\SO(3)$.

\subsection{The spin 1/2 Hamiltonian}

\subsubsection{The classical spinor Hamiltonian}

We labeled the operator $H^{(1)}_{f,v}$ in \eqref{spin1H} with a superscript $(1)$ because the derivatives $X^i$ transform in the adjoint representation of $\SU(2)$, that is with a spin 1. As a consequence (see \cite{3d-wdw}), the recursion relation obtained involves shifts on a spin by the amount $\pm 1$.

Thus, if the recursion \eqref{spin 1 wdw} is iterated from an initial condition with all spins being integers, we can solve the family of 6j-symbols with integer spins. This amounts to solving the model for $\SO(3)$. However, from this solution we cannot get information on the family of symbols with, say, three half-integer spins and the three others being integers. Hence, these two families of 6j-symbols have to be related by hand, choosing appropriately (i.e. "fine-tuning") the initial conditions.

If one could derive a recursion with spins $1/2$, one would then be in position of solving directly the model for the group $\SU(2)$. To this aim, one needs variables transforming under the fundamental representation of $\SU(2)$, in contrast with the flux variables. We have just described such variables in the previous section, the spinors.

Hence, we mimic the construction of the constraint \eqref{spin1H}. In particular, the graphical representation should be unchanged, as given in the figure \ref{fig:newH}, but with insertions of spin 1/2 instead of spin 1. Technically, the rotation matrix $R(g)$ should now be changed with the fundamental representation. Thus, we propose
\beq \label{def H1/2}
H^{(--)}_{e_6 e_1 e_2} \equiv \langle \tl z_6\vert \tl z_2]\ [\tl z_6\vert\  g_6\,g_1\,g_2\mone - \unit \ \vert \tl z_2\rangle.
\ee
Note that we could actually write four such constraint, by changing $\tl z_2, \tl z_6$ with $\varsigma\tl z_2$ and/or $\varsigma\tl z_6$,
\beq
\begin{aligned}
H^{(-+)}_{e_6 e_1 e_2} &\equiv \langle \tl z_6\vert \tl z_2\rangle\ [\tl z_6\vert\  g_6\,g_1\,g_2\mone - \unit \ \vert \tl z_2],\\
H^{(+-)}_{e_6 e_1 e_2} &\equiv [\tl z_6\vert \tl z_2]\ \langle\tl z_6\vert\  g_6\,g_1\,g_2\mone - \unit \ \vert \tl z_2\rangle,\\
H^{(++)}_{e_6 e_1 e_2} &\equiv [\tl z_6\vert \tl z_2\rangle\ \langle\tl z_6\vert\  g_6\,g_1\,g_2\mone - \unit \ \vert \tl z_2].
\end{aligned}
\ee
Some simple algebra on spinors gives that the sum (with correct signs) of these four observables leads back to the holonomy around the cycle (126) of our tetrahedral graph,
\be
H^{(-+)}_{e_6 e_1 e_2}+H^{(+-)}_{e_6 e_1 e_2}-H^{(++)}_{e_6 e_1 e_2}-H^{(--)}_{e_6 e_1 e_2}
=
\langle \tl z_2\vert \tl z_2\rangle\langle \tl z_6\vert \tl z_6\rangle\,
\tr_{\f12}(g_6\,g_1\,g_2\mone - \unit)\;,
\ee
where $\tr_{\f12}$ is the trace on 2$\times$2 matrices. Therefore, our $H^{(\alpha\beta)}$ observables provide a decomposition of the holonomy operator around the closed loop (126) into simpler components, which will translate into quantum operators generating specific shifts of the spins around the loop.

Moreover, these observables are not independent and there are some relations between them. For instance, taking into account that $[z\vert g\vert z'\rangle = -\overline{\langle z\vert g \vert z']}$ for group elements $g\in\SU(2)$, one gets $H^{(\epsilon_6 \epsilon_2)}_{e_6 e_1 e_2} = \overline{H^{(-\epsilon_6 -\epsilon_2)}_{e_6 e_1 e_2}}$, for $\epsilon_2,\epsilon_6=\pm$.
Furthermore the fact that $g=g_6\,g_1\,g_2\mone$ is unitary, $gg^\dagger=\unit$, translates into quadratic relations between the $H^{(\pm\pm)}$ observables\footnotemark. Finally, we obtain that there are only three independent real components.
More precisely, we would like to impose all the constraints $H^{(\alpha\beta)}=0$. These are $2\times 4$ real constraints. Due to the complex conjugation relations, this already reduces to 4 real constraints. Then discarding the pre-factors, $\langle \tl z_6\vert \tl z_2]$ and so on, imposing $H^{(\alpha\beta)}=0$ amounts to requiring that the four matrix elements of the 2$\times$2 matrix $g_6\,g_1\,g_2\mone - \unit$ vanish. Since $g_6\,g_1\,g_2\mone$ is in $\SU(2)$ and only depends on three independent real parameters, we do have in the end only 3 independent real constraints, which simply express that the holonomy around the closed loop (126) must be the identity, $g_6\,g_1\,g_2\mone =\unit$.
\footnotetext{
We decompose the relation $gg^\dagger=\unit$ for $g=g_6\,g_1\,g_2\mone$ into its explicit matrix elements by inserting twice the identity in term of spinors $\langle z|z \rangle\,\unit= |z\rangle\langle z|+|z][z|$ for $z=\tl z_6$ and $z=\tl z_2$.
}
%


The above form of the constraint contains the holonomies along three different links. But it can be simplified as depending of the holonomy along a single link with the following lemma.
\begin{lemma} \label{lemma:H1/2class}
The observable $H_{e_6 e_1 e_2}$ admits a form with only the holonomy along $e_1$,
\beq
H^{(--)}_{e_6 e_1 e_2} = \langle \tl z_6\vert \tl z_2]\ \sqrt{\frac{ \langle \tl z_2\vert \tl z_2\rangle\, \langle \tl z_6\vert \tl z_6\rangle}{\langle z_2\vert z_2\rangle\,\langle z_6\vert z_6\rangle}}\  [z_6\vert\ g_1\ \vert z_2\rangle - \langle \tl z_6\vert \tl z_2]\ [\tl z_6\vert \tl z_2\rangle \;,
\ee
and a form completely in terms of spinors,
\beq
H^{(--)}_{e_6 e_1 e_2} = \sqrt{\frac{ \langle \tl z_2\vert \tl z_2\rangle\, \langle \tl z_6\vert \tl z_6\rangle}{\langle z_2\vert z_2\rangle\,\langle z_6\vert z_6\rangle}}\  \langle \tl z_6\vert \tl z_2] \frac{ [z_6\vert \Bigl(\vert \tl z_1\rangle\langle z_1\vert + \vert \tl z_1][z_1\vert\Bigr)\vert z_2\rangle}{\sqrt{\langle z_1\vert z_1\rangle\,\langle \tl{z}_1\vert \tl{z}_1\rangle }} - \langle \tl z_6\vert \tl z_2]\ [\tl z_6\vert \tl z_2\rangle\;.
\ee
\end{lemma}

They simply come from applying the parallel transport relations of spinors, \eqref{spinor transport} and \eqref{holonomy spinor}, to get rid of the different holonomies.

\subsubsection{The quantum spinor Hamiltonian}

Promoting the above quantities to operators is done by using the quantization map we have discussed in the previous section. Mainly, the spinor contractions are sent to operators $\widehat{E}_{ee'}, \widehat{F}_{ee'}$ and $\widehat{F}^\dagger_{ee'}$.

At the quantum level, one faces numerous ordering ambiguities. However, several of them are resolved from the fact that we know the standard action of matrix elements of holonomies on spin networks. From this, we have shown on the equation \eqref{quantum hol} that the norm of the spinors in the expression of $g_1$, $1/\sqrt{\langle z_1\vert z_1\rangle\,\langle \tl{z}_1\vert \tl{z}_1\rangle }$, only contributes to a factor $1/d_{j_1}$ in the spin network basis. In the equation \eqref{quantum transport}, we have shown that the parallel transport $[\tl z_6\vert\  g_6$ does not bring any factor, which means that the ratio $\sqrt{\langle \tl z_2\vert \tl z_2\rangle}/{\langle z_2\vert z_2\rangle}$ is naturally set to 1 at the quantum level.

This way, one has reduced the ordering ambiguities to a single one, the ordering of the two factors in the definition \eqref{def H1/2} of the Hamiltonian. Let us define two operators, one for each ordering, say left and right. Using the forms of the classical observable displayed in the lemma \ref{lemma:H1/2class}, the quantization with the left ordering is
\begin{align}
\widehat{H}^{(--)}_{e_6 e_1 e_2|\rm L}\ s^{\{j_e\}}_{\rm tet} &= \langle \tl a_6\vert \tl a_2]\,\Bigl([a_6\vert\,g_1\,\vert a_2\rangle - [\tl a_6\vert \tl a_2\rangle\Bigr)\ s^{\{j_e\}}_{\rm tet}\;, \label{H1/2L}\\
&= \hF_{26}^\dagger\ \Bigl(\f1{d_{j_1}}\,\bigl( \hF_{61}\,\hE_{12} + \hE_{16}\,\hF_{12}\bigr) - \hF_{62}\Bigr)\ s^{\{j_e\}}_{\rm tet}\;,
\end{align}
while the right ordering is
\begin{align}
\widehat{H}^{(--)}_{e_6 e_1 e_2|\rm R}\ s^{\{j_e\}}_{\rm tet} &= \Bigl([a_6\vert\,g_1\,\vert a_2\rangle - [\tl a_6\vert \tl a_2\rangle\Bigr)\, \langle \tl a_6\vert \tl a_2]\ s^{\{j_e\}}_{\rm tet}\;, \label{H1/2R}\\
&=  \Bigl(\f1{d_{j_1}}\,\bigl( \hF_{61}\,\hE_{12} + \hE_{16}\,\hF_{12}\bigr) - \hF_{62}\Bigr)\,\hF_{26}^\dagger\ s^{\{j_e\}}_{\rm tet}\;. \label{H1/2Rbis}
\end{align}
Note that $\hF_{26}^\dagger$ commutes with $(\hF_{61}\,\hE_{12} + \hE_{16}\,\hF_{12})$, since $\hF_{26}^\dagger$ acts on the vertex where $e_2, e_6$ meet, while that $(\hF_{61}\,\hE_{12} + \hE_{16}\,\hF_{12})$ act on the two other vertices of the cycle. Hence, the true ambiguity is the ordering between $\hF_{26}^\dagger$ and $\hF_{62}$.

The quantization of $\widehat{H}^{(\alpha,\beta)}$ for $\alpha,\beta=\pm$ is obtained by changing in the above definitions $\tl a_2$ (and/or $\tl a_6$) to $\varsigma \tl a_2$ (and/or $\varsigma \tl a_6$). Then, it turns out there is a relationship between the two orderings,
\be \label{Hdagger}
\Bigl(\widehat{H}^{(\alpha, \beta)}_{e_6 e_1 e_2|\rm R}\Bigr)^\dagger = \widehat{H}^{(-\alpha, -\beta)}_{e_6 e_1 e_2|\rm L}\;.
\ee
This is the quantum equivalent to $H^{(\alpha, \beta)}_{e_6 e_1 e_2} = \overline{H^{(-\alpha, -\beta)}_{e_6 e_1 e_2}}$.

\begin{theorem} \label{prop:wdw}
Consider the 2-sphere triangulated by the boundary of a tetrahedron, where kinematical states take the form $\psi(g_1,\dotsc,g_6) = \sum_{\{j_e\}} \bigl[\prod_{e=1}^6 d_{j_e}\bigr] \psi(j_1,\dotsc,j_6)\,s_{\rm tet}^{\{j_e\}}(g_1,\dotsc,g_6)$. The Wheeler-DeWitt equations for the physical, topological state are
\beq
\widehat{H}^{(\alpha,\beta)}_{e_6 e_1 e_2|L}\ \vert \psi\rangle =0\;,
\ee
for $\alpha,\beta=\pm1/2$. In the spin representation, they give difference equations,
\begin{multline} \label{1/2coeffs}
A^{(-\alpha,-\beta)}_{+\f12}(j_1)\ \psi\bigl(j_1+1/2,j_2-\beta,\dotsc,j_6-\alpha \bigr) \,+\, A^{(-\alpha,-\beta)}_{-\f12}(j_1)\ \psi\bigl(j_1-1/2, j_2-\beta,\dotsc,j_6-\alpha \bigr)\\
+ A^{(-\alpha,-\beta)}_{0}(j_1)\ \psi(j_1,\dotsc,j_6) =0\;,
\end{multline}
which are known to come from the Biedenharn-Elliott identity and determine the 6j-symbol.
\end{theorem}

The coefficients of the difference equations are
\begin{align}
\nonumber A^{(\alpha,\beta)}_0(j_1) &= (-1)^{j_2+j_4+j_6+1}\begin{Bmatrix} j_2 & j_2+\beta &\f12 \\ j_6+\alpha &j_6 &j_4\end{Bmatrix}\;, \\
A^{(\alpha,\beta)}_{\pm \f12}(j_1) &= (-1)^{2j_1+j_2+ j_3+j_5+j_6+\f{\pm1+1}{2}+\alpha+\beta}\,d_{j_1\pm\f12}\ \begin{Bmatrix} j_1\pm\f12 & j_1 &\f12\\ j_2 &j_2+\beta &j_3\end{Bmatrix}\,\begin{Bmatrix} j_1\pm \f12 & j_1 &\f12\\ j_6 &j_6+\alpha &j_5\end{Bmatrix}\;. \label{coeffs}
\end{align}
The relation \eqref{Hdagger} between $\widehat{H}_{|\rm L}$ and the adjoint of $\widehat{H}_{|\rm R}$ translates to identities between those coefficients,
\begin{align}
\nonumber &A_0^{(\alpha, \beta)}(j_1, j_2-\alpha, \dotsc, j_6-\beta) = (-1)^{\alpha+\beta}\,A_0^{(-\alpha, -\beta)}(j_1,j_2,\dotsc,j_6)\;,\\ &d_{j_1+\f12}\,A_{-\f12}^{(\alpha, \beta)}(j_1+\f12, j_2-\alpha, j_6-\beta) = d_{j_1}\, A_{+\f12}^{(-\alpha, -\beta)}(j_1,j_2, \dotsc,j_6)\;,
\end{align}
which can also be checked directly from \eqref{coeffs}. Those identities can be used to prove \eqref{1/2coeffs} by direct calculations, which are reported in the appendix \ref{app:directproof}.

There are four possibilities for $(\alpha, \beta)$, and hence four difference equations for each plaquette. However, due to the discrete symmetries of the tetrahedral graph, only three equations are actually different. The physical reason why three equations per plaquette are independent is that the Hamiltonian is built to impose (classically) the equation $g_p=\unit$. For each face, this is a $\SU(2)$-valued equation which has three real components.

Combining three of these equations, one can reproduce \cite{SG1} the spin 1 recursion formula \eqref{rec1 6j} on the 6j-symbol. Since the latter can be solved from a single initial condition, we can obtain families of 6j-symbols where the spins differ from the initial spins by integers. Two such families are $\left\{ \begin{smallmatrix} n_1 &n_2 &j_3\\j_4 &j_5 &n_6\end{smallmatrix}\right\}$ and $\left\{ \begin{smallmatrix} n_1+\f12 &n_2+\f12 &j_3\\j_4 &j_5 &n_6+\f12\end{smallmatrix}\right\}$, where $n_1, n_2, n_6\in\N$.

To relate the different families, one has to resolve steps of half-integers (which are not integers). This is precisely done using the spin 1/2 recursions, which consistently set the relationships between the initial conditions of the different families.

We will prove the theorem for $\widehat{H}^{(--)}$, as it works similarly for the other cases. It follows from the two propositions below, together with the equation \eqref{Hdagger}.

\begin{proposition} \label{prop:method}
$\widehat{H}^{(--)}_{e_6 e_1 e_2|\rm R}\ s^{\{j_e\}}_{\rm tet}$ annihilates the tetrahedral spin network on the identity,
\beq
\langle \psi_{\rm phys}\,\vert \widehat{H}^{(--)}_{e_6 e_1 e_2|\rm R}\,\vert s^{\{j_e\}}\rangle =
\widehat{H}^{(--)}_{e_6 e_1 e_2|\rm R}\ s^{\{j_e\}}_{\rm tet}(g_1,\dotsc, g_6)_{|g_e=\unit} = 0\;.
\ee
\end{proposition}

Before showing up the calculations, let us emphasize that it is easy to understand the action of the operator. Indeed, consider the form \eqref{H1/2R} of the operator. The multiplication of the spin network by matrix elements of $g_1$ in its fundamental representation induces shifts of the spin $j_1$ to $j_1\pm 1/2$. Then the role of the operators $a_6, \tl a_6, a_2, \tl a_2$ is $i)$ to make the whole operator $\SU(2)$ invariant, and $ii)$ to shift $j_2, j_6$ so that the rules of tensor products of representations are still satisfied after the shifts of $j_1$ (i.e. it is necessary to shift $j_2$ so that $j_3$ is in the tensor product $\calH_{j_1\pm\f12}\otimes \calH_{j_2-\f12}$).

The proof of this proposition is based on a very simple method which can be applied in more generic situations than the tetrahedral spin network \cite{in prepa}.

{\bf Proof of proposition \ref{prop:method}.} First observe that the effect of $F_{26}^\dagger = \langle \tl a_6\vert \tl a_2]$, given by the equation \eqref{F26dagger}, is to shift the spins $j_2, j_6$ on the half-lines of $e_2, e_6$ meeting at $v$ to $l_2 = j_2-\f12, l_6 = j_6-\f12$. Thus, the intertwiner $\iota_{j_2^* j_6^* j_4}$ in the spin network function is changed with $\iota_{l_2^* l_6^* j_4}$. Up to some multiplicative pre-factors, we are left with the remaining action of the operator, that is $[a_6\vert\,g_1\,\vert a_2\rangle - [\tl a_6\vert \tl a_2\rangle$, using the form of the operator given in \eqref{H1/2R}. It is not hard to see that this gives zero due to the somewhat trivial following identity,
\begin{align} 
\nonumber &\sum_{A,B=\pm\f12}(-1)^{\f12-A}\,\langle 1/2, A\vert\,g_1\,\vert 1/2, B\rangle\ \langle l_2, n_2\vert\, g_2\, a^B\, \vert j_2, m_2\rangle\ \langle l_6, n_6\vert\, g_6\,a^{-A}\, \vert j_6, m_6\rangle
\Bigl. \Bigr\rvert_{\unit} \\
&= \sum_{A=\pm\f12}(-1)^{\f12-A}\,\langle l_2, n_2\vert\, a^A\,g_2\, \vert j_2, m_2\rangle\ \langle l_6, n_6\vert\, a^{-A}\,g_6\, \vert j_6, m_6\rangle
\Bigl. \Bigr\rvert_{\unit}\;. \label{spin1/2Hafter}
\end{align}
We emphasize that this is one of the key identities in this paper. It states that we can commute the creators and annihilators with the group elements, here $g_2, g_6$, when the latter are taken to be the unit of the group. Moreover, in that case, the sum of the left hand side simplifies since $\langle 1/2, A\vert\,g_1\,\vert 1/2, B\rangle_{|\unit} = \delta_{A,B}$.

To see the action of $[a_6\vert\,g_1\,\vert a_2\rangle - [\tl a_6\vert \tl a_2\rangle$ in the above equation, one has to dress it with the matrix elements $\langle j_e,n_e\vert g_e\vert j_e,m_e\rangle_{|\unit}$ on the links $e\neq e_2, e_6$, and contract them with the four intertwiners $\iota_{j_1 j_2 j_3}, \iota_{j_1^* j_5^* j_6}, \iota_{j_3^* j_4^* j_5}$ and $\iota_{l_2^* l_6^* j_4}$. That gives the equivalent form
\begin{multline} \label{keyeq}
\frac1{d_{j_1}}\sum_{n_1,\dotsc,n_6} \Bigl[\langle j_6-1/2, n_6\vert F_{61}\,\iota_{j_1^* j_5^* j_6}\vert j_1 +1/2,n_1 ; j_5,n_5\rangle\ \langle j_1+1/2,n_1; j_2-1/2,n_2;j_3,n_3\vert E_{12}\,\iota_{j_1 j_2 j_3}\vert 0\rangle \\
\begin{aligned}
&+ \langle j_6-1/2, n_6\vert E_{16}\,\iota_{j_1^* j_5^* j_6}\vert j_1 -1/2,n_1 ; j_5,n_5\rangle\,\langle j_1-1/2,n_1; j_2-1/2,n_2;j_3,n_3\vert F_{12}\,\iota_{j_1 j_2 j_3}\vert 0\rangle\Bigr]\\
&\times \langle j_5,n_5\vert \iota_{j_3^* j_4^* j_5}\vert j_3,n_3;j_4,n_4\rangle\,\langle j_4,n_4\vert \iota_{(j_2-\f12)^*(j_6-\f12)^* j_4}\vert j_2-1/2,n_2; j_6-\f12,n_6\rangle\\
=\ & \sum_{m_1,\dotsc,m_6} \Bigl[\langle j_6, m_6\vert \iota_{j_1^* j_5^* j_6}\vert j_1 ,m_1 ; j_5,m_5\rangle\ \langle j_1,m_1; j_2,m_2;j_3,m_3\vert \iota_{j_1 j_2 j_3}\vert 0\rangle\ \langle j_5,m_5\vert \iota_{j_3^* j_4^* j_5}\vert j_3,m_3;j_4,m_4\rangle\Bigr]
\end{aligned} \\
\times \langle j_4,m_4\vert F_{62}\, \iota_{(j_2-\f12)^*(j_6-\f12)^* j_4}\vert j_2,m_2; j_6,m_6\rangle\;.
\end{multline}
Both sides of that identity are contractions of four intertwiners following the pattern of the tetrahedral graph. The holonomies have been evaluated on the unit. The left hand side contains two different terms, obtained by putting for the holonomy $g_1 = ( \vert \tl{a}_e\rangle \langle a_e \vert + \vert \tl{a}_e] [a_e\vert)/d_{j_1}$ in the above \eqref{spin1/2Hafter}. These terms correspond to $[a_6\vert\,g_1\,\vert a_2\rangle = (E_{16}F_{12}+ F_{61}E_{12})/d_{j_1}$, and encode two shifts of $j_1$, by $\pm1/2$. The right hand side is the part $F_{62} = [\tl a_6\vert \tl a_2\rangle$.
\qed

\begin{proposition} \label{prop:rec1/2}
Moreover, expanding the different terms of the proposition \ref{prop:method} using re-coupling produces a spin $1/2$ recursion formula on the 6j-symbol,
\begin{multline}
\widehat{H}^{(--)}_{e_6 e_1 e_2|\rm R}\ s^{\{j_e\}}_{\rm tet}(g_1,\dotsc, g_6)_{|g_e=\unit}  \\
= A^{(--)}_{+\f12}(j_1)\ \begin{Bmatrix} j_1+\f12 &j_2-\f12 &j_3 \\j_4 &j_5 &j_6-\f12\end{Bmatrix} + A_{0}^{(--)}(j_1)\ \begin{Bmatrix} j_1 &j_2 &j_3 \\j_4 &j_5 &j_6\end{Bmatrix} + A_{-\f12}^{(--)}(j_1)\ \begin{Bmatrix} j_1-\f12 &j_2-\f12 &j_3 \\j_4 &j_5 &j_6-\f12 \end{Bmatrix}\;.
\end{multline}
\end{proposition}

{\bf Proof of proposition \ref{prop:rec1/2}.} It simply follows from considering the form of the operator $H^{(1/2)}_{e_6 e_1 e_2|R}$ displayed in the equation \eqref{H1/2Rbis}, and evaluating it using the actions of the operators $E_{ee'}, F_{ee'}, F_{ee'}^\dagger$ which are given in the section \ref{sec:grasping}.

Using \eqref{F62}, \eqref{F26dagger}, we get that $F_{62}\, F_{26}^\dagger$ is diagonal on spin networks,
\beq
-F_{62}\, F_{26}^\dagger\ s^{\{j_e\}}_{\rm tet} \,=\, d_{j_2}d_{j_2-\f12}\,d_{j_6}d_{j_6-\f12}\ \begin{Bmatrix} j_2 &j_2-\f12 &\f12 \\j_6-\f12 &j_6 &j_4\end{Bmatrix}^2\ s^{\{j_1, j_2,\dotsc,j_6\}}_{\rm tet}\;.
\ee
The term which lowers $j_1$ by $-1/2$ is obtained thanks to \eqref{E16}, \eqref{F12} and \eqref{F26dagger},
\begin{multline} \label{E16 F12 F26dagger}
\f1{d_{j_1}}\,E_{16}\,F_{12}\,F_{26}^\dagger\ s^{\{j_e\}}_{\rm tet} \,=\, (-1)^{j_2+j_4+j_6} (-1)^{j_1+j_5+j_6}(-1)^{j_1+j_2+j_3} d_{j_1-\f12}\,d_{j_2}d_{j_2-\f12}\,d_{j_6}d_{j_6-\f12}\\
\begin{Bmatrix} j_2 &j_2-\f12 &\f12 \\j_6-\f12 &j_6 &j_4\end{Bmatrix} \begin{Bmatrix} j_1 &j_1-\f12 &\f12\\ j_6-\f12 &j_6 &j_5\end{Bmatrix} \begin{Bmatrix} j_1 &j_1-\f12 &\f12\\ j_2-\f12 &j_2 &j_3\end{Bmatrix}\ s^{\{j_1-\f12, j_2-\f12,\dotsc,j_6-\f12\}}_{\rm tet}\;.
\end{multline}
The term which increases $j_1$ by $+1/2$ is obtained using \eqref{F61}, \eqref{E12} and \eqref{F26dagger} again,
\begin{multline}
\f1{d_{j_1}}\,F_{61}\,E_{12}\,F_{26}^\dagger\ s^{\{j_e\}}_{\rm tet} \,=\, (-1)^{j_2+j_4+j_6+1} (-1)^{j_1+j_5+j_6}(-1)^{j_1+j_2+j_3} d_{j_1+\f12}\,d_{j_2}d_{j_2-\f12}\,d_{j_6}d_{j_6-\f12}\\
\begin{Bmatrix} j_2 &j_2-\f12 &\f12 \\j_6-\f12 &j_6 &j_4\end{Bmatrix} \begin{Bmatrix} j_1 &j_1+\f12 &\f12\\ j_6-\f12 &j_6 &j_5\end{Bmatrix} \begin{Bmatrix} j_1 &j_1+\f12 &\f12\\ j_2-\f12 &j_2 &j_3\end{Bmatrix}\ s^{\{j_1+\f12, j_2-\f12,\dotsc,j_6-\f12\}}_{\rm tet}\;.
\end{multline}
Then, we sum the three contributions, add a phase $(-1)^{j_2+j_4+j_6+1}$, divide by a common factor, $d_{j_2}d_{j_2-\f12}\,d_{j_6}d_{j_6-\f12}\times\left\{ \begin{smallmatrix} j_2 &j_2-\f12 &\f12 \\j_6-\f12 &j_6 &j_5\end{smallmatrix}\right\}$, and finally evaluate the spin network function on the unit of the group.
\qed

\subsection{Geometric interpretation: Quantum Euclidean geometry}

Although we consider throughout this paper the group $\SU(2)$ as the local symmetry group, our methods are expected to generically apply to any compact Lie group or finite group (see \cite{baby-sf} for a recent review on spin net and spin foam models with finite groups). But $\SU(2)$ is interesting from the geometric point of view. In the three-dimensional context, it is the (universal cover of the) isometry group of tangent spaces to spacetime. This leads to an interpretation of the spin network geometry which is well-known in the context of 3d quantum gravity, revisited here with the new spin 1 and spin 1/2 Hamiltonians.

The key idea is to consider the piece of 2d triangulation dual to the spin network graph, whose triangles are dual to the 3-valent vertices, edges dual to links, and vertices dual plaquettes. By a trivial extension of the graph (with links carrying the spin zero), we can assume that it is a triangulation of the canonical surface (space), hence characterizing its topology.

The usual phase space variables, holonomies and fluxes, characterize the geometry of an embedding of the triangulation into 3-space. The Hamiltonian constraint of curvature vanishing asks for an embedding in flat 3-space. We briefly sum up how this can be seen and refer to \cite{3d-wdw} for details and references.

The three fluxes on a vertex are interpreted as the normals to the edges of the dual triangulation. This is possible due to the Gau\ss{} law \eqref{closure}, which is exactly the closure condition for the triangle dual to the node. That also means that the norm of a flux, $X_e^2$, is the squared length of the dual edge. Similarly, the dihedral angle $\phi_{ee'}$ between two dual edges $(e^*, e^{'*})$ of a triangle is contained in the dot product of $X_{e}, X_{e'}$, where the dual edges $e, e'$ meet at a vertex:
\beq
\cos \phi_{ee'} = -\eps_{ee'}\ \f{X_e\cdot X_{e'}}{\vert X_e\vert \vert X_{e'}\vert},
\ee
where $\eps_{ee'}$ is $1$ if $e$ and $e'$ are both outgoing or ingoing, and $-1$ else. Lengths (and dihedral angles), which are determined by fluxes, completely characterize some intrinsic geometry of the canonical surface, and are invariant under local rotations changing the frames in which fluxes are expressed.

From the 2d angles (and hence lengths), one can compute the dihedral angles between neighboring triangles as if they were embedded in flat 3-space. Consider triangles $t_1, t_2$ dual to vertices $v_1, v_2$ of the spin network graph, which are linked by the edge $e$. The classical angle of flat space between the two triangles $t_1, t_2$ can be computed from the three 2d angles around the node which is dual to the cycle $c$, using the formula:
\beq \label{flat angle}
\cos \Theta_{t_1 t_2}\,(X,\tl{X}) = \f{\cos\phi_{e_1 e_2} - \cos\phi_{e_1 e}\,\cos\phi_{e_2 e}}{\sin\phi_{e_1 e}\ \sin\phi_{e_2 e}}\;.
\ee
Further, the holonomies $g_e$ also captures some notion of dihedral angles, through the formula
\beq
\cos \theta_{t_1 t_2}\,(X, \tl{X}, g) = -\frac{N_{t_1}\cdot \Ad(g_e) N_{t_2}}{\vert N_{t_1}\vert\ \vert N_{t_2}\vert}\;,
\ee
where $N_t$ are normals to the triangles which are evaluated like $N_{t_1}^i = \epsilon^i_{\phantom{i}jk} \tl{X}_e^j \tl{X}_{e_1}^k$.

It turns out the Hamiltonian $H^{(1)}_{v,p}$ simply expresses the difference between the two notions of dihedral angles between neighboring triangles,
\beq
H^{(1)}_{v,p} = \sin\phi_{e_1 e}\ \sin\phi_{e_2 e}\ \Bigl( \cos \Theta_{t_1 t_2} \,-\, \cos \theta_{t_1 t_2}\Bigr)\;.
\ee
Hence, the topological equation for the ground state, $H^{(1)}_{v,p} =0$, just imposes that the dihedral angles $\theta_{t_1 t_2}$ computed from the holonomies are those of the flat embedding.

Quantizing $H^{(1)}$ hence means quantizing flat Euclidean geometry. This is what the Wheeler-DeWitt equation \eqref{spin 1 wdw} does.

At the quantum level, the re-coupling coefficients $A_{-1,0,1}$ thus provides a well-defined way to evaluate the quantum operators which correspond to the above products of trigonometric functions. (This is particularly clear in the large spin limit \cite{SG1, recursion-semiclass}). The most transparent case is that of the 2d dihedral angle, e.g. between the dual edges to $e_2,e_6$ in the figure \ref{fig:tet}. We have seen that the spin network function is an eigenvector of $\tl X_2\cdot \tl X_6$, so that
\beq
\bigl(\widehat{\cos\phi_{e_2 e_6}}\ s_{\rm tet}^{\{j_e\}}\bigr)(g_1,\dotsc,g_6) = \frac{\bigl[j_2(j_2+1)+j_6(j_6+1) - j_4(j_4+1)\bigr]}{2\sqrt{j_2(j_2+1)\ j_6(j_6+1)}}\ s_\Gamma^{\{j_e\}}(g_1,\dotsc,g_6)\;.
\ee
This exactly reproduces the classical expression of the dihedral angles of a triangle, with quantized lengths $\ell_e(j_e) = \sqrt{j_e(j_e+1)}$.

It is reasonable to think that the spin 1/2 Hamiltonian carries a similar geometric interpretation, but with half dihedral angles instead. It is however hard to make precise, since one needs operators diagonal on spin network functions, so at least quadratic quantities like $F_{62}F_{26}^\dagger$, and then one faces ordering ambiguities. So the best we can do is to check that the large spin limit is indeed the same as that of $\cos\phi_{e_2 e_6}/2$ as computed from the above operator. The latter gives
\begin{align}
\nonumber \cos^2\frac{\phi_{e_2 e_6}}{2} &= \frac{\cos\phi_{e_2 e_6} +1}{2}\;,\\
&= \frac{\bigl(\sqrt{j_2(j_2+1)} + \sqrt{j_6(j_6+1)} + \sqrt{j_4(j_4+1)}\bigr)\bigl(\sqrt{j_2(j_2+1)} + \sqrt{j_6(j_6+1)} - \sqrt{j_4(j_4+1)}\bigr)}{4\, \sqrt{j_2(j_2+1)\,j_6(j_6+1)}}\;,
\end{align}
on spin network functions. From the spin 1/2 graspings, we get
\beq
\frac{F_{62}\,F_{26}^\dagger}{E_{22}\,E_{66}} = \frac{\bigl(j_2+j_6+j_4+1\bigr)\bigl(j_2+j_6-j_4\bigr)}{(2j_2)\,(2j_6)}\;.
\ee
In the large spin limit, the square roots of Casimirs go to the spins themselves, $\sqrt{j(j+1)}\sim j$, and the two quantities above have the same limit.

\subsection{Equivalence with the flatness constraint} \label{sec:back flatness}


We would like to see exactly how the recursion can be used to find wave functions satisfying the flatness everywhere. As it was already observed in \cite{recurrence-paper}, it is clear that the recursion relations we have written so far only depend on the fact that we consider a cycle with three links. In particular, they do not depend on the fact that we have closed the graph by gluing the links $e_3, e_4, e_5$ to a node in the figure \ref{fig:tet}.

Hence, we now consider a cycle with three links and 3-valent vertices, and use the notations of the figure \ref{fig:tet}, but assuming that the links $e_3, e_4, e_5$ do not necessarily meet. We still have exactly the same recursion relations, one for each node of the cycle. This gives three independent difference equations to impose three real constraints on the holonomy $g_p$ around the cycle. With their help, we want to show that the physical state factorizes on the cycle. In the spin network basis, the state is a function of the spins. In this configuration, its dependence on the spins $j_1,j_2,j_6$ actually completely factorizes:
\beq \label{factorization 6j}
\psi_{\rm phys}(j_e) = (-1)^{2j_5}\,\begin{Bmatrix} j_1 &j_2 &j_3\\j_4 &j_5 &j_6\end{Bmatrix}\ \phi(j_e'),
\ee
where the spins $j_e'$ are the spins of all links of the graph except $e_1, e_2, e_6$. This is known in the 3d gravity case from the projector onto physical states \cite{noui-perez-ps3d}. Such a factorization property is actually very natural. Indeed, a flat connection induces on the boundary of a face with three links holonomies which are up to gauge necessarily trivial. Then, it is known that spin networks evaluated on the identity satisfy such factorizations, and the case of cycles with more links is also known \cite{recurrence-paper}.

The idea to extract the factorization is that the recursion relations (for spin 1/2 and hence spin 1) hold on $j_1, j_2, j_6$. We can thus implement them from $\psi(j_1, j_2, j_6, j_e')$ so as to reach an initial state with $j_1=0, j_2 =j_3$ and $j_6 = j_5$. The factorization \eqref{factorization 6j} is then proved with the initial condition\footnote{It will be convenient in the following to explicitly factor a phase $(-1)^{2j_5}$. The basic reason is that in our conventions $e_5$ is oriented inwards the cycle while $e_2, e_4$ are outwards.} $\psi(j_e)_{|j_1=0} = (-1)^{2j_5} \left\{ \begin{smallmatrix} 0 &j_3 &j_3 \\ j_4 &j_5 & j_5\end{smallmatrix}\right\} \phi(j_e')$.

Now we are ready to present the consequences of the factorization \eqref{factorization 6j} in the group representation of wave functions, in which we also get a factorization of the cycle,
\beq
\psi_{\rm phys}(g_1,g_2,g_3,g_4,g_5,g_6,\dotsc) = \delta(g_6\,g_1\,g_2\mone)\ \phi(g_3 g_2\mone, g_4, g_6 g_5,\dotsc)\;,
\ee
with
\beq
\phi(g_3 g_2\mone, g_4, g_6 g_5,\dotsc) = \sum_{\{j_e'\}} \Bigl[\prod_{e'} d_{j_e'}\Bigr]\ \phi(j_e')\ s^{\{j_e'\}}(g_3 g_2\mone, g_4, g_6 g_5,\dotsc)\;.
\ee
We thus see that the goal has been achieved: the wave function has support on holonomies with trivial $\SU(2)$ parallel transport on the cycle, $g_6 g_1 g_2\mone=\unit$.

To obtain the above equations, one writes the group Fourier transform of the physical topological state,
\begin{align}
\nonumber \psi_{\rm phys}(g_e) &= \sum_{\{j_e\}} \Bigl[\prod_{e} d_{j_e}\Bigr]\ \psi_{\rm phys}(j_e)\ s^{\{j_e\}}(g_e)\;,\\
&= \sum_{\{j_e'\}} \Bigl[\prod_{e'} d_{j_e'}\Bigr]\ \phi(j_e') \sum_{j_1,j_2,j_6} d_{j_1} d_{j_2} d_{j_6}\,\begin{Bmatrix} j_1 &j_2 &j_3\\j_4 &j_5 &j_6\end{Bmatrix}\ (-1)^{2j_5}\,s^{\{j_e\}}(g_e)\;.
\end{align}
Using the group averaging identity $\int dh\ D^{(j_1)}(h)\otimes D^{(j_2)}(h)\otimes D^{(j_3)}(h) = \iota_{j_1 j_2 j_3}\vert 0\rangle \langle 0\vert \iota_{j_1 j_2 j_3}$, and the Fourier expansion of the Dirac delta on the group, $\delta(g) = \sum_j d_j \chi_j(g)$, the sums over $j_1, j_2, j_6$ can be explicitly performed,
\beq
\sum_{j_1,j_2,j_6} d_{j_1} d_{j_2} d_{j_6}\,\begin{Bmatrix} j_1 &j_2 &j_3\\j_4 &j_5 &j_6\end{Bmatrix}\ (-1)^{2j_5}\,s^{\{j_e\}}(g_e) = \delta(g_6\,g_1\,g_2\mone)\ s^{\{j_e'\}}(g_3 g_2\mone, g_4, g_6 g_5,\dotsc)\;.
\ee

\section{New representations of the topological equation} \label{sec:new-bases}

We are now interested in finding alternative forms of our key equation \eqref{1/2coeffs}. In the usual spin network basis, it is a difference equation which corresponds to the well-known recursion formulae for the 6j-symbol. Beyond the result of having a new Hamiltonian, with its nice geometric interpretation, one of the major progress is that the key equation comes from an operator. Hence one can simply re-write $\widehat{H}\vert \psi\rangle =0$ in a different basis, and the physical content, i.e. the flatness constraint of the topological model, is preserved.

One of the technical interesting aspects of using the Schwinger's bosonic representation of $\SU(2)$ is that it is possible to go to bases of coherent states easily. We will first recast the recursion relation in a basis of $\SU(2)$ intertwiner, closely related to those introduced in \cite{livine-speziale-CS}, and which are detailed in \cite{fine-structure}. This basis is mixed, i.e. it contains spins and spinors. Then, we will move to a basis based on the coherent states of the Schwinger's bosons, thus getting rid of spins. The constraint equation will appear as a holomorphic differential equation solved by Schwinger's generating function of 6j-symbols.

Further representations are provided in the appendix \ref{app:addrep}.

\subsection{The recursion formula in the coherent SU(2) basis}

For $z=\left(\begin{smallmatrix} z^-\\z^+\end{smallmatrix}\right)\in\C^2$, introduce the following states,
\beq
\vert j,z\rangle = \f{\bigl( z^- a^{-\dagger} + z^+ a^{+\dagger}\bigr)^{2j}}{\sqrt{(2j)!}}\vert 0\rangle = \sqrt{(2j)!} \sum_{m=-j}^j \f{(z^-)^{j-m}\ (z^+)^{j+m}}{\sqrt{(j-m)!\,(j+m)!}}\,\vert j,m\rangle\;.
\ee
They are normalized to $\langle j,z\vert j,z\rangle = (\vert z_-\vert^2 + \vert z_+\vert^2)^{2j}$. They are just the standard coherent states for each of the two quantum harmonic oscillators, projected onto the subspace of fixed spin $j$. In particular, they are holomorphic, homogeneous polynomials of degree $(2j)$. The annihilators and creators act as
\beq
a^A\,\vert j,z\rangle = \sqrt{2j}\ z^A\,\vert j-\f12,z\rangle\;, \quad \text{and}\qquad a^{A\dagger}\,\vert j,z\rangle = \f1{\sqrt{2j+1}}\ \f{\partial\phantom{z}}{\partial z^A}\,\vert j+\f12,z\rangle\;.
\ee
In addition to lowering the spin by $1/2$, the annihilator $a^A$ multiplies the state by $z^A$, while the creator $a^{A\dagger}$ increases the spin by $1/2$ and derive the state with respect to $z^A$.

The duality map $\varsigma$, \eqref{map sigma}, allows to define anti-holomorphic states, $\vert j,z]\equiv \vert j,\varsigma z\rangle$. We will mainly use their duals on which the boson operators act as
\beq
[j,z\vert\,a^A = \f{(-1)^{\f12-A}}{\sqrt{2j+1}}\ \f{\partial \phantom{z}}{\partial z^{-A}}\,[j+\f12,z\vert\;,\quad \text{and}\qquad
[j,z\vert\,a^{A\dagger} = \sqrt{2j}\ (-1)^{\f12-A}\, z^{-A}\,[j-\f12,z\vert\;.
\ee

We construct a basis of intertwiners by group averaging tensor products of these states, naturally coined coherent intertwiners,
\beq \label{coh inter def}
\iota_{(j_e)}(z_e)\vert 0\rangle \equiv \int_{\SU(2)} dg\ \bigotimes_{e}\ g\,\vert j_e, z_e\rangle.
\ee
and from them, we define the spin network functions in this basis. On the tetrahedral graph of the figure \ref{fig:tet}, they are labeled by six spins $(j_e)$ like before, but also one spinor $z_e$ for each source node $s(e)$ of each link $e$, and another spinor $\tl z_e$ for the target node $t(e)$. Consider for example the node of the figure \ref{fig:tet} where $e_1, e_2, e_3$ meet, all three outgoing. Then the intertwiner components are proportional to the 3jm-symbol,
\beq
\langle j_1, m_1;j_2, m_2; j_3, m_3\vert \iota_{(j_e,z_e)}\vert 0\rangle = \begin{pmatrix} j_1 &j_2 &j_3\\ m_1 &m_2 &m_3\end{pmatrix} \sum_{(n_e)_{e=1,2,3}} \prod_{e=1}^3 \sqrt{(2j_e)!}\f{(z_e^-)^{j_e-n_e}\ (z_e^+)^{j_e+n_e}}{\sqrt{(j_e-n_e)!\,(j_e+n_e)!}}\,\begin{pmatrix} j_1 &j_2 &j_3\\ n_1 &n_2 &n_3\end{pmatrix}\;,
\ee
and the proportionality coefficient is an invariant holomorphic polynomial on $(z_1, z_2, z_3)$. For additional interesting aspects on these intertwiners (especially of higher valence), we refer to \cite{conrady-closure, etera-factor}. The polynomial is actually exactly known \cite{varshalovich-book}, and is the generating function for the 3jm-symbols with fixed spins. That gives
\beq \label{factorcohint}
\iota_{j_1 j_2 j_3 }(z_1, z_2, z_3) \,=\, P_{j_1 j_2 j_3}(z_1,z_2,z_3)\ \iota_{j_2 j_2 j_3}\;
\ee
with
\beq
\begin{aligned}
P_{j_1 j_2 j_3}(z_1,z_2,z_3) &= \frac{1}{\Delta(j_1 j_2 j_3)\,(J_{123}+1)!}\ \prod_{e=1}^3 \sqrt{(2j_e)!}\ [z_2\vert z_1\rangle^{J_{123}-2j_3}\ [z_3\vert z_2\rangle^{J_{123}-2j_1}\ [z_1\vert z_3\rangle^{J_{123}-2j_2}\;,\\
\Delta(j_1 j_2 j_3) &= \sqrt{\frac{(J_{123}-2j_1)!\,(J_{123}-2j_2)!\,(J_{123}-2j_3)!}{(J_{123}+1)!}}\;.
\end{aligned}
\ee
We have set $J_{123}\equiv j_1+j_2+j_3$ for the total spin. On a vertex with some incoming links, where target spinors $\tl z_e$ live, it is convenient to built the coherent intertwiner using the bras $[j_e,\tl z_e\vert$. For example, on the node where $e_1, e_5, e_6$ meet on the figure \ref{fig:tet}, we take
\beq
\iota_{j_1^* j_5^* j_6}(\tl z_1, \tl z_5, z_6) = \int_{\SU(2)} dh\ [j_1,\tl z_1\vert\,h\mone \otimes [j_5,\tl z_5\vert\,h\mone \otimes h\,\vert j_6, z_6\rangle\;.
\ee
This convention gives formula similar to the one above,
\beq
\iota_{j_1^* j_5^* j_6}(\tl z_1, \tl z_5, z_6) \,=\, P_{j_1 j_5 j_6}(\tl z_1,\tl z_5,z_6) \ \iota_{j_1^* j_5^* j_6}\;.
\ee

Equipped with these intertwiners, we can form the corresponding spin network function on the tetrahedral graph,
\begin{align} \label{cohspinnet}
s^{\{j_e, z_e, \tl z_e\}}_{\rm tet}(g_e) &= \int_{\SU(2)^4} \prod_{v=1}^4 dh_v\ \prod_{e=1}^6 [j_e, w_e \vert\  h\mone_{t(e)}\,g_e\,h_{s(e)}\ \vert j_e, z_e\rangle\;,\\
&= \bigl[P_{j_1 j_2 j_3}(z_1,z_2,z_3)\,P_{j_1 j_5 j_6}(\tl z_1, \tl z_5, z_6)\, P_{j_3 j_4 j_5}(\tl z_3, \tl z_4, z_5)\, P_{j_2 j_6 j_4}(\tl z_2, \tl z_6, z_4)\bigr]\ s^{\{j_e\}}_{\rm tet}(g_1,\dotsc, g_6)\;.
\end{align}
This factorization between the standard spin network function, which is independent on the spinor labels, and the holomorphic polynomials, which are independent of the arguments of the function but capture the whole dependence on the spinors, is specific to the case of three-valent nodes. In particular, the evaluation on the trivial holonomies, $g_e=\unit$, gives the 6j-symbol times the four polynomials, which we denote like
\beq
\begin{Bmatrix} j_1 &j_2 &j_3 \\j_4 &j_5 &j_6\end{Bmatrix}\bigl(z_e, \tl z_e\bigr) \equiv \bigl[P_{j_1 j_2 j_3}(z_1,z_2,z_3)\,P_{j_1 j_5 j_6}(\tl z_1, \tl z_5, z_6)\, P_{j_3 j_4 j_5}(\tl z_3, \tl z_4, z_5)\, P_{j_2 j_6 j_4}(\tl z_2, \tl z_6, z_4)\bigr]\ \begin{Bmatrix} j_1 &j_2 &j_3 \\j_4 &j_5 &j_6\end{Bmatrix}\;.
\ee

We are now ready to state the main result of the section.
\begin{proposition} \label{rec spin coh}
The equation $\widehat{H}_{e_6 e_1 e_2|R}^{(--)} s^{\{j_e, z_e, \tl z_e\}}_{\rm tet}(g_e)_{|\unit} = 0$ reads
\begin{multline} \label{eq:rec spin coh}
\frac{(2j_2)\,(2j_6)}{2j_1+1}\,\biggl[\sum_{A,B = \pm1/2} \Bigl(\frac{\partial\phantom{z}}{\partial{\tl z_1^A}}\otimes z_6^A\Bigr)\,\Bigl(\frac{\partial\phantom{z}}{\partial{z_1^B}}\otimes z_2^B\Bigr)\biggr]\ \begin{Bmatrix} j_1+\f12 &j_2-\f12 &j_3\\j_4 &j_5 &j_6-\f12\end{Bmatrix}\bigl(z_e,\tl z_e\bigr) \\
+ (2j_2)\,(2j_6)\,(2j_1)\ [\tl z_1\vert z_6\rangle\,[z_1\vert z_2\rangle\ \begin{Bmatrix} j_1-\f12 &j_2-\f12 &j_3\\j_4 &j_5 &j_6-\f12\end{Bmatrix}\bigl(z_e,\tl z_e\bigr) \\
-(2j_1+1)\ \Bigl[\frac{\partial\phantom{z}}{\partial{\tl z_6}}\Big\vert \frac{\partial\phantom{z}}{\partial{\tl z_2}}\Big\rangle\ \begin{Bmatrix} j_1 &j_2 &j_3\\j_4 &j_5 &j_6\end{Bmatrix}\bigl(z_e,\tl z_e\bigr) = 0\;.
\end{multline}
\end{proposition}

The remarkable feature of this equation is that it obviously generates the same shifts on the 6j-symbol as in the previous recursion, but the coefficients $A_{0,\pm1/2}$ of the recursion have been exchanged with simple multiplications and derivatives with respect to the spinors.

{\bf Proof of the proposition \ref{rec spin coh}.} First, it is clear from the argument already used in the proof of the proposition \ref{prop:method} that $\widehat{H}_{e_6 e_1 e_2|R}^{(--)} s^{\{j_e, z_e, \tl z_e\}}_{\rm tet}(g_e)_{|\unit}$ is indeed identically zero. The key relation is displayed in \eqref{spin1/2Hafter}.

To actually evaluate the action of the Hamiltonian on the coherent spin network, we use the form \eqref{H1/2Rbis}, which we recall: $\bigl(\f1{d_{j_1}}( \widehat{F}_{61}\,\widehat{E}_{12} + \widehat{E}_{16}\,\widehat{F}_{12}) - \widehat{F}_{62}\bigr)\,\widehat{F}_{26}^\dagger$. It is thus sufficient to compute the actions of the elementary operators $\widehat{E},\widehat{F},\widehat{F}^\dagger$ in the basis of coherent intertwiners. Notice that these operators are invariant under $\SU(2)$, so that they commute with the group averaging in the definition of the coherent intertwiner \eqref{coh inter def}. For example,
\beq
\widehat{F}_{26}^\dagger\ \iota_{j_2^* j_6^* j_4}\bigl(\tl z_2, \tl z_6, z_4\bigr) = \int_{\SU(2)} dh\ \sum_A\,(-1)^{\f12-A}\ [j_2,\tl z_2\vert a^{-A\dagger}\,h\mone \otimes [j_6,\tl z_6\vert a^{A\dagger}\,h\mone\otimes h\vert j_4,z_4\rangle\;.
\ee
Then, we just apply the creators and annihilators in this basis, to get
\beq
\widehat{F}_{26}^\dagger\ \iota_{j_2^* j_6^* j_4}\bigl(\tl z_2, \tl z_6, z_4\bigr) = \sqrt{2j_2\,2j_6}\ [\tl z_2\vert \tl z_6\rangle\ \iota_{(j_2-\f12)^* (j_6-\f12)^* j_4}\bigl(\tl z_2, \tl z_6, z_4\bigr)\;.
\ee
Remember that $\widehat{F}_{26}^\dagger$ is the first operator in the right ordering of the Hamiltonian. We can actually factor (and ignore) the multiplicative contribution, $\sqrt{2j_2\,2j_6}\ [\tl z_2\vert \tl z_6\rangle$, and consider that $F_{26}^\dagger$ is only here to shift the spins $j_2, j_6$ to $j_2-1/2, j_6-1/2$. Indeed, by dressing the trivial identity \eqref{spin1/2Hafter} with intertwiners, we derived the equation \eqref{keyeq}. Both \eqref{spin1/2Hafter} and \eqref{keyeq} hold in any basis, and in particular we can take the coherent intertwiners in \eqref{keyeq}. Hence, we can just apply the latter, and plug into it the following actions of the other relevant operators,
\begin{align}
\widehat{F}_{12}\,\iota_{j_1 j_2 j_3}(z_1, z_2, z_3) &= \sqrt{2j_1\,2j_2}\ [z_1\vert z_2\rangle\ \iota_{j_1-\f12 j_2-\f12 j_3}(z_1, z_2, z_3)\;, \\
\widehat{E}_{12}\,\iota_{j_1 j_2 j_3}(z_1, z_2, z_3) &= \sqrt{\frac{2j_2}{2j_1+1}}\ \sum_A \Bigl(\frac{\partial\phantom{z}}{\partial{z_1^A}}\otimes z_2^A\Bigr)
\iota_{j_1+\f12 j_2-\f12 j_3}(z_1, z_2, z_3)\;, \\
\widehat{F}_{61}\,\iota_{j_1^* j_5^* j_6} (\tl z_1, \tl z_5, z_6) &= \sqrt{\frac{2j_6}{2j_1+1}}\ \sum_A \Bigl(\frac{\partial\phantom{z}}{\partial{\tl z_1^A}}\otimes z_6^A\Bigr)\ \iota_{(j_1+\f12)^* j_5^* j_6-\f12}(\tl z_1, \tl z_5, z_6)\;,\\
\widehat{E}_{16}\,\iota_{j_1^* j_5^* j_6}(\tl z_1, \tl z_5, z_6) &= \sqrt{2j_1\,2j_6}\ [\tl z_1\vert z_6\rangle\ \iota_{(j_1-\f12)^* j_5^* j_6-\f12}(\tl z_1, \tl z_5, z_6)\;, \\
\widehat{F}_{62}\,\iota_{(j_2-\f12)^* (j_6-\f12)^* j_4}(\tl z_2, \tl z_6, z_4) &= \frac{1}{\sqrt{2j_2\,2j_6}}\ \Bigl[\frac{\partial\phantom{z}}{\partial{\tl z_6}}\Big\vert \frac{\partial\phantom{z}}{\partial{\tl z_2}}\Big\rangle\ \iota_{j_2^* j_6^* j_4}(\tl z_2, \tl z_6, z_4)\;.
\end{align}
\qed

Another proof using the special factorization \eqref{factorcohint} of the three-valent coherent intertwiners is available, though it is unclear how it could be more generally applied. The idea is that the coefficients of the recursion are taken into account as multiplications and derivatives with respect to spinors. But the dependence of the coherent 6j-symbol on the spinors is only through the holomorphic polynomials $P$. Hence, the latter satisfy the key relations which enables to get the correct coefficients. Let us just give an example. We have seen just above that $\widehat{F}_{62}$ produces a second order operator, $[\frac{\partial\phantom{z}}{\partial{\tl z_6}}\vert \frac{\partial\phantom{z}}{\partial{\tl z_2}}\rangle$. Then, it is tedious but straightforward to show that
\begin{align}
\frac1{\sqrt{2j_2\,2j_6}}\,\Bigl[\frac{\partial\phantom{z}}{\partial{\tl z_6}}\Big\vert \frac{\partial\phantom{z}}{\partial{\tl z_2}}\Big\rangle\ P_{j_2 j_6 j_4}(\tl z_2, \tl z_6, z_4) &= \sqrt{\bigl(j_2+j_6+j_4+1\bigr)\bigl(j_2+j_6-j_4\bigr)}\ P_{j_2-\f12 j_6-\f12 j_4}(\tl z_2, \tl z_6, z_4)\;,\\
&= - A_0^{(-,-)}\ \sqrt{d_{j_2} d_{j_2-\f12}\,d_{j_6} d_{j_6-\f12}}\ P_{j_2-\f12 j_6-\f12 j_4}(\tl z_2, \tl z_6, z_4)\;.
\end{align}
The coefficient $A_0$ is given in \eqref{coeffs}. All other terms can be evaluated this way, and the pre-factors, in particular $P_{j_2-\f12 j_6-\f12 j_4}$, can be factorized, as well as the other polynomials, finally leading to the spin 1/2 recursion relation on the 6j-symbol.


\subsection{The Schwinger's generating function for 6j-symbols} \label{sec:schwinger}

In this section, we will get rid of the spins, and fully transform the recursion relation to a differential equation with respect to the spinors only.
On each node of the spin network graph, we build some intertwiners in the following way. Assume that the node has its three links outgoing, like the links $1,2,3$ on the figure \ref{fig:tet}. Consider
\begin{align}
\iota(z_1,z_2,z_3) &\equiv \sum_{j_1,j_2,j_3} \frac{(j_1+j_2+j_3+1)!}{\prod_{e=1}^3 (2j_e)!} \int dh\ h\,\bigotimes_{e=1}^3 \langle a_e\vert z_e\rangle\,\vert 0\rangle_e\;,\\
&= \sum_{j_1,j_2,j_3} \frac{(j_1+j_2+j_3+1)!}{\sqrt{\prod_{e=1}^3 (2j_e)!}}\ \iota_{j_1 j_2 j_3}(z_1,z_2,z_3)\;,
\end{align}
where the intertwiners $\iota_{j_1 j_2 j_3}(z_1,z_2,z_3)$ are defined in \eqref{coh inter def}. We thus get invariant vectors on $\otimes_{e=1}^3 \left(\oplus_{j_e} \calH_{j_e}\right)$, the tensor product of the spaces of all irreducible representations on each leg. When some lines are actually ingoing on the node, we use the same convention as in the previous section, e.g. on the node of the figure \ref{fig:tet} where $2,6,4$ meet
\beq
\iota(\tl z_2,\tl z_6,z_4) = \sum_{j_2,j_4,j_6} \frac{(j_2+j_4+j_6+1)!}{\sqrt{\prod_e (2j_e)!}}\ \iota_{j_2^* j_6^* j_4}(\tl z_2,\tl z_6,z_4)\;.
\ee

On a spin network graph, we associate one such intertwiner to each node, and glue them using the holonomies on the links. This is the natural gluing, where all irreducible representations are orthogonal. This leads to the notion of {\bf spinor networks}, which we define as
\beq
\mathcal S^{\{z_e,\tl z_e\}}(g_e) = \sum_{j_1,\dotsc,j_6} \frac{\prod_{v} \bigl(J_v+1\bigr)!}{\prod_{e=1}^6 (2j_e)!}\ s^{\{j_e,z_e,\tl z_e\}}(g_e)\;,
\ee
where the coherent spin network $s^{\{j_e,z_e,\tl z_e\}}$ is defined in \eqref{cohspinnet} and $J_v$ is the sum of the three spins meeting on the node $v$,
\beq
J_{123} = j_1+j_2+j_3\;,\quad J_{156} = j_1 +j_5+j_6\;,\quad J_{264} = j_2+j_6+j_4\;,\quad J_{345}=j_3+j_4+j_5\;.
\ee

Its evaluation on the unit is a 12-spinor symbol, which is the Bargmann representation of the 6j-symbol, mostly known as the Schwinger's generating function of 6j-symbols,
\beq \label{spinor symbol 2}
\mathcal S(z_e,\tl z_e) \equiv \mathcal S^{\{z_e,\tl z_e\}}(\unit) = \sum_{j_1,\dotsc,j_6} \left[\prod_v \sqrt{\frac{\bigl(J_v+1\bigr)!}{\prod_{e\supset v} (J_v-2j_e)!}}\right]\ \begin{Bmatrix} j_1 &j_2 &j_3 \\j_4 &j_5 &j_6\end{Bmatrix}\ \prod_v \prod_{e\supset v} [z_{e''}\vert z_{e'}\rangle^{J_v-2j_{e}} \;.
\ee
The reason why it generates the standard 6j-symbols is the special factorization of the coherent three-valent intertwiners \eqref{factorcohint}. There are twelve brackets $[z_{e''}\vert z_{e'}\rangle$ in the final product over vertices and links. The notation is as follows. To each vertex $v$ we consider a cyclic ordering of the links which meet there, respectively $(123), (156), (264)$ and $(345)$\footnote{These are actually the cyclic orderings of the 3jm-symbols in the definition of the tetrahedral spin network function \eqref{defspinnet}.}. In the above formula, we have written $(e,e', e'')$ the links meeting at each vertex with the correct ordering. Since there are three choices of $e$, we have three brackets for each node, and since there are four nodes, that gives a total product of twelve brackets (and the tilde is understood for the spinor on the target vertex of each link).

This generating function is known to have a closed form \cite{varshalovich-book}, which can be derived from several methods, \`a la Bargmann (see \cite{wu} for example) using complex integrals, or \`a la Schwinger \cite{schwinger} using his boson operators to evaluate inner products of states,
\beq \label{schwinger generating}
\mathcal S(z_e,\tl z_e)= \frac1{\left[1+ \mathcal M(z_e, \tl z_e)\right]^2}\;, \qquad \mathcal M(z_e,\tl z_e) = \sum_{\rm 3-cycles} \prod_{v} [z_{e'}\vert z_{e}\rangle + \sum_{{\rm 4-cycles}} \prod_{v} [z_{e'}\vert z_{e}\rangle\;,
\ee
with
\begin{align}
\nonumber \sum_{\rm 3-cycles} \prod_{v} [z_{e'}\vert z_{e}\rangle &= [z_2\vert z_1\rangle [\tl z_1\vert z_6\rangle [\tl z_6\vert \tl z_2\rangle +
[z_3\vert z_2\rangle [\tl z_2\vert z_4\rangle [\tl z_4\vert \tl z_3\rangle +
[z_1\vert z_3\rangle [\tl z_3\vert z_5\rangle [\tl z_5\vert \tl z_1\rangle +
[z_4\vert \tl z_6\rangle [z_6\vert \tl z_5\rangle [z_5\vert \tl z_4\rangle\;,\\
\nonumber \sum_{{\rm 4-cycles}} \prod_{v} [z_{e'}\vert z_{e}\rangle &= [z_1\vert z_3\rangle [\tl z_4\vert \tl z_3\rangle [ z_4\vert \tl z_6\rangle [\tl z_1 \vert z_6\rangle +
[z_3\vert z_2\rangle [\tl z_3\vert z_5\rangle [ z_6\vert \tl z_5\rangle [\tl z_6 \vert \tl z_2\rangle +
[z_2\vert z_1\rangle [\tl z_2\vert z_4\rangle [ z_5\vert \tl z_4\rangle [\tl z_5 \vert \tl z_1\rangle\;.
\end{align}
There are two types of contributions, one from the cycles of three links, and the other from cycles with four links. There are obviously four cycles with three links, corresponding to the triangles on the tetrahedral graph. There are also three cycles with four links. Each of them is obtained by removing a couple of two opposite edges on the tetrahedral graph. That gives the cycles $(e_1 e_2 e_4 e_5), (e_1 e_3 e_4 e_6), (e_2 e_3 e_5 e_6)$ on the figure \ref{fig:tet}. On each node $v$ of each such cycle, the function $\mathcal S(z_e,\tl z_e)$ picks up a contribution from the bracket $[z_e'\vert z_e\rangle$ where $e, e'$ are the two links meeting at $v$ on the boundary of the cycle.

\begin{proposition} \label{prop:schwinger}
The equation $\widehat{H}^{(--)}_{e_6 e_1 e_2|R} \mathcal S^{(z_e,\tl z_e)}(g_e)_{|\unit} = 0$ is true and corresponds to a differential equation on the generating function,
\begin{multline}
\biggl[\Bigl[\sum_{A,B = \pm1/2} \Bigl(\frac{\partial\phantom{z}}{\partial{\tl z_1^A}}\otimes z_6^A\Bigr)\,\Bigl(\frac{\partial\phantom{z}}{\partial{z_1^B}}\otimes z_2^B\Bigr)\Bigr]\ \Bigl(2+\f12\sum_{e\supset v_{264}}\sum_C z_e^C\frac{\partial\phantom{z}}{\partial{z_e^C}}\Bigr) -\Bigl[\frac{\partial\phantom{z}}{\partial{\tl z_6}}\Big\vert \frac{\partial\phantom{z}}{\partial{\tl z_2}}\Big\rangle\ \bigl(1+\sum_A z_1^A\frac{\partial\phantom{z}}{\partial z_1^A}\bigr)\\
+ [\tl z_1\vert z_6\rangle\,[z_1\vert z_2\rangle\
\Bigl(2+\f12\sum_{e\supset v_{264}}\sum_C z_e^C\frac{\partial\phantom{z}}{\partial{z_e^C}}\Bigr) \Bigl(2+\f12\sum_{e\supset v_{123}}\sum_C z_e^C\frac{\partial\phantom{z}}{\partial{z_e^C}}\Bigr)
\Bigl(2+\f12\sum_{e\supset v_{156}}\sum_C z_e^C\frac{\partial\phantom{z}}{\partial{z_e^C}}\Bigr)
\biggr]
\mathcal S\bigl(z_e,\tl z_e\bigr) = 0\;.
\end{multline}
\end{proposition}


{\bf Proof of the proposition \ref{prop:schwinger}.} We start from the proposition \ref{rec spin coh} and the equation \eqref{eq:rec spin coh}, which we multiply by $\frac{\prod_{v} (J_v+1)!}{\prod_{e=1}^6 (2j_e)!}$ to obtain the correct normalization in the sum over spins of the spinor network. We form the quantities
$\frac{\prod_{v} (J_v+1)!}{\prod_{e=1}^6 (2j_e)!}\,s^{\{j_e,z_e,\tl z_e\}}(g_e)$ with the appropriate shifts on $j_1,j_2,j_6$. All factors $1/(2j)!$ can be absorbed this way. However, on the two terms which shift $j_1$, $\prod_{v} (J_v+1)!$ leaves some factors of the type $(j_2-1/2+j_6-1/2+j_4+2)$. The latter can be generated from derivatives with respect to spinors, $(2+\f12\sum_{e\supset v_{264}}\sum_C z_e^C\frac{\partial\phantom{z}}{\partial{z_e^C}})$, since $E_{ee} = \sum_A z_e^A\partial_{z_e^A}$ becomes in the spin representation $E_{ee} = 2j_e$. The final term in \eqref{eq:rec spin coh} displays a dimension factor $2j_1+1$, which can also be taken into account via the operator $(1+\sum_A z_1\frac{\partial\phantom{z}}{\partial z_1^A})$ (or the same with $\tl z_1$). Then, the proposition is obtained by summing over all spins $(j_1,\dotsc,j_6)$.
\qed

We have thus obtained the equation of the dynamics as a holomorphic differential equation. One can check explicitly that the Schwinger's generating function \eqref{schwinger generating} is a solution. It is however not clear to us whether other solutions can be found. In any case, looking for a solution as a power series will obviously reproduce in fine the recursion formula for the 6j-symbol.

Notice that the generating function $\mathcal S$ is most naturally a function of the classical variables $F^*_{ee'} = [z_{e'}\vert z_e\rangle$. Hence, it is interesting to recast the above equation of the dynamics with derivatives with respect to $F^*_{ee'}$. For instance, let us consider the explicit recursion given in appendix \eqref{explicit rec++}. We multiply it with $\sqrt{(j_2+j_6+j_4+2)(j_2+j_6-j_4+1)}$ and then by the appropriate factors to built from the three terms three generating functions. The spin dependent coefficients can be re-absorbed into multiplications and derivatives with respect to the $F^*$ variables. After some algebra, one gets
\begin{multline} \label{diff S}
\biggl(\frac{\partial\phantom{F}}{\partial F^*_{21}}\,\frac{\partial\phantom{F}}{\partial F^*_{16}}\,\frac{\partial\phantom{F}}{\partial F^*_{62}} \,-\,  F^*_{51}F^*_{13}\,\frac{\partial\phantom{F}}{\partial F^*_{32}}\,\frac{\partial\phantom{F}}{\partial F^*_{62}}\,\frac{\partial\phantom{F}}{\partial F^*_{65}} \\+ \Bigl(F^*_{13}\frac{\partial\phantom{F}}{\partial F^*_{13}} + F^*_{21}\frac{\partial\phantom{F}}{\partial F^*_{21}} +1\Bigr) \Bigl[F^*_{24}\frac{\partial\phantom{F}}{\partial F^*_{24}} + F^*_{46}\frac{\partial\phantom{F}}{\partial F^*_{46}} + F^*_{62}\frac{\partial\phantom{F}}{\partial F^*_{62}} +2\Bigr] \Bigl(F^*_{62}\frac{\partial\phantom{F}}{\partial F^*_{62}} +1\Bigr)\biggr)\ \mathcal S(F^*_{ee'}) \,=\,0\;.
\end{multline}
The first two terms of this equation can be understood as follows. Remember that a variable $F^*_{ee'}$ is associated to each couple of links which meet on a node of the graph. The closed expression for the generating function displays $\mathcal M$, \eqref{schwinger generating}, the products of those variables which share a cycle, and a sum over all 3-cycles and 4-cycles. Besides, the operator which generates the equation we are looking at acts on a fixed cycle of the graph, chosen to be here $(e_6 e_1 e_2)$. The first term we observe (which comes from the shift of $j_1$ by $+1/2$) contains the derivatives with respect to the three variables which live on the chosen cycle, $F^*_{21}, F^*_{16}, F^*_{62}$, which give a contribution $1$ when acting on $\mathcal M$,
\beq
\frac{\partial\phantom{F}}{\partial F^*_{21}}\,\frac{\partial\phantom{F}}{\partial F^*_{16}}\,\frac{\partial\phantom{F}}{\partial F^*_{62}} \ \mathcal M = \frac{\partial\phantom{F}}{\partial F^*_{21}}\,\frac{\partial\phantom{F}}{\partial F^*_{16}}\,\frac{\partial\phantom{F}}{\partial F^*_{62}} \bigl(F^*_{21}\,F^*_{16}\,F^*_{62}\bigr) = 1\;.
\ee
The second term is more interesting. Indeed, when the derivatives act on $\mathcal M$, all cycles which do not contain the corresponding $F^*_{ee'}$ disappear. Hence, the derivatives with respect to $F^*_{32}, F^*_{62}, F^*_{65}$ select one 4-cycle, that with the links $(e_3 e_2 e_6 e_5)$, and then
\beq
F^*_{51}F^*_{13}\,\frac{\partial\phantom{F}}{\partial F^*_{32}}\,\frac{\partial\phantom{F}}{\partial F^*_{62}}\,\frac{\partial\phantom{F}}{\partial F^*_{65}} \ \mathcal M = F^*_{51}\,F^*_{13}\,F^*_{35}\;.
\ee
The idea is thus that this contribution transforms the 4-cycle into the 3-cycle $(e_1 e_5 e_3)$.

We will discuss other equations satisfied by the generating function in the section \ref{sec:other rec}.

\section{Comparisons with other equations for the BF phase} \label{sec:other-eqs}

\subsection{Back to the plaquette operator, with spinors}

We have discussed in the section \ref{sec:tent} the plaquette operator in terms of the character operator along a cycle, in the group and in the spin picture. Let us now explain how to get the analogous equation to \eqref{recholgen} in the basis of $\SU(2)$ coherent spin networks $s^{\{j_e,z_e,\tl z_e\}}(g_e)$. We work the character in the fundamental representation, from which all others can be obtained. It is then better to write the product of holonomies around $p$ in terms of the bosonic operators,
\beq
\chi_{\f12}(g_p)\  s^{\{j_e,z_e,\tl z_e\}} = \prod_{e=1}^n \frac{ \vert \tl a_e\rangle \langle a_e\vert + \vert \tl a_e][ a_e\vert}{d_{j_e}}\ s^{\{j_e,z_e,\tl z_e\}}\;.
\ee
As we have seen, the re-coupling process in the spin basis enables to extract a 6j-symbol at each vertex. Hence, we will write the action of the character vertex by vertex, instead of a product over links. This gives
\beq
\chi_{\f12}(g_p)\  s^{\{j_e,z_e,\tl z_e\}} = \sum_{\epsilon_e=0,1} \ \prod_{v\subset p} \langle \varsigma^{\epsilon_e} a_e\vert \varsigma^{\epsilon_{e'}} \tl a_{e'}\rangle\quad \frac{1}{\prod_e d_{j_e}}\ s^{\{j_e,z_e,\tl z_e\}}\;.
\ee
Here we sum over all $\epsilon_e=0,1$ to include all combinations of the operators $\widehat{E},\widehat{F},\widehat{F}^\dagger$ at each vertex $v$. $(e,e')$ denotes a pair of links on the boundary of $p$ which meet at $v$, and so that $e$ is outgoing and $e'$ ingoing. In the usual spin basis we have already shown that the above operators produce the expected re-coupling coefficients of the equation \eqref{recholgen}. To get the equation in the $\SU(2)$ coherent spin network basis from \eqref{recholgen}, one can use the following correspondence,
\begin{alignat}{3}
&\langle a_e\vert \tl a_{e'}\rangle \ &\rightarrow &\ \begin{Bmatrix} j_{e'}+\f12& j_{e'} & \f12 \\ j_e& j_e+\f12 & l_1\end{Bmatrix} \ &\rightarrow &\ \frac1{\sqrt{d_{j_e}\,d_{j_{e'}}}}  \Big[\frac{\partial\phantom{z}}{\partial \tl z_{e'}}\big\vert \frac{\partial\phantom{z}}{\partial z_{e}}\Big\rangle\;,\\
&\langle a_e\vert \tl a_{e'}] \ &\rightarrow &\ \begin{Bmatrix} j_{e'}-\f12& j_{e'} & \f12 \\ j_e& j_e+\f12 & l_1\end{Bmatrix} \ &\rightarrow &\ -\sqrt{\frac{2j_{e'}}{d_{j_e}}} \sum_{A=\pm\f12} \frac{\partial\phantom{z}}{\partial z_e^A}\,\tl z^A_{e'}\;,\\
&[a_e\vert \tl a_{e'}\rangle \ &\rightarrow &\ \begin{Bmatrix} j_{e'}+\f12& j_{e'} & \f12 \\ j_e& j_e-\f12 & l_1\end{Bmatrix} \ &\rightarrow &\ \sqrt{\frac{2j_{e}}{d_{j_{e'}}}} \sum_{A=\pm\f12} \frac{\partial\phantom{z}}{\partial \tl z_{e'}^A}\,z^A_{e}\;,\\
&[ a_e\vert \tl a_{e'}] \ &\rightarrow &\ \begin{Bmatrix} j_{e'}-\f12& j_{e'} & \f12 \\ j_e& j_e-\f12 & l_1\end{Bmatrix} \ &\rightarrow &\ \sqrt{2j_{e'}\,2j_e}\ [\tl z_{e'}\vert z_e\rangle\;,
\end{alignat}
while ignoring the signs in \eqref{recholgen}.

\subsection{Different types of recursions and differential equations} \label{sec:other rec}

We have at our disposal the fundamental spin 1/2 recursion for the topological ground states, theorem \ref{prop:wdw}, derived from a Hamiltonian $\widehat{H}_{v,p}$, and which we have shown to reproduce the standard flatness constraint in the section \ref{sec:back flatness}. This equation implies a difference equation on a single spin, which can be solved explicitly (cycle after cycle). So it amounts to completely solve the theory. Obviously the algebra generated by the new scalar Hamiltonian we have presented can be used to derive numerous recursion relations.

Technically, the success of this Hamiltonian relies on the fact that the coefficients of the recursion, $A_{\pm,0}$, are really non-trivial coefficients. Having such non-trivial coefficients is needed to relate three consecutive terms of the 6j-symbols. But it could also be interesting to consider different recursions on the ground states which
\begin{itemize}
 \item involve numerous shifts on several spins and more terms than the usual three terms (and hence cannot be solved from a single initial condition),
 \item but, in contrast with the usual recursions, have simple, or even trivial, coefficients,
 \item and are generically much easier to derive than the fundamental spin 1/2 (or spin 1) recursion on the 6j-symbol.
\end{itemize}
Such recursions should be considered analogous to Ward identities or Schwinger-Dyson equations, which encode very interesting properties of the theory and are simpler to deal with than the equations for the effective action itself.

Since one can construct numerous recursion relations from successive applications of the Hamiltonian, we need some criteria to select the `nice' and useful ones. We can distinguish the following cases
\begin{itemize}
\item the recursion is generated by a natural operator, like the plaquette operator (and from which a nice geometric interpretation can be found),
\item the recursion has trivial coefficients (this is only possible if the recursion contains a large number of terms),
\item the recursion comes from a simple differential equation on the generating function.
\end{itemize}

We have already encountered such recursions, that for the plaquette operators, interpreted as tent move evolutions in the section \eqref{sec:tent}. Clearly this operator is simpler that the new Hamiltonian, since this is just a $\SU(2)$ character, but it shifts all the spins around a face. We will now give other examples. Some of them were found in \cite{recurrence-paper, yetanother}. In particular, from an integral representation of the isosceles 6j-symbol, $\left\{\begin{smallmatrix} j_1 & a& b\\ j_2 &a &b\end{smallmatrix}\right\}$, one can derive the following equation,
\beq \label{recurrenceiso6j}
\sum_{\eps_a=\pm}\begin{Bmatrix} j_1 &a+\f{\eps_a}2 &b\\j_2 &a +\f{\eps_a}2 &b\end{Bmatrix} + \sum_{\eps_b=\pm} \begin{Bmatrix} j_1 &a &b+\f{\eps_b}2\\j_2 &a &b+\f{\eps_b}2\end{Bmatrix} +
\sum_{\eta_1,\eta_2=\pm} \begin{Bmatrix} j_1+\f{\eta_1}{2} &a &b\\j_2+\f{\eta_2}2 &a  &b\end{Bmatrix} = 0\;.
\ee
Unlike the standard recurrence relations found in the literature, this one has trivial coefficients and all the spins are shifted. These features are achieved at the price of increasing the number of the terms in the relation, from the usual three or four to eight. To understand the status of this recursion, one should take the perspective of the spin 1/2 recursions. Since the latter can be used to evaluate all 6j-symbols, the above equation should also be derived from them. That was done in \cite{recurrence-paper}. The main result of the derivation is that one has to $i)$ take combinations of spin networks on which different Hamiltonians will act, $ii)$ use spin 1/2 recursions on both $j_1, j_2$ successively, and $iii)$ sum over different cycles and all values of the free parameters $\alpha, \beta$ in \eqref{1/2coeffs}. More precisely
\beq
\sum_{\alpha, \beta=\pm1/2} H^{(1/2)\alpha, \beta}_{e_2 e_1 e_6} \ H^{(1/2)\alpha, -\beta}_{e_5 e_4 e_6}\ s^{\{j_1,j_2-\alpha,j_3,j_4,j_5-\alpha,j_6\}}(\unit) + H^{(1/2)\beta, \alpha}_{e_3 e_1 e_5}\ H^{(1/2)-\alpha,\beta}_{e_5 e_4 e_6}\ s^{\{j_1,j_2,j_3-\beta,j_4,j_5,j_6-\beta\}}(\unit) = 0\;,
\ee
is a recursion relation on the 6j-symbol, which shifts all the spins, and whose coefficients trivialize when choosing $j_2=j_5=a, j_3=j_6=b$, so as to get \eqref{recurrenceiso6j}.

Another example was found in \cite{yetanother} from the integral representation of the squared 6j-symbol, and it was interpreted in the large spin limit as encoding the closure of the tetrahedron dual to the spin network graph. It also corresponds to a recursion on the amplitude of the model on the 3-sphere triangulated by two tetrahedra, in the spin representation (known as the spin foam representation of the partition function), and this is the first and only recursion of this type that we know so far.

Finally, let us discuss the relationship between recursion relations in the spin network basis and differential equations on the associated generating function. The idea is simply that the coefficients of the recursion relation come from acting with derivatives on the generating function. The reason why the equation of the proposition \ref{prop:schwinger} on the generating function is relatively complicated is that it encodes the fundamental recursion, so that several sums of partial derivatives are needed to create the non-trivial coefficients of the recursion. However, one can also observe that the generating function satisfies other equations which are simpler.

As noticed in \cite{labarthe}, the special structure of the generating function, $\mathcal S=(1+\mathcal M)^{-2}$ with $\mathcal M$ being a sum over closed diagrams (here 3-cycles and 4-cycles) of the graph induces some interesting relations. Denote the operator
\beq
L_{ee'} = F^*_{ee'}\,\frac{\partial \phantom{F}}{\partial F^*_{ee'}}\;,
\ee
and consider the fact that $L_{ee'}$ acts on $\mathcal M$ by projecting it to only those cycles which contain $F^*_{ee'}$, which we denote $\mathcal M_{ee'}\equiv L_{ee'}\mathcal M(F^*)$. In the case of the 6j-symbol, each $F^*_{ee'}$ appears once in a 3-cycle and once in a 4-cycle (it can be read directly in \eqref{schwinger generating}. Choose two couples of links $(ab)$ and $(ij)$ which meet on the same or on two different vertices. Then,
\beq \label{simple eq}
\Bigl[ \mathcal M_{ab}\,L_{ij} - \mathcal M_{ij}\,L_{ab}\Bigr]\,\mathcal S(F^*) = 0\;.
\ee
This equation induces a recursion on the 6j-symbol, by expanding $\mathcal S$ as a power series and identifying the different orders. Its coefficients keep a simple form since we have only two first derivatives. However, the multiplication by $\mathcal M_{ab}$ and $\mathcal M_{ij}$ produces numerous shifts on the 6j-symbols (and we prefer not to write it, the above form as a differential equation being much nicer).

\subsection{Generalization of the new scalar Hamiltonians to arbitrary lattices with 3-valent vertices} \label{sec:gen}

It is also worth comparing the case of the triangular lattice which we have now studied in details with the case of a generic lattice. The scalar Hamiltonians, with spin 1 or 1/2, we have introduced keep the same form when working with an arbitrary lattice. Denote the links on the boundary of of a plaquette $p$ by $(e_1,\dotsc,e_n)$ and consider any vertex on it, say $v$ where $e_1, e_n$ meet, both ingoing, and take $v$ as the reference vertex for the holonomy $g_p$ around $p$. Then, the spin 1 Hamiltonian is
\beq
H^{(1)}_{v,p} = \tl X_n\cdot \bigl(\id - \Ad(g_p)\bigr)\,\tl X_1 = \tl X_n\cdot \tl X_1 - X_n\cdot \Ad(g_{n-1} \dotsm g_2)\,X_1\;,
\ee
where we have used the fact that $X_e$ and $\tl X_e$ are related by parallel transport along $e$. Similarly, it is natural to extend the spin 1/2 Hamiltonian to
\beq
H^{(1/2)}_{v,p} = \langle \tl z_n\vert \tl z_1]\ [\tl z_n\vert\,g_n\dotsm g_2 g_1\mone - \unit \vert \tl z_1\rangle = \sqrt{\frac{\langle \tl z_n\vert \tl z_n\rangle \langle \tl z_1\vert \tl z_1\rangle}{\langle z_n\vert z_n\rangle \langle z_1\vert z_1\rangle}}\,\langle \tl z_n\vert \tl z_1]\ [z_n\vert g_{n-1}\dotsm g_2\vert z_1\rangle - \langle \tl z_n\vert \tl z_1]\ [\tl z_n\vert \tl z_1\rangle\;.
\ee
For a fixed $p$, important differences with the plaquette operator described in the previous section are
\begin{itemize}
 \item there is a single plaquette operator $\chi_{\f12}(g_p)$, but $n$ Hamiltonians $H^{(1)}_{v,p}$, one for each node (this was a key point in the triangular case, to show that the Hamiltonians enable to get back to the flatness constraint in the section \ref{sec:back flatness}).
 \item each of them only affects $(n-2)$ spins, since the holonomies on the two edges meeting on the reference vertex do not enter the expression.
\end{itemize}
Quite clearly, evaluating the actions of these operators on spin network states on the identity will produce recursion relations for Wigner 3nj-symbols, which are solutions to the topological phase. Recursion for the 9j and 12j-symbols are known from textbooks \cite{varshalovich-book}, and recursions for 15-symbols have been recently found in \cite{recurrence-paper}. They were discovered using an invariance under a four-dimensional Pachner move of 4d triangulations. We strongly think that the above Hamiltonians provide us with a good tool to generically understand the whole set of recursions for arbitrary Wigner symbols (this is work in progress \cite{in prepa}). Among the remaining issues is the fact that we only expect three constraints to be independent (since the curvature vanishing has three real components), and it is not clear why this is the case from the point of view of recursion relations.

In the basis of coherent intertwiners used in the section \ref{sec:schwinger}, where there are no spins anymore, solutions to the topological phase should be given by generating functions of Wigner 3nj-symbols, i.e. in the Bargmann representation. These are holomorphic functions of $2L$ spinors, if $L$ is the number of links. These generating functions are known from \cite{labarthe} (see also more recently \cite{schnetz}), but we have not analyzed the differential equations they satisfy so far.

\section{Conclusion}

In this paper, we have given a very detailed account for the topological phase of the $\SU(2)$ BF model. While the continuum formulation is quite well-understood at a formal level (functional determinants in the path integral), a full lattice description which can be explicitly handled was missing. It is important because the BF model is at the core of recent developments concerning topological aspects in condensed matter, quantum information and background independent quantization such as loop quantum gravity.

Solutions are given by evaluations of spin networks, producing $\SU(2)$ Wigner coefficients. While that was expected, our approach, through the new Hamiltonian, goes deeper than previous works into the structure of the theory, by generating the basic recursions on Wigner 6j-symbols, completing this way the program of \cite{3d-wdw}.


In 2+1 dimensions, the holonomy-flux algebra representation of the new Hamiltonian emphasizes the geometric content of 2+1 gravity. As an outcome, we get a quantization of flat, Euclidean geometry in terms of recursions from group representation theory. We have given several kinds of such equations in the last section. Other equations which turn out to quantize simple geometric properties of flat space have appeared in \cite{yetanother, recurrence-paper}.

As already noticed in \cite{3d-wdw}, a surprising result is that there is not a single time reparametrization as expected from geometrodynamics, but instead different directions of evolutions which mix spatial diffeomorphisms and time reparametrization.

In this topological model, the long-standing issue of the ambiguity on the spin of the Wilson loops used to regularize the curvature in the Hamiltonian gets a precise answer. The fundamental representation is the only one which enables to solve the model for the group $\SU(2)$ without having to supplement it with additional initial conditions by hand.

The method we have used can be easily applied to other compact Lie groups. The cases of finite or Abelian Lie groups are even simpler. The point is that the Biedenharn-Elliott identity is a non-trivial relation which is part of the machinery of representation theory. It is always possible to get from it recursion relations like those we studied here, and they will always be generated by our new Hamiltonian. As we have argued, they are better suited than the recursions which come from the plaquette operators considered in \cite{levin-wen-condensation}, though part of the same mathematical framework.

Our Hamiltonian can also be used on plaquettes with an arbitrary number of links, generating this way recursion relations which are satisfied by generic Wigner symbols. Some of them are known from some kind of Biedenharn-Elliott identity satisfied by the 9j-symbol \cite{varshalovich-book}, but the generic case is actually new. It would definitely be interesting to study further those recursions, which certainly have a lot to do with integrability of Wigner coefficients as discussed in \cite{3nj-marzuoli}.

When written with spinors, coherent states quantizations are available which change the recursions to partial derivative equations on holomorphic functions. Some of the most interesting features of those representations are the existence of a closed formula for the generating function of 6j-symbols \eqref{schwinger generating} and the simple equation \eqref{simple eq} it satisfies. This is a special case of an analysis performed in \cite{labarthe, schnetz} for generic Wigner coefficients. The spinor variables have also been very useful in recent mathematics \cite{costantino-generating, Yu}. We hope in the future to apply these tools in the context of topological order and the dynamics of loop quantum gravity.

A key physical property which we have ignored here is the question of the excitations which violate the topological conditions. In the continuum, the BF model is known to feature exotic statistics \cite{bergeron-fractional-bf, baez-fractional-bf}, which depending on the dimension involve particles, strings and/or branes. The dynamics is presented in \cite{baez-perez-strings-bf,fairbairn-perez-strings-bf}. It would certainly be interesting to introduce them in our framework, and to compare with lattice models exhibiting similar phenomena (topological order and exotic statistics involving branes) like \cite{bombin-branes} where the model is exactly solvable. More generically, we think our model is robust enough to allow the study of theories expanded around the topological BF phase, as proposed in \cite{freidel-action-principle} and realized in the continuum, Yang-Mills theory in \cite{cattaneo-4dYM}.


\section*{Acknowledgements}

The authors also thank Laurent Freidel, who shared with us his ideas to extract the spin 1/2 recursions, which were basically the same as those we have used.

Research at Perimeter Institute is supported by the Government of Canada through Industry Canada and by the Province of Ontario through the Ministry of Research and Innovation.

\appendix
\section{Explicit formulae}

The explicit graspings
\begin{gather}
\widehat{F}_{12}\,\iota_{j_1 j_2 j_3}\vert 0\rangle \equiv [a_1\vert a_2\rangle\,\iota_{j_1 j_2 j_3}\vert 0\rangle = - \delta_{k_1, j_1-\f12} \delta_{k_2, j_2-\f12}\ \sqrt{\bigl(j_1+j_2+j_3+1\bigr)\bigl(j_1+j_2-j_3\bigr)}\ \iota_{k_1 k_2 j_3}\vert 0\rangle\;, \\
\widehat{E}_{12}\,\iota_{j_1 j_2 j_3}\vert 0\rangle \equiv \langle a_1\vert a_2\rangle\,\iota_{j_1 j_2 j_3}\vert 0\rangle = - \delta_{k_1, j_1+\f12} \delta_{k_2, j_2-\f12}\ \sqrt{\bigl(j_1-j_2+j_3+1\bigr)\bigl(-j_1+j_2+j_3\bigr)}\ \iota_{k_1 k_2 j_3}\vert 0\rangle\;, \\
\widehat{F}_{61}\,\iota_{j_1^* j_5^* j_6} \equiv [a_6\vert \tl a_1\rangle\,\iota_{j_1^* j_5^* j_6} = - \delta_{k_1,j_1+\f12} \delta_{k_6,j_6-\f12}\ \sqrt{\bigl(j_1+j_5-j_6+1\bigr)\bigl(-j_1+j_5+j_6\bigr)}\ \iota_{k_1^* j_5^* k_6}\;,\\
\widehat{F}_{16}^\dagger\,\iota_{j_1^* j_5^* j_6} \equiv \langle a_6\vert \tl{a}_1]\,\iota_{j_1^* j_5^* j_6} = \delta_{k_1,j_1-\f12} \delta_{k_6,j_6+\f12}\ \sqrt{\bigl(-j_1+j_5+j_6+1\bigr)\bigl(j_1+j_5-j_6\bigr)}\ \iota_{k_1^* j_5^* k_6}\;, \\
\widehat{E}_{61}\,\iota_{j_1^* j_5^* j_6} \equiv \langle a_6\vert \tl{a}_1\rangle \,\iota_{j_1^* j_5^* j_6} = \delta_{k_1,j_1+\f12} \delta_{k_6,j_6+\f12}\ \sqrt{\bigl(j_1+j_5+j_6+2\bigr)\bigl(j_1-j_5+j_6+1\bigr)}\ \iota_{k_1^* j_5^* k_6}\;, \\
\widehat{E}_{16}\,\iota_{j_1^* j_5^* j_6} \equiv \langle \tl a_1\vert a_6\rangle \,\iota_{j_1^* j_5^* j_6} = \delta_{k_1,j_1-\f12} \delta_{k_6,j_6-\f12}\ \sqrt{\bigl(j_1+j_5+j_6+1\bigr)\bigl(j_1-j_5+j_6\bigr)}\ \iota_{k_1^* j_5^* k_6}\;, \\
\widehat{F}_{62}\,\iota_{j_2^* j_6^* j_4} \equiv [\tl{a}_6\vert \tl{a}_2\rangle\,\iota_{j_2^* j_6^* j_4} = \delta_{k_2,j_2+\f12} \delta_{k_6,j_6+\f12}\ \sqrt{\bigl(j_2+j_6+j_4+2\bigr)\bigl(j_2+j_6-j_4+1\bigr)}\ \iota_{k_2^* k_6^* j_4}\;, \\
\widehat{F}_{26}^\dagger\,\iota_{j_2^* j_6^* j_4} \equiv \langle \tl{a}_6\vert \tl{a}_2]\,\iota_{j_2^* j_6^* j_4} = -\delta_{l_2,j_2-\f12} \delta_{l_6,j_6-\f12}\ \sqrt{\bigl(j_2+j_6+j_4+1\bigr)\bigl(j_2+j_6-j_4\bigr)}\ \iota_{l_2^* l_6^* j_4}\;.
\end{gather}

The spin 1/2 recursions,
\beq
\begin{aligned}
&\sqrt{\bigl(-j_1+j_5+j_6\bigr)\bigl(j_1+j_5-j_6+1\bigr)\bigl(-j_1+j_2+j_3\bigr)\bigl(j_1-j_2+j_3+1\bigr)} \begin{Bmatrix} j_1+\f12 &j_2-\f12 &j_3\\j_4 &j_5& j_6-\f12\end{Bmatrix} \\
&- d_{j_1} \sqrt{\bigl(j_2+j_6+j_4+1\bigr)\bigl(j_2+j_6-j_4\bigr)} \begin{Bmatrix} j_1 &j_2 &j_3\\j_4 &j_5& j_6\end{Bmatrix}\\
&- \sqrt{\bigl(j_1-j_5+j_6\bigr)\bigl(j_1+j_5+j_6+1\bigr)\bigl(j_1+j_2-j_3\bigr)\bigl(j_1+j_2+j_3+1\bigr)} \begin{Bmatrix} j_1-\f12 &j_2-\f12 &j_3\\j_4 &j_5& j_6-\f12\end{Bmatrix} = 0\;.
\end{aligned}
\ee

\beq \label{explicit rec++}
\begin{aligned}
&\sqrt{\bigl(j_1-j_5+j_6+1\bigr)\bigl(j_1+j_5+j_6+2\bigr)\bigl(j_1+j_2-j_3+1\bigr)\bigl(j_1+j_2+j_3+2\bigr)} \begin{Bmatrix} j_1+\f12 &j_2+\f12 &j_3\\j_4 &j_5& j_6+\f12\end{Bmatrix} \\
&+ d_{j_1} \sqrt{\bigl(j_2+j_6+j_4+2\bigr)\bigl(j_2+j_6-j_4+1\bigr)} \begin{Bmatrix} j_1 &j_2 &j_3\\j_4 &j_5& j_6\end{Bmatrix}\\
&- \sqrt{\bigl(j_1+j_5-j_6\bigr)\bigl(-j_1+j_5+j_6+1\bigr)\bigl(j_1-j_2+j_3\bigr)\bigl(-j_1+j_2+j_3+1\bigr)} \begin{Bmatrix} j_1-\f12 &j_2+\f12 &j_3\\j_4 &j_5& j_6+\f12\end{Bmatrix} = 0\;.
\end{aligned}
\ee

\beq
\begin{aligned}
&\sqrt{\bigl(j_1-j_5+j_6+1\bigr)\bigl(j_1+j_5+j_6+2\bigr)\bigl(-j_1+j_2+j_3\bigr)\bigl(j_1-j_2+j_3+1\bigr)} \begin{Bmatrix} j_1+\f12 &j_2-\f12 &j_3\\j_4 &j_5& j_6+\f12\end{Bmatrix} \\
&- d_{j_1} \sqrt{\bigl(j_2-j_6+j_4\bigr)\bigl(-j_2+j_6+j_4+1\bigr)} \begin{Bmatrix} j_1 &j_2 &j_3\\j_4 &j_5& j_6\end{Bmatrix}\\
&+ \sqrt{\bigl(j_1+j_5-j_6\bigr)\bigl(-j_1+j_5+j_6+1\bigr)\bigl(j_1+j_2-j_3\bigr)\bigl(j_1+j_2+j_3+1\bigr)} \begin{Bmatrix} j_1-\f12 &j_2-\f12 &j_3\\j_4 &j_5& j_6+\f12\end{Bmatrix} = 0\;.
\end{aligned}
\ee

\section{Direct proof of the Wheeler-DeWitt equation} \label{app:directproof}

We derive the difference equation \eqref{1/2coeffs} by direct evaluation in the spin network basis. Since $\widehat{F}_{26}^\dagger$ commutes with $\widehat{E}_{16} \widehat{F}_{12}+ \widehat{F}_{61}\widehat{E}_{12}$, we can use the result of the equation \eqref{E16 F12 F26dagger},
\begin{align}
\nonumber &\sum_{\{j_e\}} \bigl[\prod_{e=1}^6 d_{j_e}\bigr] \psi(j_e) \f1{d_{j_1}}\,\widehat{E}_{16}\,\widehat{F}_{12}\,\widehat{F}_{26}^\dagger\ s^{\{j_e\}}_{\rm tet} \\
\nonumber &=\, \sum_{\{j_e\}} d_{j_2}d_{j_2-\f12}\,d_{j_6}d_{j_6-\f12} \bigl[\prod_{e=1}^6 d_{j_e}\bigr] \psi(j_e)\ A_0^{(-,-)}(j_e)\,A_{-\f12}^{(-,-)}(j_e)\ s^{\{j_1-\f12, j_2-\f12,\dotsc,j_6-\f12\}}_{\rm tet}\;,\\
&=\, \sum_{\{j_e\}} d_{j_2}d_{j_2+\f12}\,d_{j_6}d_{j_6+\f12}\ \bigl[\prod_{e=1}^6 d_{j_e}\bigr]\ A_0^{(+,+)}(j_e) \ A_{-\f12}^{(+,+)}(j_e)
\psi\Bigl(j_1+\f12, j_2+\f12, \dotsc,j_6+\f12\Bigr)\ s^{\{j_1, j_2,\dotsc,j_6\}}_{\rm tet} \;.
\end{align}
On the second step, we have just re-indexed the sum, by $j_1\rightarrow j_1+1/2, j_2\rightarrow j_2+1/2, j_6\rightarrow j_6+1/2$, and we have used the following identities on the coefficients,
\begin{align}
\nonumber &A_0^{(\alpha, \beta)}(j_1, j_2-\alpha, \dotsc, j_6-\beta) = (-1)^{\alpha+\beta}\,A_0^{(-\alpha, -\beta)}(j_1,j_2,\dotsc,j_6)\;,\\
\nonumber &d_{j_1+\f12}\,A_{-\f12}^{(\alpha, \beta)}(j_1+\f12, j_2-\alpha, j_6-\beta) = d_{j_1}\, A_{+\f12}^{(-\alpha, -\beta)}(j_1,j_2, \dotsc,j_6)\;.
\end{align}
Obviously, we have a similar equation for the action of $\widehat{F}_{61}\widehat{E}_{12} \widehat{F}_{26}^\dagger$.

The key point is the action $\widehat{F}_{26}^\dagger \widehat{F}_{62}$ since we now have a different ordering. Basically, $\widehat{F}_{62}$ will shift the spins $j_2, j_6$ by $+1/2$ on the half-legs at the vertex $v$ (figure \ref{fig:tet}), and generate the coefficient $A_0^{(+,+)}$. Then, $\widehat{F}_{26}^\dagger$ lowers these spins so that they get back to their initial value. The difference with the calculations of the proposition \ref{prop:rec1/2} is thus that the coefficient we will extract involves shifts of $j_2, j_6$ by $+1/2$ instead of $-1/2$.

More precisely, we use the spin 1/2 graspings of the equations \eqref{F26dagger}, \eqref{F62} to get
\beq
-\widehat{F}_{26}^\dagger \widehat{F}_{62}\ \psi(g_1,\dotsc,g_6) = \sum_{\{j_e\}} d_{j_2}d_{j_2+\f12}\,d_{j_6}d_{j_6+\f12}\ \bigl[\prod_{e=1}^6 d_{j_e}\bigr]\ \bigl[ A_0^{(+,+)}(j_e)\bigr]^2\ \psi(j_e)\ s^{\{j_1, j_2,\dotsc,j_6\}}_{\rm tet} \;.
\ee
We can now sum the three contributions. They all have a common factor, $d_{j_2}d_{j_2+\f12}\,d_{j_6}d_{j_6+\f12}\ \bigl[\prod_{e=1}^6 d_{j_e}\bigr]  A_0^{(+,+)}(j_e)$,
\begin{multline}
\nonumber \widehat{H}^{(1/2)}_{e_6 e_1 e_2|L} \ \psi(g_1, \dotsc,g_6)
= \sum_{\{j_e\}} d_{j_2}d_{j_2+\f12}\,d_{j_6}d_{j_6+\f12}\ \bigl[\prod_{e=1}^6 d_{j_e}\bigr]\  A_0^{(+,+)}(j_e)\ s^{\{j_1, j_2,\dotsc,j_6\}}_{\rm tet}\\ \Bigl[A^{(+,+)}_{+\f12}\ \psi\Bigl(j_1+\f12,j_2+\f12,\dotsc,j_6+\f12\Bigr) \,+\, A^{(+,+)}_{-\f12}(j_1)\ \psi\Bigl(j_1-\f12, j_2+\f12,\dotsc,j_6+\f12\bigr)
+ A^{(+,+)}_{0}(j_1)\ \psi(j_1,\dotsc,j_6) \Bigr]\;.
\end{multline}
The last step is to project the above equation on an arbitrary spin network state, using their natural inner product. It is well-known that they are orthogonal, and given their normalization \eqref{ps}, we get
\begin{multline}
\langle s^{\{j_e\}}_{\rm tet}\vert\ \widehat{H}^{(1/2)}_{e_6 e_1 e_2|L}\ \vert \psi\rangle = A^{(+,+)}_{0}(j_1)\ \psi(j_1,\dotsc,j_6)\\
+A^{(+,+)}_{+\f12}\ \psi\Bigl(j_1+\f12,j_2+\f12,\dotsc,j_6+\f12\Bigr) \,+\, A^{(+,+)}_{-\f12}(j_1)\ \psi\Bigl(j_1-\f12, j_2+\f12,\dotsc,j_6+\f12\bigr)\;.
\end{multline}

\section{Additional representations} \label{app:addrep}

\subsection{Another generating function for 6j-symbols}

In this section, we use coherent intertwiners obtained from the coherent states for each Schwinger's boson. This enables to get rid of the spins, and to fully transform the recursion relation to a differential equation with respect to spinors only. For a spinor $z$, we introduce on the space of irreducible $\SU(2)$ representations the following state
\beq
\sum_{j\in\f\N2} \frac1{\sqrt{(2j)!}}\ \vert j,z\rangle = \sum_{j\in\f\N2} \sum_{m=-j}^j \f{(z^-)^{j-m}\ (z^+)^{j+m}}{\sqrt{(j-m)!\,(j+m)!}}\,\vert j,m\rangle\;.
\ee
It is an eigenstate of $a^A$ with eigenvalue $z^A$, and $a^{A\dagger}$ acts like $\partial_{z^A}$ as usual. Just like in the previsou section, coherent intertwiners are built by tensor products and group averaging of such states. Consider the node of the figure \ref{fig:tet} where $e_1, e_2, e_3$ meet and are all outgoing. The three half-lines carry spinors $(z_1, z_2, z_3)$, and the corresponding intertwiner is
\beq
i(z_1, z_2, z_3) \equiv \sum_{j_1,j_2,j_3} \frac1{\sqrt{(2j_1)!\,(2j_2)!\,(2j_3)!}}\ P_{j_1 j_2 j_3}(z_1,z_2,z_3)\ \iota_{j_1 j_2 j_3}\;.
\ee
On a spin network graph, we associate one such intertwiner to each node, and glue them using the holonomies on the links. This is the natural gluing, where all irreducible representations are orthogonal. This leads to the notion of {\bf spinor networks}, which we define for the moment as
\beq
s^{\{z_e,\tl z_e\}}(g_e) = \sum_{j_1,\dotsc,j_6} \prod_{e=1}^6\frac1{(2j_e)!}\ s^{\{j_e,z_e,\tl z_e\}}(g_e)\;,
\ee
where the coherent spin network $s^{\{j_e,z_e,\tl z_e\}}$ is defined in \eqref{cohspinnet}. Its evaluation on the unit gives
\beq \label{spinor symbol 1}
s(z_e,\tl z_e) \equiv s^{\{z_e,\tl z_e\}}(\unit) = \sum_{j_1,\dotsc,j_6} \Bigl[\prod_{e=1}^6\frac1{(2j_e)!}\Bigr]\ \left[\begin{Bmatrix} j_1 &j_2 &j_3 \\j_4 &j_5 &j_6\end{Bmatrix}\bigl(z_e, \tl z_e\bigr)\right]\;.
\ee

\begin{proposition} \label{prop:spinor symbol 1}
The Hamiltonian equation $\widehat{H}^{(1/2)}_{e_6 e_1 e_2|R} s^{\{z_e,\tl z_e\}}(g_e)_{|\unit} =0$ reads in this basis
\beq
\biggl[\biggl[\sum_{A,B = \pm1/2} \Bigl(\frac{\partial\phantom{z}}{\partial{\tl z_1^A}}\otimes z_6^A\Bigr)\,\Bigl(\frac{\partial\phantom{z}}{\partial{z_1^B}}\otimes z_2^B\Bigr)\biggr]\ + [\tl z_1\vert z_6\rangle\,[z_1\vert z_2\rangle -  \Bigl[\frac{\partial\phantom{z}}{\partial{\tl z_6}}\Big\vert \frac{\partial\phantom{z}}{\partial{\tl z_2}}\Big\rangle\ \Bigl(1+\sum_A z_1^A\,\frac{\partial\phantom{z}}{\partial z_1^A}\Bigr)\biggr] s(z_e,\tl z_e) = 0\;.
\ee
\end{proposition}

{\bf Proof of the proposition \ref{prop:spinor symbol 1}.} This equation is easily obtained from the proposition \ref{rec spin coh}. First correct the normalization to match that of the spinor symbol \eqref{spinor symbol 1}. This removes all spin dimension factors from the recursion \eqref{eq:rec spin coh} except $(2j_1+1)$ on the term which has no shift. Nevertheless, this factor is simply generated by the operator $(1+E_{11})$. In the coherent spinor basis, the latter is the operator $(1+\sum_A z_1^A\,\frac{\partial\phantom{z}}{\partial z_1^A})$. One then has just to sum over all spins to get the proposition.
\qed

\subsection{U(N) coherent states}

The fact that the operators $(\widehat{E}_{ee'})$ satisfy a $\u(N)$ algebra naturally leads to the introduction of $\U(N)$ coherent intertwiners \cite{U(N)coherent}
\begin{align}
\nonumber \iota_{J_{123}}(z_1,z_2,z_3) &= \sum_{j_1+j_2+j_3=J_{123}\in\N} \frac1{\sqrt{\prod_{e=1}^3 (2j_e)!}}\ \iota_{j_1 j_2 j_3}(z_1,z_2,z_3)\;,\\
&= \frac1{(2J_{123})!} \int_{\SU(2)} dh\ h\  \Bigl(\sum_{e=1}^3 \langle a_e\vert z_e\rangle\Bigr)^{2J_{123}} \ \otimes_e\vert 0\rangle_e\;. \label{U(N)intertwiner}
\end{align}
Those vectors correspond to the decomposition of the space of intertwiners with $N$ legs (above $N=3$) $\calH_N$ into sectors $\calH_N^{(J)}$ in which the sum of all spins meeting at the node is fixed to $\sum_ej_e = J\in\N$,
\beq
\calH_N = \bigoplus_{J\in\N} \calH_N^{(J)}\;,\quad \text{with} \qquad \calH_N^{(J)} = \bigoplus_{\sum_e j_e=J} \Inv\bigl[\otimes_e \calH_{j_e}\bigr]\;.
\ee
Those $\U(N)$ coherent states are eigenvectors of the total spin operator,
\beq
\Bigl(\sum_e \widehat{E}_{ee}\Bigr)\ \iota_{J_{v}}(z_e) = \sum_e \Bigl( \sum_A z^A_e \frac{\partial \phantom{z}}{\partial z^A_e}\Bigr)\ \iota_{J_{v}}(z_e) = 2J_v\,\ \iota_{J_{v}}(z_e)\;,
\ee
and are covariant with respect to the natural action of $\SL(N,\C)$ acting on the set of $N$ spinors.

Further, still assuming that all links are outgoing at $v$, we now that the operator $\widehat{E}_{ee'}$ exchanges a quantum $1/2$ between the leg $e$ and $e'$, hence the total spin is kept fixed. The operator $\widehat{F}_{ee'}$ (respectively $\widehat{F}_{ee'}^\dagger$) destroys (respectively creates) a spin $1/2$ on $e$ and $e'$, and thus changes the total spin by $-1/2$ (respectively $+1/2$),
\begin{align}
\widehat{E}_{12}\,\iota_{J_{123}}(z_1,z_2,z_3) &= \sum_A z_2^A\,\frac{\partial\phantom{z}}{\partial z_1^A}\,\iota_{J_{123}}(z_1,z_2,z_3)\;,\\
\widehat{F}_{12}\,\iota_{J_{123}}(z_1,z_2,z_3) &= [z_1\vert z_2\rangle\ \iota_{J_{123}-1}(z_1,z_2,z_3)\;,\\
\widehat{F}_{12}^\dagger\,\iota_{J_{123}}(z_1,z_2,z_3) &= \Bigl[\frac{\partial\phantom{z}}{\partial z_1}\Big\vert \frac{\partial\phantom{z}}{\partial z_2}\Big\rangle\ \iota_{J_{123}+1}(z_1,z_2,z_3)\;.
\end{align}

One can obviously glue these intertwiners via holonomies, and get a {\bf $\U(N)$ coherent 4J-symbol} by setting the holonomies to the identity,
\beq
\mathcal S(J_v;z_e,\tl z_e) = \frac1{\sqrt{\prod_v (J_v+1)!}} \sum_{\substack{j_1,\dotsc,j_6 \\ \sum_{e\supset v} j_e = J_v}} \frac1{\sqrt{\prod_{e,v} (J_v-2j_e)!}}\ \begin{Bmatrix} j_1 &j_2 &j_3 \\j_4 &j_5 &j_6\end{Bmatrix}\ \prod_v \prod_{e\supset v} [z_{e''}\vert z_{e'}\rangle^{J_v-2j_{e}} \;,
\ee
with the same notations as for the equation \eqref{spinor symbol 2}.

\begin{proposition}
The equation of the topological dynamics becomes in this basis
\begin{multline}
\biggl[\sum_{A,B = \pm1/2} \Bigl(\frac{\partial\phantom{z}}{\partial{\tl z_1^A}}\otimes z_6^A\Bigr)\,\Bigl(\frac{\partial\phantom{z}}{\partial{z_1^B}}\otimes z_2^B\Bigr)\biggr]\ \mathcal S(J_{123},J_{156},J_{264}-1,J_{345};z_e,\tl z_e)\\
+ [\tl z_1\vert z_6\rangle\,[z_1\vert z_2\rangle\ \mathcal S(J_{123}-1,J_{156}-1,J_{264}-1,J_{345};z_e,\tl z_e) -  \Bigl[\frac{\partial\phantom{z}}{\partial{\tl z_6}}\Big\vert \frac{\partial\phantom{z}}{\partial{\tl z_2}}\Big\rangle\ \Bigl(1+\sum_A z_1^A\,\frac{\partial\phantom{z}}{\partial z_1^A}\Bigr)\ \mathcal S(J_v;z_e,\tl z_e) = 0\;.
\end{multline}
\end{proposition}

The proof proceeds like in the previous sections, either by direct calculations from the Hamiltonian $\widehat{H}^{(--)}_{e_6 e_1 e_2|R}$ in terms of the operators $\widehat{E}, \widehat{F},\widehat{F}^\dagger$, or starting from the same equation in another basis such as \eqref{eq:rec spin coh}.

Let us describe the three terms of this equation. The first comes from that of the recursion formula for the 6j-symbol which shifts $j_1$ by $+1/2$, and  $j_2, j_6$ by $-1/2$, so that at the level of the total spin on each vertex, only $J_{264}$ is shifted. The second term is due to the term of the recursion which shifts $j_1, j_2, j_6$ all by $-1/2$. This affects three total spins, $J_{123}, J_{156}, J_{264}$. Finally, the last contribution is that without any shifts on the 6j-symbol arguments.

It is also possible to obtain the $\U(N)$ coherent 4J-symbol from the generating function of 6j-symbols $\mathcal S(z_e,\tl z_e)$. Indeed, introduce four real positive parameters $(t_v)$, one for each vertex, and substitute each spinor $z_e$ with $\sqrt{t_{s(e)}}z_e$ and $\tl z_e$ with $\sqrt{t_{t(e)}}\tl z_e$, where $s(e), t(e)$ are the source and target vertices of the link $e$. Then, we get the desired function by differentiating $\mathcal S(z_e,\tl z_e)$ $(J_v+1)$ times with respect to each $t_v$ before setting them to 0,
\beq
\mathcal S(J_v;z_e,\tl z_e) = \Bigr[\prod_v \frac1{(J_v+1)!J_v!}\,\frac{d^{J_v}\phantom{t}}{dt_v^{J_v}}\Bigr]\ \mathcal S\bigl(\sqrt{t_{s(e)}}\,z_e, \sqrt{t_{s(e)}}\,\tl z_e\bigr)\,\Big\vert_{t_v=0}\;.
\ee
To see that precisely, it is sufficient to focus on an intertwiner. From the definition \eqref{U(N)intertwiner} and the fact that the $\SU(2)$ coherent intertwiner $\iota_{j_e}(z_e)$ is a homogeneous polynomial of degree $2(\sum_e j_e) = 2J_v$, we get
\beq
\iota(\sqrt{t}z_1,\sqrt{t}z_2, \sqrt{t}z_3) = \sum_{K\in\N} (K+1)!\ t^K\ \iota_{K}(z_1,z_2,z_3)\;,
\ee
so that
\beq
\frac1{(J+1)!J!}\,\frac{d^{J}\phantom{t}}{dt^{J}}\ \iota(\sqrt{t}z_1,\sqrt{t}z_2, \sqrt{t}z_3)\,\big\vert_{t=0}  = \iota_{J}(z_1,z_2,z_3)\;.
\ee

Finally notice that the generating functions we have observed so far are actually power series in the classical variables $F^*_{ee'} = [z_{e'}\vert z_e\rangle$. This strengthens the idea of \cite{return-spinor} of quantizing the model with wave functions on the $F^*$ variables, i.e. representing the kinematical space of states on each $N$-valent node $\calH_N^{(J)}$ as
\beq
\calH_J^{(F^*)} \equiv \{ P\in \mathbbm P[F^*],\ \forall t\in\C\quad P(t\,F^*_{ee'}) = t^J\,P(F^*_{ee'}) \}\;,
\ee
the space of holomorphic, homogeneous polynomials of degree $J$ in the $F^*$ variables. It could be interesting to study this proposal deeper, and to rewrite the Hamiltonian in this language, similarly to the equation \eqref{diff S} obtained in the previous section on the generating function of 6j-symbols. However we will stop here the description of the Wheeler-DeWitt equation in different bases, and we will sketch different directions to generalize our present work.



\begin{thebibliography}{1}

\bibitem{wen-top-orders}
  X.~G.~Wen,
  ``Topological orders and edge excitations in fractional quantum Hall states,''
  Advances in Physics {\bf 44}, 405 (1995).

\bibitem{wen-niu-fqhe}
  X.~G.~Wen and Q.~Niu,
  ``Ground-state degeneracy of the fractional quantum Hall states in the presence of a random potential and on high-genus Riemann surfaces,''
  Phys.\ Rev.\  B {\bf 41}, 9377 (1990).

\bibitem{frohlich-kerler}
  J.~Frohlich and T.~Kerler,
  ``Universality in quantum Hall systems,''
  Nucl.\ Phys.\  B {\bf 354}, 369 (1991).

\bibitem{wen-spin-liquid91}
  X.~G.~Wen
  ``Mean-field theory of spin-liquid states with finite energy gap and topological orders,''
  Phys.\ Rev.\ B {\bf 44}, 2664 (1991).

\bibitem{kalmeyer-fqhe}
  V.~Kalmeyer and R.~B.~Laughlin,
  ``Equivalence of the resonating valence bond and fractional quantum Hall states,''
  Phys.\ Rev.\ Lett.\  {\bf 59}, 2095 (1987).

\bibitem{wen-wilczek-zee}
  X.~G.~Wen, F.~Wilczek and A.~Zee,
  ``Chiral spin states and superconductivity,''
  Phys.\ Rev.\  B {\bf 39}, 11413 (1989).

\bibitem{wen-spin-liquid02}
  X.~G.~Wen,
  ``Quantum orders and symmetric spin liquids,''
  Phys.\ Rev.\ B {\bf 65}, 165113 (2002).

\bibitem{hasan-colloquium}
  M.~Z.~Hasan, and C.~L.~Kane,
  ``Topological Insulators,''
  Rev.\ Mod.\ Phys.\ {\bf 82}, 3045 (2010).

\bibitem{cho-bf-insulators}
  G.~Y.~Cho and J.~E.~Moore,
  ``Topological BF field theory description of topological insulators,''
  Annals Phys.\  {\bf 326}, 1515 (2011)
  [arXiv:1011.3485 [cond-mat.str-el]].

\bibitem{hansson-superconductors}
  T.~H.~Hansson, V.~Oganesyan and S.~L.~Sondhi,
  ``Superconductors are topologically ordered,''
  Annals of Physics {\bf 313}, 497 (2004).

\bibitem{top-superconductors}
  M.~C.~Diamantini, P.~Sodano and C.~A.~Trugenberger,
  ``Superconductors with topological order,''
  Eur.\ Phys.\ J.\  B {\bf 53}, 19 (2006)
  [arXiv:hep-th/0511192].

\bibitem{horowitz-bf}
  G.~T.~Horowitz,
  ``Exactly Soluble Diffeomorphism Invariant Theories,''
  Commun.\ Math.\ Phys.\  {\bf 125}, 417 (1989).

\bibitem{blau-thompson-bf}
  M.~Blau and G.~Thompson,
  ``Topological Gauge Theories of Antisymmetric Tensor Fields,''
  Annals Phys.\  {\bf 205}, 130 (1991),\\
  M.~Blau and G.~Thompson,
  ``A new class of topological field theories and the Ray-Singer torsion,''
  Phys.\ Lett.\  B {\bf 228}, 64 (1989).

\bibitem{cattaneo-3d-4dBF}
  A.~S.~Cattaneo, P.~Cotta-Ramusino, J.~Frohlich and M.~Martellini,
  ``Topological BF theories in three-dimensions and four-dimensions,''
  J.\ Math.\ Phys.\  {\bf 36}, 6137 (1995)
  [arXiv:hep-th/9505027].

\bibitem{maggiore-sorella-perturbative-4dbf}
  N.~Maggiore and S.~P.~Sorella,
  ``Perturbation theory for antisymmetric tensor fields in four-dimensions,''
  Int.\ J.\ Mod.\ Phys.\  A {\bf 8}, 929 (1993)
  [arXiv:hep-th/9204044].

\bibitem{symmetries-bf-bv}
  F.~Gieres, J.~M.~Grimstrup, H.~Nieder, T.~Pisar and M.~Schweda,
  ``Symmetries of topological field theories in the BV framework,''
  Phys.\ Rev.\  D {\bf 66}, 025027 (2002)
  [arXiv:hep-th/0111258].

\bibitem{renormalisability-bf}
  C.~Lucchesi, O.~Piguet and S.~P.~Sorella,
  ``Renormalization and finiteness of topological BF theories,''
  Nucl.\ Phys.\  B {\bf 395}, 325 (1993)
  [arXiv:hep-th/9208047].

\bibitem{witten-2dym}
  E.~Witten,
  ``On quantum gauge theories in two-dimensions,''
  Commun.\ Math.\ Phys.\  {\bf 141}, 153 (1991).

\bibitem{cattaneo-4dYM}
  A.~S.~Cattaneo, P.~Cotta-Ramusino, F.~Fucito, M.~Martellini, M.~Rinaldi, A.~Tanzini and M.~Zeni,
  ``Four-dimensional Yang-Mills theory as a deformation of topological BF theory,''
  Commun.\ Math.\ Phys.\  {\bf 197}, 571 (1998)
  [arXiv:hep-th/9705123].

\bibitem{witten-3d}
  E.~Witten,
  ``(2+1)-Dimensional Gravity as an Exactly Soluble System,''
  Nucl.\ Phys.\  B {\bf 311}, 46 (1988).\\
  E.~Witten,
  ``Topology Changing Amplitudes in (2+1)-Dimensional Gravity,''
  Nucl.\ Phys.\  B {\bf 323}, 113 (1989).

\bibitem{freidel-depietri}
  R.~De Pietri, L.~Freidel,
  ``so(4) Plebanski action and relativistic spin foam model,''
  Class.\ Quant.\ Grav.\  {\bf 16}, 2187-2196 (1999).
  [gr-qc/9804071].

\bibitem{bf-mdm}
  L.~Freidel, A.~Starodubtsev,
  ``Quantum gravity in terms of topological observables,''
  [hep-th/0501191].


\bibitem{dennis-top-memory}
  E.~Dennis, A.~Kitaev, A.~Landahl and J.~Preskill,
  ``Topological quantum memory,''
  J.\ Math.\ Phys.\  {\bf 43}, 4452 (2002)
  [arXiv:quant-ph/0110143].


\bibitem{kitaev-code}
  A.~Y.~Kitaev,
  ``Fault tolerant quantum computation by anyons,''
  Annals Phys.\  {\bf 303}, 2 (2003)
  [arXiv:quant-ph/9707021].

\bibitem{baez-spinnets}
  J.~C.~Baez,
  ``Spin network states in gauge theory,''
  Adv.\ Math.\  {\bf 117}, 253 (1996)
  [arXiv:gr-qc/9411007].

\bibitem{baez-intro-bf}
  J.~C.~Baez,
  ``An Introduction to spin foam models of quantum gravity and BF theory,''
  Lect.\ Notes Phys.\  {\bf 543}, 25 (2000)
  [arXiv:gr-qc/9905087].

\bibitem{perez-intro-lqg}
  A.~Perez,
  ``Introduction to loop quantum gravity and spin foams,''
  arXiv:gr-qc/0409061.

\bibitem{ashtekar-status-report}
  A.~Ashtekar and J.~Lewandowski,
  ``Background independent quantum gravity: A Status report,''
  Class.\ Quant.\ Grav.\  {\bf 21}, R53 (2004)
  [arXiv:gr-qc/0404018].

\bibitem{levin-wen-condensation}
  M.~A.~Levin and X.~G.~Wen,
  ``String net condensation: A Physical mechanism for topological phases,''
  Phys.\ Rev.\  B {\bf 71}, 045110 (2005)
  [arXiv:cond-mat/0404617].

\bibitem{spinnets-marzuoli}
  V.~Aquilanti, A.~C.~P.~Bitencourt, C.~d.~S.~Ferreira, A.~Marzuoli and M.~Ragni,
  ``Quantum and semiclassical spin networks: From atomic and molecular physics
  to quantum computing and gravity,''
  Phys.\ Scripta {\bf 78}, 058103 (2008)
  [arXiv:0901.1074 [quant-ph]].

\bibitem{3d-wdw}
  V.~Bonzom and L.~Freidel,
  ``The Hamiltonian constraint in 3d Riemannian loop quantum gravity,''
  Class.\ Quant.\ Grav.\  {\bf 28}, 195006 (2011).
  [arXiv:1101.3524 [gr-qc]].

\bibitem{freidel-louapre-PR1}
  L.~Freidel and D.~Louapre,
  ``Ponzano-Regge model revisited I: Gauge fixing, observables and interacting
  spinning particles,''
  Class.\ Quant.\ Grav.\  {\bf 21}, 5685 (2004)
  [arXiv:hep-th/0401076].

\bibitem{barrett-naish}
  J.~W.~Barrett and I.~Naish-Guzman,
  ``The Ponzano-Regge model,''
  Class.\ Quant.\ Grav.\  {\bf 26}, 155014 (2009)
  [arXiv:0803.3319 [gr-qc]].

\bibitem{baby-sf}
  B.~Bahr, B.~Dittrich and J.~P.~Ryan,
  ``Spin foam models with finite groups,''
  arXiv:1103.6264 [gr-qc].

\bibitem{garoufalidis}
  S.~Garoufalidis, R.~van der Veen and w.~a.~a.~Zagier,
  ``Asymptotics of classical spin networks,''
  arXiv:0902.3113 [math.GT].

\bibitem{costantino-generating}
  F.~Costantino and J.~March\'e
  ``Generating series and asymptotics of classical spin networks,''
  arXiv:1103.5644 [math.GT].

\bibitem{qm6j}
  V.~Aquilanti, H.~M.~Haggard, A.~Hedeman, N.~Jeevanjee, R.~G.~Littlejohn and L.~Yu,
  ``Semiclassical Mechanics of the Wigner $6j$-Symbol,''
  arXiv:1009.2811 [math-ph].

\bibitem{roberts}
  J.~Roberts,
  ``Classical 6j-symbols and the tetrahedron,''
  Geom. Topol. {\bf 3} (1999), 21-6
  [arXiv:math-ph/9812013].

\bibitem{6jnlo}
  V.~Bonzom, E.~R.~Livine, M.~Smerlak and S.~Speziale,
  ``Towards the graviton from spinfoams: The Complete perturbative expansion of the 3d toy model,''
  Nucl.\ Phys.\  B {\bf 804}, 507 (2008)
  [arXiv:0802.3983 [gr-qc]].

\bibitem{pushing6j}
  M.~Dupuis and E.~R.~Livine,
  ``Pushing Further the Asymptotics of the 6j-symbol,''
  Phys.\ Rev.\  D {\bf 80}, 024035 (2009)
  [arXiv:0905.4188 [gr-qc]].

\bibitem{barrett-asym-summary}
  J.~W.~Barrett, R.~J.~Dowdall, W.~J.~Fairbairn, H.~Gomes, F.~Hellmann and R.~Pereira,
  ``Asymptotics of 4d spin foam models,'' (2010).
  arXiv:1003.1886 [gr-qc].

\bibitem{Yu}
  R.~G.~Littlejohn and L.~Yu,
  ``Semiclassical Analysis of the Wigner $9J$-Symbol with Small and Large
  Angular Momenta,''
  Phys.\ Rev.\  A {\bf 83}, 052114 (2011)
  [arXiv:1104.1499 [math-ph]].\\
  L.~Yu,
  ``Semiclassical Analysis of the Wigner $12J$-Symbol with One Small Angular Momentum: Part I,''
  arXiv:1104.3275 [math-ph].\\
  L.~Yu,
  ``Asymptotic Limits of the Wigner $15J$-Symbol with Small Quantum Numbers,''
  arXiv:1104.3641 [math-ph].

\bibitem{3nj-marzuoli}
  R.~W.~Anderson, V.~Aquilanti and A.~Marzuoli,
  ``3nj Morphogenesis and Semiclassical Disentangling,''
  J.\ Phys.\ Chem.\ {\bf A 113} (2009) 15106.
  arXiv:1001.4386 [quant-ph].

\bibitem{3njsmall}
  V.~Bonzom, P.~Fleury,
  ``Asymptotics of Wigner 3nj-symbols with Small and Large Angular Momenta: An Elementary Method,''
  [arXiv:1108.1569 [quant-ph]].

\bibitem{dowdall-handlebodies}
  R.~J.~Dowdall, H.~Gomes and F.~Hellmann,
  ``Asymptotic analysis of the Ponzano-Regge model for handlebodies,''
  J.\ Phys.\ A  {\bf 43}, 115203 (2010)
  [arXiv:0909.2027 [gr-qc]].

\bibitem{SG1}
  K.~Schulten and R.~G.~Gordon,
  ``SEMICLASSICAL APPROXIMATIONS TO 3J AND 6J COEFFICIENTS FOR QUANTUM
  MECHANICAL COUPLING OF ANGULAR MOMENTA,''
  J.\ Math.\ Phys.\  {\bf 16}, 1971 (1975).

\bibitem{6jmaite}
  M.~Dupuis and E.~R.~Livine,
  ``The 6j-symbol: Recursion, Correlations and Asymptotics,''
  Class.\ Quant.\ Grav.\  {\bf 27}, 135003 (2010)
  [arXiv:0910.2425 [gr-qc]].

\bibitem{yetanother}
  V.~Bonzom and E.~R.~Livine,
  ``Yet Another Recursion Relation for the 6j-Symbol,''
  arXiv:1103.3415 [gr-qc].

\bibitem{recursion-semiclass}
  V.~Bonzom,
  ``Spin foam models and the Wheeler-DeWitt equation for the quantum 4-simplex,''
  Phys.\ Rev.\  {\bf D84}, 024009 (2011).
  [arXiv:1101.1615 [gr-qc]].

\bibitem{recurrence-paper}
  V.~Bonzom, E.~R.~Livine and S.~Speziale,
  ``Recurrence relations for spin foam vertices,''
  Class.\ Quant.\ Grav.\  {\bf 27}, 125002 (2010)
  [arXiv:0911.2204 [gr-qc]].

\bibitem{varshalovich-book}
  D.~A.~Varshalovich, A.~N.~Moskalev and V.~K.~Khersonsky,
  ``QUANTUM THEORY OF ANGULAR MOMENTUM: IRREDUCIBLE TENSORS, SPHERICAL
  HARMONICS, VECTOR COUPLING COEFFICIENTS, 3NJ SYMBOLS,''
{\it  SINGAPORE, SINGAPORE: WORLD SCIENTIFIC (1988) 514p}

\bibitem{noui-perez-ps3d}
  K.~Noui and A.~Perez,
  ``Three-dimensional loop quantum gravity: Physical scalar product and spin
  foam models,''
  Class.\ Quant.\ Grav.\  {\bf 22}, 1739 (2005)
  [arXiv:gr-qc/0402110].

\bibitem{smolin-ultralocality}
  L.~Smolin,
  ``The Classical limit and the form of the Hamiltonian constraint in nonperturbative quantum general relativity,''
  arXiv:gr-qc/9609034.

\bibitem{fine-structure}
  L.~Freidel and E.~R.~Livine,
  ``The Fine Structure of SU(2) Intertwiners from U(N) Representations,''
  J.\ Math.\ Phys.\  {\bf 51}, 082502 (2010)
  [arXiv:0911.3553 [gr-qc]].

\bibitem{U(N)coherent}
  L.~Freidel and E.~R.~Livine,
  ``U(N) Coherent States for Loop Quantum Gravity,''
  J.\ Math.\ Phys.\  {\bf 52}, 052502 (2011)
  [arXiv:1005.2090 [gr-qc]].

\bibitem{return-spinor}
  E.~F.~Borja, L.~Freidel, I.~Garay and E.~R.~Livine,
  ``U(N) tools for Loop Quantum Gravity: The Return of the Spinor,''
  Class.\ Quant.\ Grav.\  {\bf 28}, 055005 (2011)
  [arXiv:1010.5451 [gr-qc]].

\bibitem{freidel-speziale-spinors}
  L.~Freidel and S.~Speziale,
  ``From twistors to twisted geometries,''
  Phys.\ Rev.\  D {\bf 82}, 084041 (2010)
  [arXiv:1006.0199 [gr-qc]].

\bibitem{johannes-spinor3}
  E.~R.~Livine, J.~Tambornino,
  ``Loop gravity in terms of spinors,''
  [arXiv:1109.3572 [gr-qc]].

\bibitem{johannes-spinor2}
  E.~R.~Livine, S.~Speziale, J.~Tambornino,
  ``Twistor Networks and Covariant Twisted Geometries,''
  [arXiv:1108.0369 [gr-qc]].

\bibitem{johannes-spinor1}
  E.~R.~Livine, J.~Tambornino,
  ``Spinor Representation for Loop Quantum Gravity,''
  [arXiv:1105.3385 [gr-qc]].

\bibitem{ooguri-3d}
  H.~Ooguri,
  ``Partition functions and topology changing amplitudes in the 3-D lattice gravity of Ponzano and Regge,''
  Nucl.\ Phys.\  B {\bf 382}, 276 (1992)
  [arXiv:hep-th/9112072].

\bibitem{twisted}
  V.~Bonzom and M.~Smerlak,
  ``Bubble divergences from twisted cohomology,''
  arXiv:1008.1476 [math-ph].

\bibitem{freidel-speziale-twisted-geom}
  L.~Freidel and S.~Speziale,
  ``Twisted geometries: A geometric parametrisation of SU(2) phase space,''
  Phys.\ Rev.\  D {\bf 82}, 084040 (2010)
  [arXiv:1001.2748 [gr-qc]].

\bibitem{mathese}
  V.~Bonzom,
  ``Geometrie quantique dans les mousses de spins: De la theorie topologique BF
  vers la relativite generale,''
  arXiv:1009.5100 [gr-qc].

\bibitem{bahr-broken-sym}
  B.~Bahr and B.~Dittrich,
  ``(Broken) Gauge Symmetries and Constraints in Regge Calculus,''
  Class.\ Quant.\ Grav.\  {\bf 26}, 225011 (2009)
  [arXiv:0905.1670 [gr-qc]].

\bibitem{dittrich-ryan1}
  B.~Dittrich and J.~P. Ryan,
  ``Phase space descriptions for simplicial 4d geometries,''
  Class.\ Quant.\ Grav.\  {\bf 28}, 065006 (2011)
  [arXiv:0807.2806 [gr-qc]].

\bibitem{in prepa}
  V.~Bonzom and P.~Fleury,
  ``A geometric approach to the evaluation of classical spin networks,''
  in preparation.

\bibitem{livine-speziale-CS}
  E.~R.~Livine and S.~Speziale,
  ``A New spinfoam vertex for quantum gravity,''
  Phys.\ Rev.\  D {\bf 76}, 084028 (2007)
  [arXiv:0705.0674 [gr-qc]].



\bibitem{conrady-closure}
  F.~Conrady and L.~Freidel,
  ``Quantum geometry from phase space reduction,''
  J.\ Math.\ Phys.\  {\bf 50}, 123510 (2009)
  [arXiv:0902.0351 [gr-qc]].

\bibitem{etera-factor}
  L.~Freidel, K.~Krasnov and E.~R.~Livine,
  ``Holomorphic Factorization for a Quantum Tetrahedron,''
  Commun.\ Math.\ Phys.\  {\bf 297}, 45 (2010)
  [arXiv:0905.3627 [hep-th]].

\bibitem{wu}
  A.C.T.~Wu,
  ``Structure of the Wigner 9j Coefficients in the Bargmann Approach,''
  J.\ Math.\ Phys.\ {\bf 13}, 84--90 (1972).

\bibitem{schwinger}
  J.~Schwinger,
  ``On angular momentum,'' Report US AEC NYO-3071 (1952),
  in Quantum Theory of Angular Momentum, eds. LC Biedenharn and H. van Dam (New York: Academic, 1965).

\bibitem{labarthe}
  J.~J.~Labarthe,
  ``Generating Functions for the Coupling Recoupling Coefficients of SU(2),''
  J.\ Phys.\ A  {\bf 8}, 1543 (1975).

\bibitem{schnetz}
  O.~Schnetz,
  ``Generating Functions for Multi-j-Symbols,''
  [arXiv:math-ph/9805027].

\bibitem{bergeron-fractional-bf}
  M.~Bergeron, G.~W.~Semenoff and R.~J.~Szabo,
  ``Canonical bf type topological field theory and fractional statistics of strings,''
  Nucl.\ Phys.\  B {\bf 437}, 695 (1995)
  [arXiv:hep-th/9407020].

\bibitem{baez-fractional-bf}
  J.~C.~Baez, D.~K.~Wise and A.~S.~Crans,
  ``Exotic statistics for strings in 4d BF theory,''
  Adv.\ Theor.\ Math.\ Phys.\  {\bf 11}, 707 (2007)
  [arXiv:gr-qc/0603085].

\bibitem{baez-perez-strings-bf}
  J.~C.~Baez and A.~Perez,
  ``Quantization of strings and branes coupled to BF theory,''
  Adv.\ Theor.\ Math.\ Phys.\  {\bf 11}, 3 (2007)
  [arXiv:gr-qc/0605087].

\bibitem{fairbairn-perez-strings-bf}
  W.~J.~Fairbairn, A.~Perez,
  ``Extended matter coupled to BF theory,''
  Phys.\ Rev.\  {\bf D78}, 024013 (2008).
  [arXiv:0709.4235 [gr-qc]].

\bibitem{bombin-branes}
  H.~Bombin and M.~A.~Martin-Delgado,
  ``Exact Topological Quantum Order in D=3 and Beyond: Branyons and Brane-Net Condensates,''
  Phys.\ Rev.\  B {\bf 75}, 075103 (2007)
  [arXiv:cond-mat/0607736].

\bibitem{freidel-action-principle}
  L.~Freidel and K.~Krasnov,
  ``Spin foam models and the classical action principle,''
  Adv.\ Theor.\ Math.\ Phys.\  {\bf 2}, 1183 (1999)
  [arXiv:hep-th/9807092].

\end{thebibliography}




\end{document}